\documentclass[aps,prb,floatfix,reprint,showpacs,superscriptaddress,longbibliography]{revtex4-2}
\usepackage{mathtools,amssymb,graphicx}
\usepackage[plainpages=false,pdfpagelabels,colorlinks=true,linkcolor=red,urlcolor=blue,citecolor=blue,pdftitle={Title},pdfauthor={},pdfdisplaydoctitle=true,pdfduplex=DuplexFlipLongEdge]{hyperref}
\usepackage{soul} 
\usepackage[dvipsnames,usenames]{color}
\usepackage[normalem]{ulem}
\usepackage{graphicx}
\usepackage{booktabs}
\usepackage{xcolor}
\usepackage{bbold, verbatim} 
\usepackage{float}
\usepackage{subcaption}
\emergencystretch=\maxdimen
\usepackage[colorinlistoftodos,prependcaption,textsize=tiny]{todonotes}
\usepackage{cprotect}
\usepackage{siunitx}
\graphicspath{ {./images/} }

\begin{document}
\title{Principal deuterium Hugoniot via Quantum Monte Carlo and $\Delta$-learning}

\author{Giacomo Tenti} 
\email{gtenti@sissa.it}
\affiliation{International School for Advanced Studies (SISSA),
Via Bonomea 265, 34136 Trieste, Italy}
\author{Kousuke Nakano} 
\email{kousuke_1123@icloud.com}
\affiliation{Center for Basic Research on Materials, National Institute for Materials Science (NIMS), Tsukuba, Ibaraki 305-0047, Japan}
\author{Andrea Tirelli} 
\affiliation{International School for Advanced Studies (SISSA),
Via Bonomea 265, 34136 Trieste, Italy}

\author{Sandro Sorella} 

\affiliation{International School for Advanced Studies (SISSA),
Via Bonomea 265, 34136 Trieste, Italy}

\author{Michele Casula}
\affiliation{Institut de Minéralogie, de Physique des Matériaux et de Cosmochimie (IMPMC), Sorbonne Université, CNRS UMR 7590, MNHN, 4 Place Jussieu, 75252 Paris, France}

\begin{abstract}
We present a study of the principal deuterium Hugoniot for pressures up to $150$ GPa, using Machine Learning potentials (MLPs) trained with Quantum Monte Carlo (QMC) energies, forces and pressures. In particular, we adopted a recently proposed workflow based on the combination of Gaussian kernel regression and $\Delta$-learning.
By fully taking advantage of this method, we explicitly considered finite-temperature electrons in the dynamics, whose effects are highly relevant for temperatures above $10$ kK.
The Hugoniot curve obtained by our MLPs shows a good agreement with the most recent experiments, particularly in the region below 60 GPa. At larger pressures, our Hugoniot curve is slightly more compressible than the one yielded by experiments, whose uncertainties generally increase, however, with pressure. 
Our work demonstrates that QMC can be successfully combined with $\Delta$-learning to deploy reliable MLPs for complex extended systems across different thermodynamic conditions, by keeping the QMC precision at the computational cost of a mean-field calculation. 
\end{abstract}
\date{\today}
\maketitle

\section{Introduction}
The study of hydrogen under extreme conditions has been a very active topic in condensed matter physics. Hydrogen is the most abundant element in the universe and the accurate knowledge of its phase diagram at pressures of the order of hundreds of GPa is extremely important for a variety of applications, such as modeling the interior of stars and giant gas planets \cite{Saumon1995,Fortney2009,Miguel2016}, the
inertial-confinement fusion \cite{Hu2015}, and 
the high-$T_c$ hydrogen-based superconductors \cite{Drozdov2015,Somayazulu2019}. Nevertheless, several properties of this system are still highly debated, even at the qualitative level \cite{McMahon2012,Cheng2020,Karasiev2021,Cheng2021}. 
  
One of the main reasons that hamper our full understanding of high-pressure hydrogen is the difficulty of reproducing extreme pressures in a laboratory. Typical shock-wave experiments \cite{Nellis2006} make use of accelerated flyer plates to compress a material sample in a very short time, thus allowing the study of specimens at high temperatures and pressures. In particular, the set of possible end-states that the system can reach from some given initial conditions, also named the \emph{principal Hugoniot}, must satisfy a set of equations, known as the Rankine-Hugoniot (RH) relations \cite{Duvall1977}, linking the thermodynamic properties of the final shocked state with those of the starting one. 
During the years, the principal deuterium Hugoniot has been measured for a wide range of pressures and with a great degree of accuracy \cite{Nellis1983,Knudson2001,Boriskov2005,Knudson2004,Hicks2009,Loubeyre2012,Knudson2017,Fernandez2019}, particularly in the lower pressure ($\lesssim$ 60 GPa) region, where the relative error on the density is as small as $2\%$ in recent experiments.

In this context, numerical approaches, in particular \emph{Ab Initio} Molecular Dynamics (AIMD) simulations, are extremely valuable, since they are not constrained by any experimental setup and can thus give further insight into this part of the phase diagram \cite{Knudson2021}. The Hugoniot region is particularly important because of the availability of experimental data that can be used to benchmark different 
theoretical methods. Among them, Density Functional Theory (DFT) has been extensively applied to compute the Hugoniot curve \cite{Lenosky2000,Galli2000,Bagnier2000,Bonev2004,Holst2008,Caillabet2011,Karasiev2019}. In this framework, the approximations behind the particular exchange-correlation functional often produce discrepancies across existing DFT schemes, whose accuracy varies according to the thermodynamic conditions. Quantum Monte Carlo (QMC) simulations, which depend on more controllable approximations, have also been performed \cite{Tubman2015,Ruggeri2020}. 
Although in principle more accurate and systematically improvable, these calculations have a much larger computational cost than DFT, and they are thus limited in system size and simulation length.
Moreover, previous QMC calculations \cite{Tubman2015} seem to give results for the principal Hugoniot in disagreement with the most recent experimental data, with the possible origin of this discrepancy being  recently debated \cite{Clay2019}. 

To overcome the large computational cost of \emph{ab initio} simulations, machine learning techniques, aimed at constructing accurate potential energy surfaces, have become increasingly popular. Within this approach, one uses a dataset of configurations, i.e. the \emph{training set}, to build a machine learning potential (MLP) that is able to reproduce energies and forces calculated with the given target method \cite{Behler2007}. Unlike DFT MLPs, the QMC ones are relatively less common, given the larger computational cost and the consequent difficulty of generating large datasets, usually necessary to construct accurate MLPs. 

In this work, we have built an accurate MLP using QMC energies, forces and pressures in the region of the deuterium Hugoniot, using the so-called $\Delta$-learning approach.
The Hugoniot curve computed by our QMC-MLP shows an excellent agreement with the most recent experiments at low density, while presenting a slightly larger compressibility for temperatures above $10$~kK, where the experimental points are also affected by larger error bars.

\begin{figure*}[!bht]
    \centering
    \begin{subfigure}{.45\textwidth}
  \centering
  \includegraphics[width=1.0\linewidth]{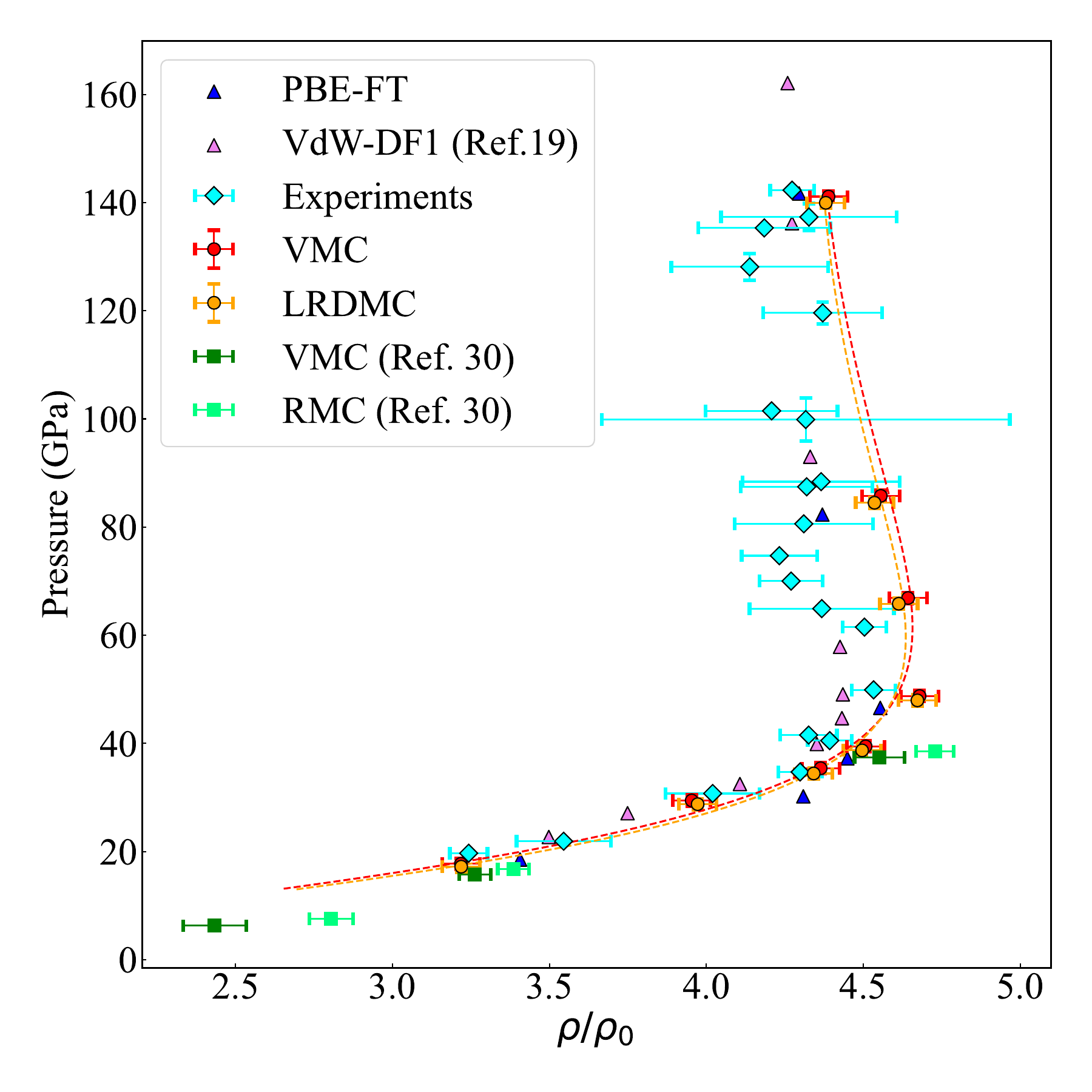}
  \caption{}
  \label{subfig:Hugoniot rho-p}
\end{subfigure}
    \begin{subfigure}{.45\textwidth}
  \centering
  \includegraphics[width=1.0\linewidth]{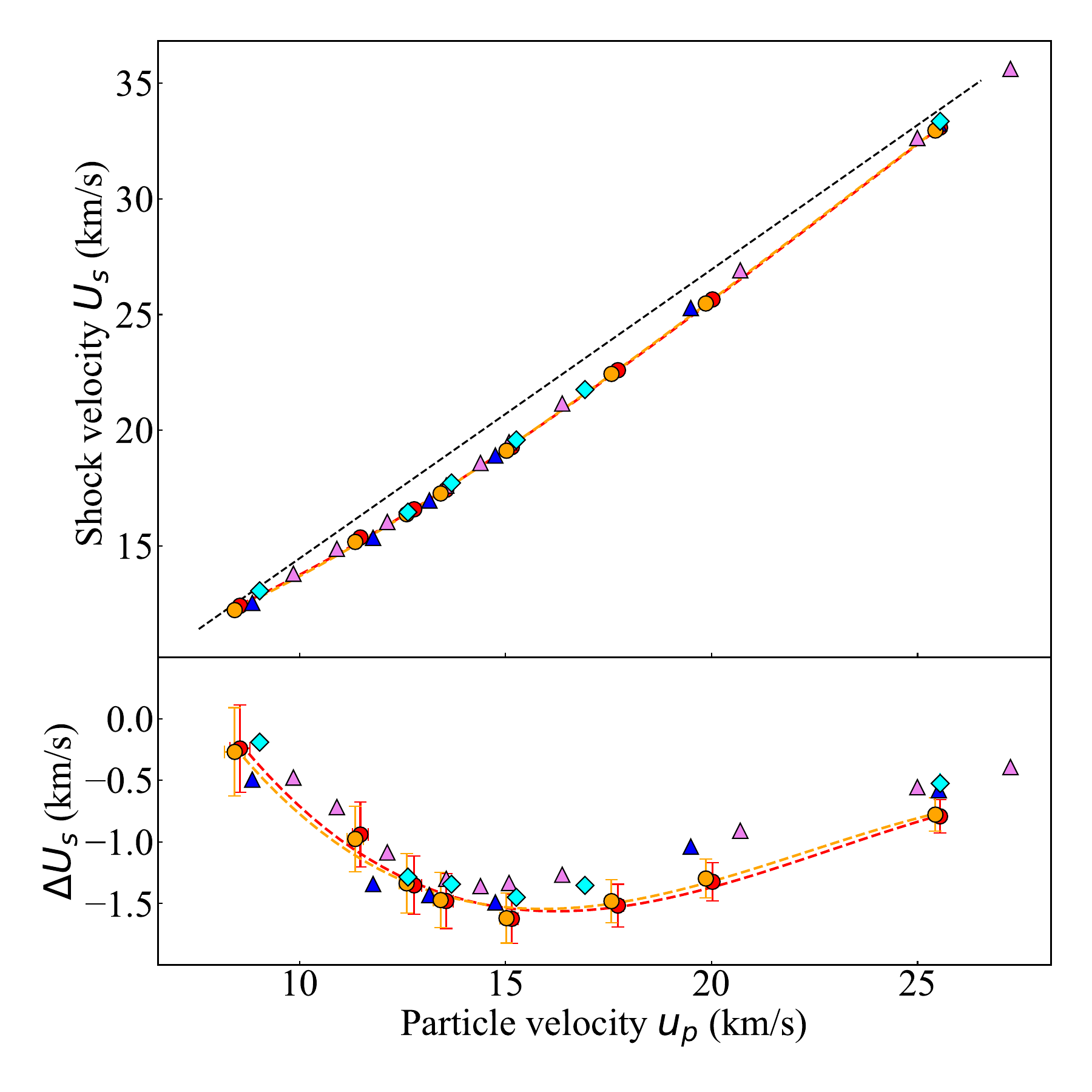}
  \caption{}
  \label{subfig: Hugoniot shock velocity}
\end{subfigure}
\caption{(\ref{subfig:Hugoniot rho-p}) 
\small{
Principal Hugoniot in the density-pressure space. Red and yellow circles are the results obtained with our MLPs trained on VMC and LRDMC datapoints, respectively, and a PBE baseline. Blue and pink triangles are the PBE result calculated in this work and the VdW-DF1 result of Ref.~\onlinecite{Knudson2017} respectively. CEIMC results of Ref.~\onlinecite{Ruggeri2020} based on Variational Monte Carlo (VMC) and Reptation Monte Carlo (RMC) are reported in green squares. Cyan diamonds are the experimental results of Refs.~\onlinecite{Knudson2004,Knudson2017,Fernandez2019}. Dashed lines are guides for the eye. (\ref{subfig: Hugoniot shock velocity}) [top panel] Hugoniot in the $u_p$--$U_s$ space. Black-dashed line is the re-analyzed gas-gun fit reported in Ref.~\onlinecite{Knudson2017}, i.e., the shock velocity extrapolated from measures of molecular deuterium at low pressure  \cite{Nellis1983}. [bottom panel] Relative shock velocity with respect to the gas-gun fit. Only the experimental points of Ref.~\onlinecite{Knudson2017} are reported.
}
}
\label{fig: Hugoniot}
\end{figure*}

{\vspace{3mm}}
\section{Computational details}
In order to build an MLP with QMC references,
we employed a combination of Gaussian Kernel Regression (GKR), Smooth Overlap of Atomic Positions (SOAP) descriptors \cite{De2016}, and $\Delta$-learning.
The same approach has been recently proposed in Ref.~\onlinecite{Tirelli2022}, where it was applied to the study of high-pressure hydrogen in similar thermodynamic conditions. Following the $\Delta$-learning approach, an MLP is trained on the difference between the target method and a usually much cheaper \emph{baseline} potential.
Here, we trained several MLPs, using Variational Monte Carlo (VMC) and Lattice Regularized Diffusion Monte Carlo (LRDMC) \cite{Casula2005, Nakano2020b} datapoints as targets, and two different DFT baselines, with the Perdew-Zunger Local Density Approximation (PZ-LDA) {\cite{Perdew1981}} and the Perdew-Burke-Ernzerhof (PBE) \cite{Perdew1996} functionals.

To determine the principal Hugoniot, we made use of the RH jump equation:
\begin{align}
    H(\rho,T) = e(\rho,T) - e_0 + \frac{1}{2} (\rho^{-1} - \rho_0^{-1})  \left[ p(\rho,T) + p_0 \right] = 0, \label{eq: Rankine-Hugoniot 1}
\end{align}
where $\rho$, $T$, $e(\rho,T)$, $p(\rho,T)$ and $\rho_0$, $T_0$, $e_0$, $p_0$ are the density, temperature, energy per particle and pressure of the final and initial states, respectively.
In particular, we ran a first set of $NVT$ simulations for a system of $N=128$ atoms at several temperatures in the [4~kK : 8~kK] range, and Wigner-Seitz radii between $1.80$ Bohr and $2.24$ Bohr, corresponding to the range where the zero of $H(\rho,T)$ was expected.  These simulations were performed considering classical nuclei and ground-state electrons, as quantum corrections and thermal effects have been shown to be negligible for these temperatures \cite{Ruggeri2020}.  At each step, the energy, forces and pressure were calculated using the {\textsc{Quantum Espresso}} package in its GPU accelerated version \cite{Giannozzi2009,Giannozzi2017,Giannozzi2020} with the chosen functional (PBE in most cases), and then corrected with our MLP trained on the difference between QMC and DFT data. The resulting dynamics has the same efficiency as a standard DFT AIMD simulation, which is roughly 100 times faster than the original QMC one.  
For the DFT simulations, we considered a 60~Ry plane-wave cutoff with a Projector Augmented Wave (PAW) pseudopotential \cite{PAWpseudo} and a $4\times 4\times4$ Monkhorst-Pack $k$-point grid, while for the dynamics we used a time step of 0.25~fs and a Langevin thermostat \cite{Ricci2003,Attaccalite2008} with damping $\gamma$ = 13~ps$^{-1}$. 
For each temperature, the Hugoniot $(\rho^*, p^*)$ coordinates are determined by fitting the Hugoniot function $H(\rho,T)$ and the pressure $p(\rho, T)$ with a spline function, and by numerically finding $\rho^*$ and the corresponding $p^*$.  

The QMC calculations were performed using the {\textsc{TurboRVB}} package~{\cite{Nakano2020, 2023turbogenius}}.
When generating our QMC-MLP model, we took particular care to
the training set construction, based on energies, forces and pressures all computed at the QMC (VMC or LRDMC) level.
In general, QMC forces and pressure are not guaranteed to be consistent with the relative potential energy surface, due to the practical difficulty of optimizing all the parameters within a given ansatz for the VMC wave function (WF). This is often called the self-consistency error \cite{Tiihonen2021,Nakano2022,Nakano2023}. Even if ML frameworks generally satisfy the consistency property by construction, the presence of biased forces and pressure in the training can spoil the accuracy of the model and produce, in principle, unintended results. Therefore, to avoid these issues, we mitigated
the self-consistency error by directly optimizing not only the Jastrow factor but also the determinantal part of the VMC WF.
The details of our QMC simulations are reported in the Supplemental Material (SM)~\cite{SM}.

Within our approach, we can fully take advantage of the $\Delta$-learning method by estimating the effect of thermalized electrons in our calculations. 
To do so, we ran simulations at temperatures $T = 10$~kK, $15$~kK, $20$~kK and $35$~kK considering the effect of finite $T$ in the underlying PBE energies and forces.
In this way, we can include the effects of thermally excited electrons in our MLP without changing it, at least at the DFT level of theory.
Here, the effect of the explicit dependence of the DFT functional on $T$ was not considered, since it has been shown to be negligible for hydrogen systems and the temperature range analyzed here \cite{Karasiev2014,Karasiev2018,Karasiev2019}.
We remark that this approach is also applicable when a DFT-MLP is used as a baseline in place of an ab initio calculation, where finite temperature effects can be estimated from the DFT density of states \cite{Mahmoud2022}.

{\vspace{3mm}}
\section{Results and Discussion}
Fig.~\ref{subfig:Hugoniot rho-p} shows our results together with several experimental values for pressures below $150$ GPa \cite{Knudson2004,Knudson2017,Fernandez2019}. We also report the principal Hugoniot obtained by directly using the PBE baseline and the Coupled Electron Ion Monte Carlo (CEIMC) results of Ref.~\onlinecite{Ruggeri2020} for comparison.  
For temperatures larger than $T = 10$~kK the results refer to the MDs obtained using finite temperature DFT as a baseline.
Both the VMC and LRDMC models give a very similar Hugoniot line, well reproducing the experimental points in the low density - low pressure region.  With respect to the most accurate data of Ref.~\onlinecite{Knudson2017}, our estimate of the relative density $\rho / \rho_0$ at the compressibility peak is $\sim 3-4\%$ larger, still within the error bars. For larger pressures, we predict a Hugoniot mostly compatible within the experiments but systematically more compressible. However, in this regime the correspondingly larger uncertainties in the measures prevent a clear-cut assessment of our outcome.
Our results also agree with the low-temperature CEIMC ones reported in Ref.~\onlinecite{Ruggeri2020} within the statistical accuracy.

Fig.~\ref{subfig: Hugoniot shock velocity} displays the same points in the $u_p-U_s$ space, where $u_p$ is the particle velocity and $U_s$ is the shock velocity, the two being calculated using the following RH relations: 
\begin{align*}
    u_p = \sqrt{(p + p_0)(\rho_0^{-1} - \rho^{-1})}, \\
    U_s = \rho_0^{-1} \sqrt{\frac{p + p_0}{\rho_0^{-1} - \rho^{-1}}}.
\end{align*}
The difference $\Delta U_s$ between these points and the linear fit on the gas-gun data re-analyzed in Ref.~\onlinecite{Knudson2017} is also shown (bottom panel of Fig.~\ref{subfig: Hugoniot shock velocity}). Notice that the drop in the slope of $U_s$ relative to $u_p$ coincides with the onset of the molecular-atomic (MA) transition, while the magnitude of the $\Delta U_s$ minimum relates to the position of the relative compression peak. In particular, the PBE Hugoniot curve manifests a premature start of the dissociation, while it predicts correctly the magnitude of the compressibility maximum. 
Our QMC results correctly predict the position of the peak and starting slope, while showing some discrepancies for $u_p \gtrsim  10 $~km/s with respect to the data of Ref.~\onlinecite{Knudson2017}. In this regime, DFT, and in particular the result obtained with the VdW-DF1 functional \cite{Dion2004,Berland2015}, seems to be in better agreement with experiments, thanks to a favorable error cancellation in the Hugoniot~\cite{Clay2019}. We noticed how, in this study, the discrepancy with the experiments is much milder than the value reported by previous QMC calculations at densities and pressures close to the compressibility peak \cite{Tubman2015}. This can be due to the explicit optimization of the WF nodal surface provided by our WF ansatz, which reduces the fixed node error mentioned in Ref.~\onlinecite{Clay2019}, the only approximation left in any projective Monte Carlo calculation, such as LRDMC and RMC.
The difference between the various methods is also apparent in their equations of state, reported in the SM~\cite{SM}.

\begin{figure}[!htb]
    \centering
    \includegraphics[width=1.0\linewidth]{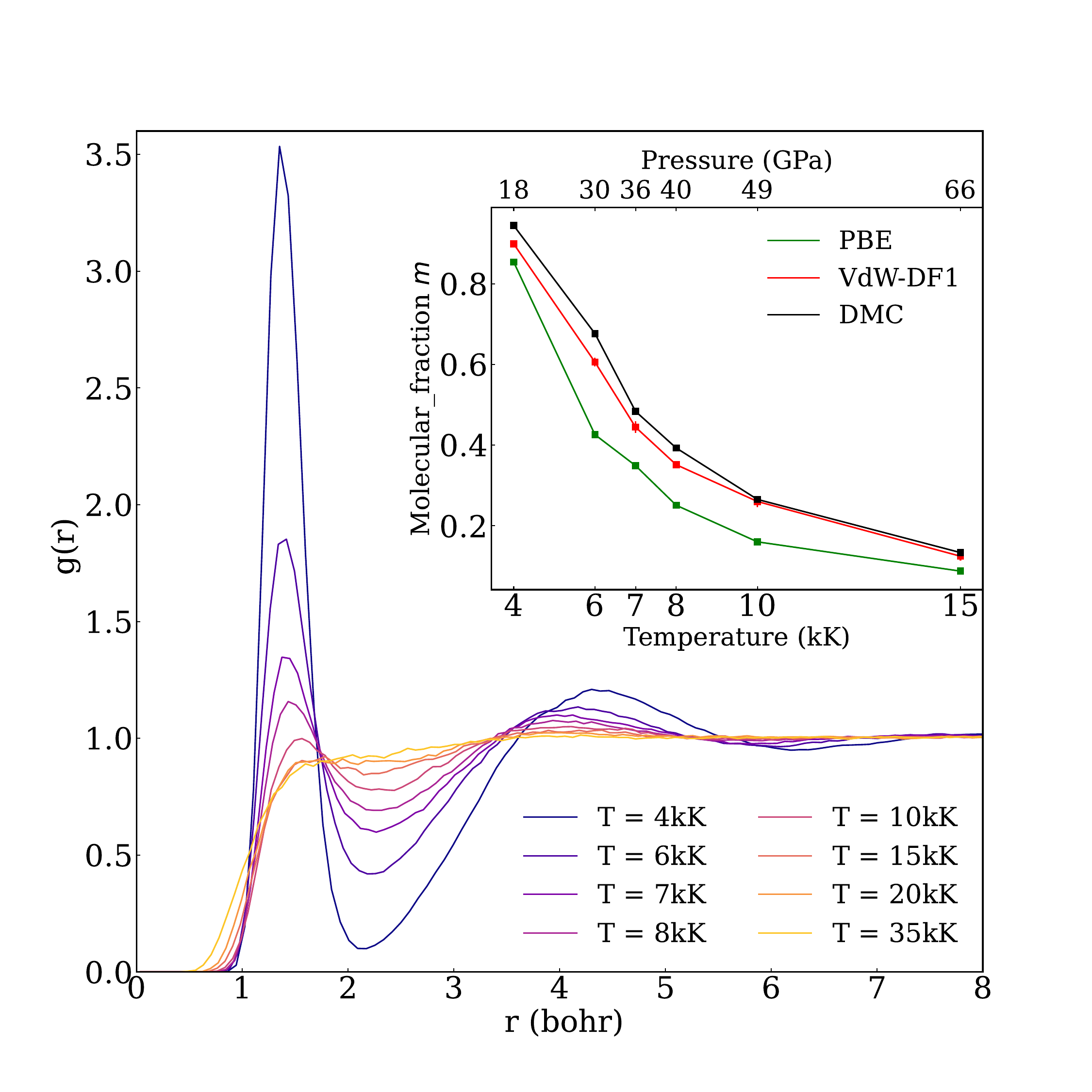}
    \caption{
    \small{
    $g(r)$ for several temperatures and densities close to the principal Hugoniot, obtained using the LRDMC model. The molecular fraction value, $m$, is reported in the inset for each value of temperature up to $15$kK. On the top axis the corresponding pressure at the Hugoniot is also shown. The values obtained with an ab initio DFT dynamics using the PBE and VdW-DF1 functionals are reported for comparison at the same temperatures (notice that a pressure and density mismatch between methods can be present in this case due to different equations of state).}}
    
    \label{fig: g(r) vs temperature}
\end{figure}

The presence of an MA transition is investigated in Fig.~\ref{fig: g(r) vs temperature}, where we report the radial distribution function, $g(r)$, calculated on trajectories obtained with the LRDMC model for several temperatures at densities close to the Hugoniot curve. The inset of Fig.~\ref{fig: g(r) vs temperature} displays the value of the molecular fraction $m$, defined as the percentage of atoms that stay within a distance of $2$ Bohr (roughly corresponding to the first $g(r)$ minimum after the molecular peak) from another particle for longer than a characteristic time, here set to $6$ fs. 
The results indicate a distinct atomic character for $T \geq 10$~kK and a clear molecular peak at lower temperatures.  
The LRDMC model shows slower decay of the molecular fraction with temperature than the PBE and VdW-DF1 ones, being compatible with the latter for temperatures above $10$~kK. 

{\vspace{3mm}}
\section{Error analysis}
To assess the quality of our principal Hugoniot determination, we analyzed the possible sources of errors in relation to our machine learning scheme. There are three main sources of errors: the uncertainties in the fit of $H(\rho,T)$, the prediction error of the MLP, and the uncertainties in the reference state energy estimate, i.e. $e_0$ in Eq.\eqref{eq: Rankine-Hugoniot 1}. We verified that, in our case, the error produced by the fit is negligible compared to the other two sources, which we will discuss next. 

As mentioned before, we followed Ref.~\onlinecite{Tirelli2022} to construct our MLPs and used a GKR model based on a modified version of the SOAP kernel \cite{De2016}.
Our final dataset, including both training and test sets, comprises $561$ configurations selected through an iterative procedure with $128$ hydrogen atoms each, where we calculated energies, pressures and forces at the VMC and LRDMC levels. These configurations correspond to  temperatures from $4$~kK up to $20$~kK and Wigner-Seitz radii from $1.80$ Bohr to $2.12$ Bohr. Finite size corrections (FSC) have also been estimated using the  KZK functional \cite{Kwee2008}.
Details on the training set construction and the QMC calculations, together with the performances of all MLP models can be found in the SM~\cite{SM}.  In particular we found a final root mean square error (RMSE), calculated on the test set, of the order of $7$ meV/atom for the energy, $250$ meV/\AA~ for the forces, and $0.75$ GPa for the pressures.

At this point, it is worth to highlight some favourable features of our machine learning approach, especially in applications where it is coupled with computationally expensive methods such as QMC. They can be itemized as follows:
\begin{itemize}
    \item {\it transferability}: the total energy of the system is expressed as a sum of local terms \cite{Behler2007}, therefore our models are capable of making accurate predictions on configurations whose size has never been encountered in the training set. In particular, our MLPs find their applicability to systems with an arbitrary number of atoms $N$.
    \item {\it efficiency and accuracy}: within the $\Delta$-learning framework, the machine learning task becomes easier. 
    Indeed, we obtained very accurate QMC potentials, by training models on small datasets and, thus, by reducing the amount of calculations needed. Moreover, since the computational cost of the ML inference is negligible compared to the baseline DFT calculation, we were able to perform QMC-driven MD simulations at the cost of a DFT dynamics. 
    \item {\it overfitting prevention}: using a {\it local sparsification} technique based on the farthest point sampling (see SM of Ref.~\onlinecite{Tirelli2022}), we discarded from each configuration a possibly large fraction of the corresponding $N$ local environments, preventing overfitting and allowing for an increased predictive power of the model on unseen data. Since the computational cost of the predictions scales with the size of the training set, this procedure drastically improves the efficiency of the final model. 
\end{itemize}

We further validated the accuracy of our MLPs by comparing the Hugoniot curve obtained using two potentials,  independently trained with the same target, e.g. VMC, but with two different baselines. 
Taking into account these results and the RMSE of the models, we can estimate an uncertainty on the prediction of $~0.06$ on the relative density $\rho / \rho_0$ and $~1$~GPa on pressure.

We now turn to the last source of error we identified, i.e. the one related to the calculation of $e_0$ and $p_0$.
To estimate the reference state energy and pressure, we followed a procedure similar to Ref. \onlinecite{Ruggeri2020}. We performed a path integral molecular dynamics (PIMD) simulation \cite{Mouhat2017} on a system of $N = 64$ deuterium atoms at a temperature $T$ = 22~K and density $\rho_0$ = 0.167 g/cm$^{3}$ (corresponding to the initial conditions reported in Ref.~\onlinecite{Knudson2017}), using DFT-PBE energy and forces. Details of this simulation are reported in the SM~\cite{SM}. From the PIMD trajectory, we extracted $170$ configurations and we calculated energies and pressures with both DFT-PBE and QMC at VMC and LRDMC levels, adding the necessary
finite size corrections.
The reference sample was generated by extracting atomic positions from one of the $128$ beads taken at random, belonging to de-correlated snapshots of the trajectory. Results for $e_0$ for the various methods are reported in Tab.~\ref{tab:reference_state_energy}. The reference state pressure $p_0$ is not reported, since it is two orders of magnitude smaller than the shocked pressure, and thus irrelevant for the Hugoniot determination. Also in this case, we studied the effect of varying $e_0$ within its confidence interval on the Hugoniot density and pressure. 
In doing so, we also took into account the possible uncertainty on the energy difference $e(\rho,T) - e_0$ originating by the finite batch size we used for estimating energy gradients in the WF optimization.
We estimated this uncertainty by running optimizations of increasing batch size on three different $128$-atom configurations. The results indicate an error $\lesssim 0.5$ mHa/atom on $e(\rho,T) - e_0$. 
Taking everything into account, varying the energy within standard deviation leads to shifts in the final principal Hugoniot which still fall in the error bars
of our predictions estimated previously.

\begin{table}[b]
    \centering
    \begin{tabular}{l|cc}
    \toprule
    & $e_{{\rm{pot}}}$ (Ha/atom)  & $e_0$ (Ha/atom)\\
    \midrule
      PBE   & -0.58217(2) & -0.58055(2)\\
      VMC   &  -0.58622(2) & -0.58460(2) \\
      LRDMC   & -0.58660(2) & -0.58498(2) \\
      VMC + FSC  & -0.58503(2)  & -0.58342(2)  \\
      LRDMC + FSC  &  -0.58542(2) & -0.58380(2) \\      
     \bottomrule
    \end{tabular}
    \caption{Estimated potential ($e_{\textrm{pot}}$) and total ($e_0$) energies per atom of the reference state at $\rho_0$ = 0.167 g/cm$^{3}$ and $T$ = 22~K for different methods, with and without finite size corrections.}
    \label{tab:reference_state_energy}
\end{table}

\color{black}

To summarize, we estimated the MLP prediction error to be the most relevant source of uncertainty for the Hugoniot, yielding, as discussed before, an absolute error of $0.06$ and $1$~GPa on the relative density and pressure, respectively, reflected on the error bars reported in Fig.~\ref{fig: Hugoniot}. 

{\vspace{3mm}}
\section{Conclusions}
In conclusion, using our recently proposed workflow for the construction of MLPs, we have been able to run reliable VMC- and LRDMC-based MD simulations and study the principal deuterium Hugoniot, in a pressure range relevant for experiments. The accuracy of the MLPs employed here has been extensively tested, supporting the validity of our calculations and estimating their uncertainty. 
The resulting Hugoniot curve shows generally good agreement with the most recent experimental measures, especially in the low temperature molecular regime. 
Exploiting the $\Delta$-learning framework, we have also been able to treat FT electrons effects in a QMC-MLP, and we have thus managed to perform accurate simulations at higher temperatures. For these temperatures, the results systematically show a more compressible Hugoniot curve than experiments, although the experimental error bars are large in this regime. The aforementioned discrepancy is milder than previously reported QMC calculations~\cite{Tubman2015,Ruggeri2020}, and falls within the measures uncertainty. In particular, this suggests that the use of optimized and more refined WFs has a key role for obtaining good results in high-pressure hydrogen. We thus believe that our results will be useful for both future experimental research and numerical investigations of the Hugoniot.
The efficiency of our computational approach could be further improved, e.g., by using cheaper baseline potentials than DFT.
Longer simulations and larger systems will then be at reach. Other many-body methods, even more expensive than QMC, can also be used as targets for this type of MLPs, since the required size of the dataset is at least one order of magnitude smaller compared to other ML approaches. Finally, our MLPs, and in particular those trained on LRDMC data points, are promising for exploring the hydrogen phase diagram by keeping a high level of accuracy across a wide range of thermodynamic conditions.

\section*{Data availability}
The machine learning code used in this work is available upon request. Additional information, such as datasets, models, and detailed results of the simulations are available at \url{https://github.com/giacomotenti/QMC_hugoniot}.
{\vspace{3mm}}
\begin{acknowledgements}

The computations in this work have mainly been performed using the  Fugaku supercomputer provided by RIKEN through the HPCI System Research Project (Project ID: hp210038 and hp220060) and Marconi100 provided by CINECA through the ISCRA project No.~{HP10BGJH1X} and the SISSA three-year agreement 2022.
A.T. acknowledges financial  support  from the  MIUR  Progetti  di  Ricerca  di  Rilevante  Interesse Nazionale  (PRIN)  Bando  2017  -  grant  2017BZPKSZ.
K.N. acknowledges financial support from the JSPS Overseas Research Fellowships, from Grant-in-Aid for Early Career Scientists (Grant No. JP21K17752), from Grant-in-Aid for Scientific Research (Grant No. JP21K03400), and from MEXT Leading Initiative for Excellent Young Researchers (Grant No. JPMXS0320220025).
This work is also supported by the European Centre of Excellence in Exascale Computing TREX - Targeting Real Chemical Accuracy at the Exascale. This project has received funding from the European Union’s Horizon 2020 - Research and Innovation program - under grant agreement no. 952165.
The authors thank Dr. Guglielmo Mazzola and Cesare Cozza for the helpful suggestions and discussion.
We dedicate this paper to the memory of Prof. Sandro Sorella (SISSA), who tragically passed away during this project, remembering him as one of the most influential contributors to the quantum Monte Carlo community, and in particular for deeply inspiring this work with the development of the \emph{ab initio} QMC code, {\textsc{TurboRVB}}.

\end{acknowledgements}

\bibliography{Hugoniot_paper.bib}

\begin{thebibliography}{69}%
\makeatletter
\providecommand \@ifxundefined [1]{%
 \@ifx{#1\undefined}
}%
\providecommand \@ifnum [1]{%
 \ifnum #1\expandafter \@firstoftwo
 \else \expandafter \@secondoftwo
 \fi
}%
\providecommand \@ifx [1]{%
 \ifx #1\expandafter \@firstoftwo
 \else \expandafter \@secondoftwo
 \fi
}%
\providecommand \natexlab [1]{#1}%
\providecommand \enquote  [1]{``#1''}%
\providecommand \bibnamefont  [1]{#1}%
\providecommand \bibfnamefont [1]{#1}%
\providecommand \citenamefont [1]{#1}%
\providecommand \href@noop [0]{\@secondoftwo}%
\providecommand \href [0]{\begingroup \@sanitize@url \@href}%
\providecommand \@href[1]{\@@startlink{#1}\@@href}%
\providecommand \@@href[1]{\endgroup#1\@@endlink}%
\providecommand \@sanitize@url [0]{\catcode `\\12\catcode `\$12\catcode `\&12\catcode `\#12\catcode `\^12\catcode `\_12\catcode `\%12\relax}%
\providecommand \@@startlink[1]{}%
\providecommand \@@endlink[0]{}%
\providecommand \url  [0]{\begingroup\@sanitize@url \@url }%
\providecommand \@url [1]{\endgroup\@href {#1}{\urlprefix }}%
\providecommand \urlprefix  [0]{URL }%
\providecommand \Eprint [0]{\href }%
\providecommand \doibase [0]{https://doi.org/}%
\providecommand \selectlanguage [0]{\@gobble}%
\providecommand \bibinfo  [0]{\@secondoftwo}%
\providecommand \bibfield  [0]{\@secondoftwo}%
\providecommand \translation [1]{[#1]}%
\providecommand \BibitemOpen [0]{}%
\providecommand \bibitemStop [0]{}%
\providecommand \bibitemNoStop [0]{.\EOS\space}%
\providecommand \EOS [0]{\spacefactor3000\relax}%
\providecommand \BibitemShut  [1]{\csname bibitem#1\endcsname}%
\let\auto@bib@innerbib\@empty
\bibitem [{\citenamefont {Saumon}\ \emph {et~al.}(1995)\citenamefont {Saumon}, \citenamefont {Chabrier},\ and\ \citenamefont {van Horn}}]{Saumon1995}%
  \BibitemOpen
  \bibfield  {author} {\bibinfo {author} {\bibfnamefont {D.}~\bibnamefont {Saumon}}, \bibinfo {author} {\bibfnamefont {G.}~\bibnamefont {Chabrier}},\ and\ \bibinfo {author} {\bibfnamefont {H.~M.}\ \bibnamefont {van Horn}},\ }\href@noop {} {\bibfield  {journal} {\bibinfo  {journal} {The astrophysical journal supplement series}\ }\textbf {\bibinfo {volume} {99}},\ \bibinfo {pages} {713} (\bibinfo {year} {1995})}\BibitemShut {NoStop}%
\bibitem [{\citenamefont {Fortney}\ and\ \citenamefont {Nettelmann}(2009)}]{Fortney2009}%
  \BibitemOpen
  \bibfield  {author} {\bibinfo {author} {\bibfnamefont {J.~J.}\ \bibnamefont {Fortney}}\ and\ \bibinfo {author} {\bibfnamefont {N.}~\bibnamefont {Nettelmann}},\ }\href {https://doi.org/10.1007/s11214-009-9582-x} {\bibfield  {journal} {\bibinfo  {journal} {Space Science Reviews}\ }\textbf {\bibinfo {volume} {152}},\ \bibinfo {pages} {423} (\bibinfo {year} {2009})}\BibitemShut {NoStop}%
\bibitem [{\citenamefont {Miguel}\ \emph {et~al.}(2016)\citenamefont {Miguel}, \citenamefont {Guillot},\ and\ \citenamefont {Fayon}}]{Miguel2016}%
  \BibitemOpen
  \bibfield  {author} {\bibinfo {author} {\bibfnamefont {Y.}~\bibnamefont {Miguel}}, \bibinfo {author} {\bibfnamefont {T.}~\bibnamefont {Guillot}},\ and\ \bibinfo {author} {\bibfnamefont {L.}~\bibnamefont {Fayon}},\ }\href@noop {} {\bibfield  {journal} {\bibinfo  {journal} {Astronomy \& Astrophysics}\ }\textbf {\bibinfo {volume} {596}},\ \bibinfo {pages} {A114} (\bibinfo {year} {2016})}\BibitemShut {NoStop}%
\bibitem [{\citenamefont {Hu}\ \emph {et~al.}(2015)\citenamefont {Hu}, \citenamefont {Goncharov}, \citenamefont {Boehly}, \citenamefont {McCrory}, \citenamefont {Skupsky}, \citenamefont {Collins}, \citenamefont {Kress},\ and\ \citenamefont {Militzer}}]{Hu2015}%
  \BibitemOpen
  \bibfield  {author} {\bibinfo {author} {\bibfnamefont {S.}~\bibnamefont {Hu}}, \bibinfo {author} {\bibfnamefont {V.}~\bibnamefont {Goncharov}}, \bibinfo {author} {\bibfnamefont {T.}~\bibnamefont {Boehly}}, \bibinfo {author} {\bibfnamefont {R.}~\bibnamefont {McCrory}}, \bibinfo {author} {\bibfnamefont {S.}~\bibnamefont {Skupsky}}, \bibinfo {author} {\bibfnamefont {L.~A.}\ \bibnamefont {Collins}}, \bibinfo {author} {\bibfnamefont {J.~D.}\ \bibnamefont {Kress}},\ and\ \bibinfo {author} {\bibfnamefont {B.}~\bibnamefont {Militzer}},\ }\href@noop {} {\bibfield  {journal} {\bibinfo  {journal} {Physics of Plasmas}\ }\textbf {\bibinfo {volume} {22}},\ \bibinfo {pages} {056304} (\bibinfo {year} {2015})}\BibitemShut {NoStop}%
\bibitem [{\citenamefont {Drozdov}\ \emph {et~al.}(2015)\citenamefont {Drozdov}, \citenamefont {Eremets}, \citenamefont {Troyan}, \citenamefont {Ksenofontov},\ and\ \citenamefont {Shylin}}]{Drozdov2015}%
  \BibitemOpen
  \bibfield  {author} {\bibinfo {author} {\bibfnamefont {A.~P.}\ \bibnamefont {Drozdov}}, \bibinfo {author} {\bibfnamefont {M.~I.}\ \bibnamefont {Eremets}}, \bibinfo {author} {\bibfnamefont {I.~A.}\ \bibnamefont {Troyan}}, \bibinfo {author} {\bibfnamefont {V.}~\bibnamefont {Ksenofontov}},\ and\ \bibinfo {author} {\bibfnamefont {S.~I.}\ \bibnamefont {Shylin}},\ }\href {https://doi.org/10.1038/nature14964} {\bibfield  {journal} {\bibinfo  {journal} {Nature}\ }\textbf {\bibinfo {volume} {525}},\ \bibinfo {pages} {73} (\bibinfo {year} {2015})}\BibitemShut {NoStop}%
\bibitem [{\citenamefont {Somayazulu}\ \emph {et~al.}(2019)\citenamefont {Somayazulu}, \citenamefont {Ahart}, \citenamefont {Mishra}, \citenamefont {Geballe}, \citenamefont {Baldini}, \citenamefont {Meng}, \citenamefont {Struzhkin},\ and\ \citenamefont {Hemley}}]{Somayazulu2019}%
  \BibitemOpen
  \bibfield  {author} {\bibinfo {author} {\bibfnamefont {M.}~\bibnamefont {Somayazulu}}, \bibinfo {author} {\bibfnamefont {M.}~\bibnamefont {Ahart}}, \bibinfo {author} {\bibfnamefont {A.~K.}\ \bibnamefont {Mishra}}, \bibinfo {author} {\bibfnamefont {Z.~M.}\ \bibnamefont {Geballe}}, \bibinfo {author} {\bibfnamefont {M.}~\bibnamefont {Baldini}}, \bibinfo {author} {\bibfnamefont {Y.}~\bibnamefont {Meng}}, \bibinfo {author} {\bibfnamefont {V.~V.}\ \bibnamefont {Struzhkin}},\ and\ \bibinfo {author} {\bibfnamefont {R.~J.}\ \bibnamefont {Hemley}},\ }\href {https://doi.org/10.1103/PhysRevLett.122.027001} {\bibfield  {journal} {\bibinfo  {journal} {Phys. Rev. Lett.}\ }\textbf {\bibinfo {volume} {122}},\ \bibinfo {pages} {027001} (\bibinfo {year} {2019})}\BibitemShut {NoStop}%
\bibitem [{\citenamefont {McMahon}\ \emph {et~al.}(2012)\citenamefont {McMahon}, \citenamefont {Morales}, \citenamefont {Pierleoni},\ and\ \citenamefont {Ceperley}}]{McMahon2012}%
  \BibitemOpen
  \bibfield  {author} {\bibinfo {author} {\bibfnamefont {J.~M.}\ \bibnamefont {McMahon}}, \bibinfo {author} {\bibfnamefont {M.~A.}\ \bibnamefont {Morales}}, \bibinfo {author} {\bibfnamefont {C.}~\bibnamefont {Pierleoni}},\ and\ \bibinfo {author} {\bibfnamefont {D.~M.}\ \bibnamefont {Ceperley}},\ }\href {https://doi.org/10.1103/RevModPhys.84.1607} {\bibfield  {journal} {\bibinfo  {journal} {Rev. Mod. Phys.}\ }\textbf {\bibinfo {volume} {84}},\ \bibinfo {pages} {1607} (\bibinfo {year} {2012})}\BibitemShut {NoStop}%
\bibitem [{\citenamefont {Cheng}\ \emph {et~al.}(2020)\citenamefont {Cheng}, \citenamefont {Mazzola}, \citenamefont {Pickard},\ and\ \citenamefont {Ceriotti}}]{Cheng2020}%
  \BibitemOpen
  \bibfield  {author} {\bibinfo {author} {\bibfnamefont {B.}~\bibnamefont {Cheng}}, \bibinfo {author} {\bibfnamefont {G.}~\bibnamefont {Mazzola}}, \bibinfo {author} {\bibfnamefont {C.~J.}\ \bibnamefont {Pickard}},\ and\ \bibinfo {author} {\bibfnamefont {M.}~\bibnamefont {Ceriotti}},\ }\href {https://doi.org/10.1038/s41586-020-2677-y} {\bibfield  {journal} {\bibinfo  {journal} {Nature}\ }\textbf {\bibinfo {volume} {585}},\ \bibinfo {pages} {217} (\bibinfo {year} {2020})}\BibitemShut {NoStop}%
\bibitem [{\citenamefont {Karasiev}\ \emph {et~al.}(2021)\citenamefont {Karasiev}, \citenamefont {Hinz}, \citenamefont {Hu},\ and\ \citenamefont {Trickey}}]{Karasiev2021}%
  \BibitemOpen
  \bibfield  {author} {\bibinfo {author} {\bibfnamefont {V.~V.}\ \bibnamefont {Karasiev}}, \bibinfo {author} {\bibfnamefont {J.}~\bibnamefont {Hinz}}, \bibinfo {author} {\bibfnamefont {S.~X.}\ \bibnamefont {Hu}},\ and\ \bibinfo {author} {\bibfnamefont {S.~B.}\ \bibnamefont {Trickey}},\ }\href {https://doi.org/10.1038/s41586-021-04078-x} {\bibfield  {journal} {\bibinfo  {journal} {Nature}\ }\textbf {\bibinfo {volume} {600}},\ \bibinfo {pages} {E12} (\bibinfo {year} {2021})}\BibitemShut {NoStop}%
\bibitem [{\citenamefont {Cheng}\ \emph {et~al.}(2021)\citenamefont {Cheng}, \citenamefont {Mazzola}, \citenamefont {Pickard},\ and\ \citenamefont {Ceriotti}}]{Cheng2021}%
  \BibitemOpen
  \bibfield  {author} {\bibinfo {author} {\bibfnamefont {B.}~\bibnamefont {Cheng}}, \bibinfo {author} {\bibfnamefont {G.}~\bibnamefont {Mazzola}}, \bibinfo {author} {\bibfnamefont {C.~J.}\ \bibnamefont {Pickard}},\ and\ \bibinfo {author} {\bibfnamefont {M.}~\bibnamefont {Ceriotti}},\ }\href {https://doi.org/10.1038/s41586-021-04079-w} {\bibfield  {journal} {\bibinfo  {journal} {Nature}\ }\textbf {\bibinfo {volume} {600}},\ \bibinfo {pages} {E15} (\bibinfo {year} {2021})}\BibitemShut {NoStop}%
\bibitem [{\citenamefont {Nellis}(2006)}]{Nellis2006}%
  \BibitemOpen
  \bibfield  {author} {\bibinfo {author} {\bibfnamefont {W.~J.}\ \bibnamefont {Nellis}},\ }\href {https://doi.org/10.1088/0034-4885/69/5/r05} {\bibfield  {journal} {\bibinfo  {journal} {Reports on Progress in Physics}\ }\textbf {\bibinfo {volume} {69}},\ \bibinfo {pages} {1479} (\bibinfo {year} {2006})}\BibitemShut {NoStop}%
\bibitem [{\citenamefont {Duvall}\ and\ \citenamefont {Graham}(1977)}]{Duvall1977}%
  \BibitemOpen
  \bibfield  {author} {\bibinfo {author} {\bibfnamefont {G.~E.}\ \bibnamefont {Duvall}}\ and\ \bibinfo {author} {\bibfnamefont {R.~A.}\ \bibnamefont {Graham}},\ }\href {https://doi.org/10.1103/RevModPhys.49.523} {\bibfield  {journal} {\bibinfo  {journal} {Rev. Mod. Phys.}\ }\textbf {\bibinfo {volume} {49}},\ \bibinfo {pages} {523} (\bibinfo {year} {1977})}\BibitemShut {NoStop}%
\bibitem [{\citenamefont {Nellis}\ \emph {et~al.}(1983)\citenamefont {Nellis}, \citenamefont {Mitchell}, \citenamefont {van Thiel}, \citenamefont {Devine}, \citenamefont {Trainor},\ and\ \citenamefont {Brown}}]{Nellis1983}%
  \BibitemOpen
  \bibfield  {author} {\bibinfo {author} {\bibfnamefont {W.~J.}\ \bibnamefont {Nellis}}, \bibinfo {author} {\bibfnamefont {A.~C.}\ \bibnamefont {Mitchell}}, \bibinfo {author} {\bibfnamefont {M.}~\bibnamefont {van Thiel}}, \bibinfo {author} {\bibfnamefont {G.~J.}\ \bibnamefont {Devine}}, \bibinfo {author} {\bibfnamefont {R.~J.}\ \bibnamefont {Trainor}},\ and\ \bibinfo {author} {\bibfnamefont {N.}~\bibnamefont {Brown}},\ }\href {https://doi.org/10.1063/1.445938} {\bibfield  {journal} {\bibinfo  {journal} {The Journal of Chemical Physics}\ }\textbf {\bibinfo {volume} {79}},\ \bibinfo {pages} {1480} (\bibinfo {year} {1983})},\ \Eprint {https://arxiv.org/abs/https://doi.org/10.1063/1.445938} {https://doi.org/10.1063/1.445938} \BibitemShut {NoStop}%
\bibitem [{\citenamefont {Knudson}\ \emph {et~al.}(2001)\citenamefont {Knudson}, \citenamefont {Hanson}, \citenamefont {Bailey}, \citenamefont {Hall}, \citenamefont {Asay},\ and\ \citenamefont {Anderson}}]{Knudson2001}%
  \BibitemOpen
  \bibfield  {author} {\bibinfo {author} {\bibfnamefont {M.~D.}\ \bibnamefont {Knudson}}, \bibinfo {author} {\bibfnamefont {D.~L.}\ \bibnamefont {Hanson}}, \bibinfo {author} {\bibfnamefont {J.~E.}\ \bibnamefont {Bailey}}, \bibinfo {author} {\bibfnamefont {C.~A.}\ \bibnamefont {Hall}}, \bibinfo {author} {\bibfnamefont {J.~R.}\ \bibnamefont {Asay}},\ and\ \bibinfo {author} {\bibfnamefont {W.~W.}\ \bibnamefont {Anderson}},\ }\href {https://doi.org/10.1103/PhysRevLett.87.225501} {\bibfield  {journal} {\bibinfo  {journal} {Phys. Rev. Lett.}\ }\textbf {\bibinfo {volume} {87}},\ \bibinfo {pages} {225501} (\bibinfo {year} {2001})}\BibitemShut {NoStop}%
\bibitem [{\citenamefont {Boriskov}\ \emph {et~al.}(2005)\citenamefont {Boriskov}, \citenamefont {Bykov}, \citenamefont {Il'kaev}, \citenamefont {Selemir}, \citenamefont {Simakov}, \citenamefont {Trunin}, \citenamefont {Urlin}, \citenamefont {Shuikin},\ and\ \citenamefont {Nellis}}]{Boriskov2005}%
  \BibitemOpen
  \bibfield  {author} {\bibinfo {author} {\bibfnamefont {G.~V.}\ \bibnamefont {Boriskov}}, \bibinfo {author} {\bibfnamefont {A.~I.}\ \bibnamefont {Bykov}}, \bibinfo {author} {\bibfnamefont {R.~I.}\ \bibnamefont {Il'kaev}}, \bibinfo {author} {\bibfnamefont {V.~D.}\ \bibnamefont {Selemir}}, \bibinfo {author} {\bibfnamefont {G.~V.}\ \bibnamefont {Simakov}}, \bibinfo {author} {\bibfnamefont {R.~F.}\ \bibnamefont {Trunin}}, \bibinfo {author} {\bibfnamefont {V.~D.}\ \bibnamefont {Urlin}}, \bibinfo {author} {\bibfnamefont {A.~N.}\ \bibnamefont {Shuikin}},\ and\ \bibinfo {author} {\bibfnamefont {W.~J.}\ \bibnamefont {Nellis}},\ }\href {https://doi.org/10.1103/PhysRevB.71.092104} {\bibfield  {journal} {\bibinfo  {journal} {Phys. Rev. B}\ }\textbf {\bibinfo {volume} {71}},\ \bibinfo {pages} {092104} (\bibinfo {year} {2005})}\BibitemShut {NoStop}%
\bibitem [{\citenamefont {Knudson}\ \emph {et~al.}(2004)\citenamefont {Knudson}, \citenamefont {Hanson}, \citenamefont {Bailey}, \citenamefont {Hall}, \citenamefont {Asay},\ and\ \citenamefont {Deeney}}]{Knudson2004}%
  \BibitemOpen
  \bibfield  {author} {\bibinfo {author} {\bibfnamefont {M.~D.}\ \bibnamefont {Knudson}}, \bibinfo {author} {\bibfnamefont {D.~L.}\ \bibnamefont {Hanson}}, \bibinfo {author} {\bibfnamefont {J.~E.}\ \bibnamefont {Bailey}}, \bibinfo {author} {\bibfnamefont {C.~A.}\ \bibnamefont {Hall}}, \bibinfo {author} {\bibfnamefont {J.~R.}\ \bibnamefont {Asay}},\ and\ \bibinfo {author} {\bibfnamefont {C.}~\bibnamefont {Deeney}},\ }\href {https://doi.org/10.1103/PhysRevB.69.144209} {\bibfield  {journal} {\bibinfo  {journal} {Phys. Rev. B}\ }\textbf {\bibinfo {volume} {69}},\ \bibinfo {pages} {144209} (\bibinfo {year} {2004})}\BibitemShut {NoStop}%
\bibitem [{\citenamefont {Hicks}\ \emph {et~al.}(2009)\citenamefont {Hicks}, \citenamefont {Boehly}, \citenamefont {Celliers}, \citenamefont {Eggert}, \citenamefont {Moon}, \citenamefont {Meyerhofer},\ and\ \citenamefont {Collins}}]{Hicks2009}%
  \BibitemOpen
  \bibfield  {author} {\bibinfo {author} {\bibfnamefont {D.~G.}\ \bibnamefont {Hicks}}, \bibinfo {author} {\bibfnamefont {T.~R.}\ \bibnamefont {Boehly}}, \bibinfo {author} {\bibfnamefont {P.~M.}\ \bibnamefont {Celliers}}, \bibinfo {author} {\bibfnamefont {J.~H.}\ \bibnamefont {Eggert}}, \bibinfo {author} {\bibfnamefont {S.~J.}\ \bibnamefont {Moon}}, \bibinfo {author} {\bibfnamefont {D.~D.}\ \bibnamefont {Meyerhofer}},\ and\ \bibinfo {author} {\bibfnamefont {G.~W.}\ \bibnamefont {Collins}},\ }\href {https://doi.org/10.1103/PhysRevB.79.014112} {\bibfield  {journal} {\bibinfo  {journal} {Phys. Rev. B}\ }\textbf {\bibinfo {volume} {79}},\ \bibinfo {pages} {014112} (\bibinfo {year} {2009})}\BibitemShut {NoStop}%
\bibitem [{\citenamefont {Loubeyre}\ \emph {et~al.}(2012)\citenamefont {Loubeyre}, \citenamefont {Brygoo}, \citenamefont {Eggert}, \citenamefont {Celliers}, \citenamefont {Spaulding}, \citenamefont {Rygg}, \citenamefont {Boehly}, \citenamefont {Collins},\ and\ \citenamefont {Jeanloz}}]{Loubeyre2012}%
  \BibitemOpen
  \bibfield  {author} {\bibinfo {author} {\bibfnamefont {P.}~\bibnamefont {Loubeyre}}, \bibinfo {author} {\bibfnamefont {S.}~\bibnamefont {Brygoo}}, \bibinfo {author} {\bibfnamefont {J.}~\bibnamefont {Eggert}}, \bibinfo {author} {\bibfnamefont {P.~M.}\ \bibnamefont {Celliers}}, \bibinfo {author} {\bibfnamefont {D.~K.}\ \bibnamefont {Spaulding}}, \bibinfo {author} {\bibfnamefont {J.~R.}\ \bibnamefont {Rygg}}, \bibinfo {author} {\bibfnamefont {T.~R.}\ \bibnamefont {Boehly}}, \bibinfo {author} {\bibfnamefont {G.~W.}\ \bibnamefont {Collins}},\ and\ \bibinfo {author} {\bibfnamefont {R.}~\bibnamefont {Jeanloz}},\ }\href {https://doi.org/10.1103/PhysRevB.86.144115} {\bibfield  {journal} {\bibinfo  {journal} {Phys. Rev. B}\ }\textbf {\bibinfo {volume} {86}},\ \bibinfo {pages} {144115} (\bibinfo {year} {2012})}\BibitemShut {NoStop}%
\bibitem [{\citenamefont {Knudson}\ and\ \citenamefont {Desjarlais}(2017)}]{Knudson2017}%
  \BibitemOpen
  \bibfield  {author} {\bibinfo {author} {\bibfnamefont {M.~D.}\ \bibnamefont {Knudson}}\ and\ \bibinfo {author} {\bibfnamefont {M.~P.}\ \bibnamefont {Desjarlais}},\ }\href {https://doi.org/10.1103/PhysRevLett.118.035501} {\bibfield  {journal} {\bibinfo  {journal} {Physical Review Letters}\ }\textbf {\bibinfo {volume} {118}},\ \bibinfo {pages} {1} (\bibinfo {year} {2017})}\BibitemShut {NoStop}%
\bibitem [{\citenamefont {Fernandez-Pa\~nella}\ \emph {et~al.}(2019)\citenamefont {Fernandez-Pa\~nella}, \citenamefont {Millot}, \citenamefont {Fratanduono}, \citenamefont {Desjarlais}, \citenamefont {Hamel}, \citenamefont {Marshall}, \citenamefont {Erskine}, \citenamefont {Sterne}, \citenamefont {Haan}, \citenamefont {Boehly}, \citenamefont {Collins}, \citenamefont {Eggert},\ and\ \citenamefont {Celliers}}]{Fernandez2019}%
  \BibitemOpen
  \bibfield  {author} {\bibinfo {author} {\bibfnamefont {A.}~\bibnamefont {Fernandez-Pa\~nella}}, \bibinfo {author} {\bibfnamefont {M.}~\bibnamefont {Millot}}, \bibinfo {author} {\bibfnamefont {D.~E.}\ \bibnamefont {Fratanduono}}, \bibinfo {author} {\bibfnamefont {M.~P.}\ \bibnamefont {Desjarlais}}, \bibinfo {author} {\bibfnamefont {S.}~\bibnamefont {Hamel}}, \bibinfo {author} {\bibfnamefont {M.~C.}\ \bibnamefont {Marshall}}, \bibinfo {author} {\bibfnamefont {D.~J.}\ \bibnamefont {Erskine}}, \bibinfo {author} {\bibfnamefont {P.~A.}\ \bibnamefont {Sterne}}, \bibinfo {author} {\bibfnamefont {S.}~\bibnamefont {Haan}}, \bibinfo {author} {\bibfnamefont {T.~R.}\ \bibnamefont {Boehly}}, \bibinfo {author} {\bibfnamefont {G.~W.}\ \bibnamefont {Collins}}, \bibinfo {author} {\bibfnamefont {J.~H.}\ \bibnamefont {Eggert}},\ and\ \bibinfo {author} {\bibfnamefont {P.~M.}\ \bibnamefont {Celliers}},\ }\href {https://doi.org/10.1103/PhysRevLett.122.255702} {\bibfield  {journal} {\bibinfo  {journal} {Phys. Rev. Lett.}\ }\textbf
  {\bibinfo {volume} {122}},\ \bibinfo {pages} {255702} (\bibinfo {year} {2019})}\BibitemShut {NoStop}%
\bibitem [{\citenamefont {Knudson}\ and\ \citenamefont {Desjarlais}(2021)}]{Knudson2021}%
  \BibitemOpen
  \bibfield  {author} {\bibinfo {author} {\bibfnamefont {M.~D.}\ \bibnamefont {Knudson}}\ and\ \bibinfo {author} {\bibfnamefont {M.~P.}\ \bibnamefont {Desjarlais}},\ }\bibfield  {journal} {\bibinfo  {journal} {Journal of Applied Physics}\ }\textbf {\bibinfo {volume} {129}},\ \href {https://doi.org/10.1063/5.0050878} {10.1063/5.0050878} (\bibinfo {year} {2021})\BibitemShut {NoStop}%
\bibitem [{\citenamefont {Lenosky}\ \emph {et~al.}(2000)\citenamefont {Lenosky}, \citenamefont {Bickham}, \citenamefont {Kress},\ and\ \citenamefont {Collins}}]{Lenosky2000}%
  \BibitemOpen
  \bibfield  {author} {\bibinfo {author} {\bibfnamefont {T.~J.}\ \bibnamefont {Lenosky}}, \bibinfo {author} {\bibfnamefont {S.~R.}\ \bibnamefont {Bickham}}, \bibinfo {author} {\bibfnamefont {J.~D.}\ \bibnamefont {Kress}},\ and\ \bibinfo {author} {\bibfnamefont {L.~A.}\ \bibnamefont {Collins}},\ }\href {https://doi.org/10.1103/PhysRevB.61.1} {\bibfield  {journal} {\bibinfo  {journal} {Phys. Rev. B}\ }\textbf {\bibinfo {volume} {61}},\ \bibinfo {pages} {1} (\bibinfo {year} {2000})}\BibitemShut {NoStop}%
\bibitem [{\citenamefont {Galli}\ \emph {et~al.}(2000)\citenamefont {Galli}, \citenamefont {Hood}, \citenamefont {Hazi},\ and\ \citenamefont {Gygi}}]{Galli2000}%
  \BibitemOpen
  \bibfield  {author} {\bibinfo {author} {\bibfnamefont {G.}~\bibnamefont {Galli}}, \bibinfo {author} {\bibfnamefont {R.~Q.}\ \bibnamefont {Hood}}, \bibinfo {author} {\bibfnamefont {A.~U.}\ \bibnamefont {Hazi}},\ and\ \bibinfo {author} {\bibfnamefont {F.~m.~c.}\ \bibnamefont {Gygi}},\ }\href {https://doi.org/10.1103/PhysRevB.61.909} {\bibfield  {journal} {\bibinfo  {journal} {Phys. Rev. B}\ }\textbf {\bibinfo {volume} {61}},\ \bibinfo {pages} {909} (\bibinfo {year} {2000})}\BibitemShut {NoStop}%
\bibitem [{\citenamefont {Bagnier}\ \emph {et~al.}(2000)\citenamefont {Bagnier}, \citenamefont {Blottiau},\ and\ \citenamefont {Cl\'erouin}}]{Bagnier2000}%
  \BibitemOpen
  \bibfield  {author} {\bibinfo {author} {\bibfnamefont {S.}~\bibnamefont {Bagnier}}, \bibinfo {author} {\bibfnamefont {P.}~\bibnamefont {Blottiau}},\ and\ \bibinfo {author} {\bibfnamefont {J.}~\bibnamefont {Cl\'erouin}},\ }\href {https://doi.org/10.1103/PhysRevE.63.015301} {\bibfield  {journal} {\bibinfo  {journal} {Phys. Rev. E}\ }\textbf {\bibinfo {volume} {63}},\ \bibinfo {pages} {015301} (\bibinfo {year} {2000})}\BibitemShut {NoStop}%
\bibitem [{\citenamefont {Bonev}\ \emph {et~al.}(2004)\citenamefont {Bonev}, \citenamefont {Militzer},\ and\ \citenamefont {Galli}}]{Bonev2004}%
  \BibitemOpen
  \bibfield  {author} {\bibinfo {author} {\bibfnamefont {S.~A.}\ \bibnamefont {Bonev}}, \bibinfo {author} {\bibfnamefont {B.}~\bibnamefont {Militzer}},\ and\ \bibinfo {author} {\bibfnamefont {G.}~\bibnamefont {Galli}},\ }\href {https://doi.org/10.1103/PhysRevB.69.014101} {\bibfield  {journal} {\bibinfo  {journal} {Phys. Rev. B}\ }\textbf {\bibinfo {volume} {69}},\ \bibinfo {pages} {014101} (\bibinfo {year} {2004})}\BibitemShut {NoStop}%
\bibitem [{\citenamefont {Holst}\ \emph {et~al.}(2008)\citenamefont {Holst}, \citenamefont {Redmer},\ and\ \citenamefont {Desjarlais}}]{Holst2008}%
  \BibitemOpen
  \bibfield  {author} {\bibinfo {author} {\bibfnamefont {B.}~\bibnamefont {Holst}}, \bibinfo {author} {\bibfnamefont {R.}~\bibnamefont {Redmer}},\ and\ \bibinfo {author} {\bibfnamefont {M.~P.}\ \bibnamefont {Desjarlais}},\ }\href {https://doi.org/10.1103/PhysRevB.77.184201} {\bibfield  {journal} {\bibinfo  {journal} {Phys. Rev. B}\ }\textbf {\bibinfo {volume} {77}},\ \bibinfo {pages} {184201} (\bibinfo {year} {2008})}\BibitemShut {NoStop}%
\bibitem [{\citenamefont {Caillabet}\ \emph {et~al.}(2011)\citenamefont {Caillabet}, \citenamefont {Mazevet},\ and\ \citenamefont {Loubeyre}}]{Caillabet2011}%
  \BibitemOpen
  \bibfield  {author} {\bibinfo {author} {\bibfnamefont {L.}~\bibnamefont {Caillabet}}, \bibinfo {author} {\bibfnamefont {S.}~\bibnamefont {Mazevet}},\ and\ \bibinfo {author} {\bibfnamefont {P.}~\bibnamefont {Loubeyre}},\ }\href {https://doi.org/10.1103/PhysRevB.83.094101} {\bibfield  {journal} {\bibinfo  {journal} {Phys. Rev. B}\ }\textbf {\bibinfo {volume} {83}},\ \bibinfo {pages} {094101} (\bibinfo {year} {2011})}\BibitemShut {NoStop}%
\bibitem [{\citenamefont {Karasiev}\ \emph {et~al.}(2019)\citenamefont {Karasiev}, \citenamefont {Hu}, \citenamefont {Zaghoo},\ and\ \citenamefont {Boehly}}]{Karasiev2019}%
  \BibitemOpen
  \bibfield  {author} {\bibinfo {author} {\bibfnamefont {V.~V.}\ \bibnamefont {Karasiev}}, \bibinfo {author} {\bibfnamefont {S.~X.}\ \bibnamefont {Hu}}, \bibinfo {author} {\bibfnamefont {M.}~\bibnamefont {Zaghoo}},\ and\ \bibinfo {author} {\bibfnamefont {T.~R.}\ \bibnamefont {Boehly}},\ }\href {https://doi.org/10.1103/PhysRevB.99.214110} {\bibfield  {journal} {\bibinfo  {journal} {Physical Review B}\ }\textbf {\bibinfo {volume} {99}},\ \bibinfo {pages} {1} (\bibinfo {year} {2019})}\BibitemShut {NoStop}%
\bibitem [{\citenamefont {Tubman}\ \emph {et~al.}(2015)\citenamefont {Tubman}, \citenamefont {Liberatore}, \citenamefont {Pierleoni}, \citenamefont {Holzmann},\ and\ \citenamefont {Ceperley}}]{Tubman2015}%
  \BibitemOpen
  \bibfield  {author} {\bibinfo {author} {\bibfnamefont {N.~M.}\ \bibnamefont {Tubman}}, \bibinfo {author} {\bibfnamefont {E.}~\bibnamefont {Liberatore}}, \bibinfo {author} {\bibfnamefont {C.}~\bibnamefont {Pierleoni}}, \bibinfo {author} {\bibfnamefont {M.}~\bibnamefont {Holzmann}},\ and\ \bibinfo {author} {\bibfnamefont {D.~M.}\ \bibnamefont {Ceperley}},\ }\href {https://doi.org/10.1103/PhysRevLett.115.045301} {\bibfield  {journal} {\bibinfo  {journal} {Physical Review Letters}\ }\textbf {\bibinfo {volume} {115}},\ \bibinfo {pages} {1} (\bibinfo {year} {2015})}\BibitemShut {NoStop}%
\bibitem [{\citenamefont {Ruggeri}\ \emph {et~al.}(2020)\citenamefont {Ruggeri}, \citenamefont {Holzmann}, \citenamefont {Ceperley},\ and\ \citenamefont {Pierleoni}}]{Ruggeri2020}%
  \BibitemOpen
  \bibfield  {author} {\bibinfo {author} {\bibfnamefont {M.}~\bibnamefont {Ruggeri}}, \bibinfo {author} {\bibfnamefont {M.}~\bibnamefont {Holzmann}}, \bibinfo {author} {\bibfnamefont {D.~M.}\ \bibnamefont {Ceperley}},\ and\ \bibinfo {author} {\bibfnamefont {C.}~\bibnamefont {Pierleoni}},\ }\href {https://doi.org/10.1103/PhysRevB.102.144108} {\bibfield  {journal} {\bibinfo  {journal} {Physical Review B}\ }\textbf {\bibinfo {volume} {102}},\ \bibinfo {pages} {144108} (\bibinfo {year} {2020})},\ \Eprint {https://arxiv.org/abs/2008.00269} {2008.00269} \BibitemShut {NoStop}%
\bibitem [{\citenamefont {Clay}\ \emph {et~al.}(2019)\citenamefont {Clay}, \citenamefont {Desjarlais},\ and\ \citenamefont {Shulenburger}}]{Clay2019}%
  \BibitemOpen
  \bibfield  {author} {\bibinfo {author} {\bibfnamefont {R.~C.}\ \bibnamefont {Clay}}, \bibinfo {author} {\bibfnamefont {M.~P.}\ \bibnamefont {Desjarlais}},\ and\ \bibinfo {author} {\bibfnamefont {L.}~\bibnamefont {Shulenburger}},\ }\href {https://doi.org/10.1103/PhysRevB.100.075103} {\bibfield  {journal} {\bibinfo  {journal} {Physical Review B}\ }\textbf {\bibinfo {volume} {100}},\ \bibinfo {pages} {75103} (\bibinfo {year} {2019})}\BibitemShut {NoStop}%
\bibitem [{\citenamefont {Behler}\ and\ \citenamefont {Parrinello}(2007)}]{Behler2007}%
  \BibitemOpen
  \bibfield  {author} {\bibinfo {author} {\bibfnamefont {J.}~\bibnamefont {Behler}}\ and\ \bibinfo {author} {\bibfnamefont {M.}~\bibnamefont {Parrinello}},\ }\href {https://doi.org/10.1103/PhysRevLett.98.146401} {\bibfield  {journal} {\bibinfo  {journal} {Physical Review Letters}\ }\textbf {\bibinfo {volume} {98}},\ \bibinfo {pages} {1} (\bibinfo {year} {2007})}\BibitemShut {NoStop}%
\bibitem [{\citenamefont {De}\ \emph {et~al.}(2016)\citenamefont {De}, \citenamefont {Bart{\'{o}}k}, \citenamefont {Cs{\'{a}}nyi},\ and\ \citenamefont {Ceriotti}}]{De2016}%
  \BibitemOpen
  \bibfield  {author} {\bibinfo {author} {\bibfnamefont {S.}~\bibnamefont {De}}, \bibinfo {author} {\bibfnamefont {A.~P.}\ \bibnamefont {Bart{\'{o}}k}}, \bibinfo {author} {\bibfnamefont {G.}~\bibnamefont {Cs{\'{a}}nyi}},\ and\ \bibinfo {author} {\bibfnamefont {M.}~\bibnamefont {Ceriotti}},\ }\href {https://doi.org/10.1039/C6CP00415F} {\bibfield  {journal} {\bibinfo  {journal} {Phys. Chem. Chem. Phys.}\ }\textbf {\bibinfo {volume} {18}},\ \bibinfo {pages} {13754} (\bibinfo {year} {2016})}\BibitemShut {NoStop}%
\bibitem [{\citenamefont {Tirelli}\ \emph {et~al.}(2022)\citenamefont {Tirelli}, \citenamefont {Tenti}, \citenamefont {Nakano},\ and\ \citenamefont {Sorella}}]{Tirelli2022}%
  \BibitemOpen
  \bibfield  {author} {\bibinfo {author} {\bibfnamefont {A.}~\bibnamefont {Tirelli}}, \bibinfo {author} {\bibfnamefont {G.}~\bibnamefont {Tenti}}, \bibinfo {author} {\bibfnamefont {K.}~\bibnamefont {Nakano}},\ and\ \bibinfo {author} {\bibfnamefont {S.}~\bibnamefont {Sorella}},\ }\href {https://doi.org/10.1103/PhysRevB.106.L041105} {\bibfield  {journal} {\bibinfo  {journal} {Phys. Rev. B}\ }\textbf {\bibinfo {volume} {106}},\ \bibinfo {pages} {L041105} (\bibinfo {year} {2022})}\BibitemShut {NoStop}%
\bibitem [{\citenamefont {Casula}\ \emph {et~al.}(2005)\citenamefont {Casula}, \citenamefont {Filippi},\ and\ \citenamefont {Sorella}}]{Casula2005}%
  \BibitemOpen
  \bibfield  {author} {\bibinfo {author} {\bibfnamefont {M.}~\bibnamefont {Casula}}, \bibinfo {author} {\bibfnamefont {C.}~\bibnamefont {Filippi}},\ and\ \bibinfo {author} {\bibfnamefont {S.}~\bibnamefont {Sorella}},\ }\href {https://doi.org/10.1103/PhysRevLett.95.100201} {\bibfield  {journal} {\bibinfo  {journal} {Phys. Rev. Lett.}\ }\textbf {\bibinfo {volume} {95}},\ \bibinfo {pages} {100201} (\bibinfo {year} {2005})}\BibitemShut {NoStop}%
\bibitem [{\citenamefont {Nakano}\ \emph {et~al.}(2020{\natexlab{a}})\citenamefont {Nakano}, \citenamefont {Maezono},\ and\ \citenamefont {Sorella}}]{Nakano2020b}%
  \BibitemOpen
  \bibfield  {author} {\bibinfo {author} {\bibfnamefont {K.}~\bibnamefont {Nakano}}, \bibinfo {author} {\bibfnamefont {R.}~\bibnamefont {Maezono}},\ and\ \bibinfo {author} {\bibfnamefont {S.}~\bibnamefont {Sorella}},\ }\href {https://doi.org/10.1103/PhysRevB.101.155106} {\bibfield  {journal} {\bibinfo  {journal} {Phys. Rev. B}\ }\textbf {\bibinfo {volume} {101}},\ \bibinfo {pages} {155106} (\bibinfo {year} {2020}{\natexlab{a}})}\BibitemShut {NoStop}%
\bibitem [{\citenamefont {Perdew}\ and\ \citenamefont {Zunger}(1981)}]{Perdew1981}%
  \BibitemOpen
  \bibfield  {author} {\bibinfo {author} {\bibfnamefont {J.~P.}\ \bibnamefont {Perdew}}\ and\ \bibinfo {author} {\bibfnamefont {A.}~\bibnamefont {Zunger}},\ }\href {https://doi.org/10.1103/PhysRevB.23.5048} {\bibfield  {journal} {\bibinfo  {journal} {Phys. Rev. B}\ }\textbf {\bibinfo {volume} {23}},\ \bibinfo {pages} {5048} (\bibinfo {year} {1981})}\BibitemShut {NoStop}%
\bibitem [{\citenamefont {Perdew}\ \emph {et~al.}(1996)\citenamefont {Perdew}, \citenamefont {Burke},\ and\ \citenamefont {Ernzerhof}}]{Perdew1996}%
  \BibitemOpen
  \bibfield  {author} {\bibinfo {author} {\bibfnamefont {J.~P.}\ \bibnamefont {Perdew}}, \bibinfo {author} {\bibfnamefont {K.}~\bibnamefont {Burke}},\ and\ \bibinfo {author} {\bibfnamefont {M.}~\bibnamefont {Ernzerhof}},\ }\href {https://doi.org/10.1103/PhysRevLett.77.3865} {\bibfield  {journal} {\bibinfo  {journal} {Phys. Rev. Lett.}\ }\textbf {\bibinfo {volume} {77}},\ \bibinfo {pages} {3865} (\bibinfo {year} {1996})}\BibitemShut {NoStop}%
\bibitem [{\citenamefont {Giannozzi}\ \emph {et~al.}(2009)\citenamefont {Giannozzi}, \citenamefont {Baroni}, \citenamefont {Bonini}, \citenamefont {Calandra}, \citenamefont {Car}, \citenamefont {Cavazzoni}, \citenamefont {Ceresoli}, \citenamefont {Chiarotti}, \citenamefont {Cococcioni}, \citenamefont {Dabo}, \citenamefont {Corso}, \citenamefont {de~Gironcoli}, \citenamefont {Fabris}, \citenamefont {Fratesi}, \citenamefont {Gebauer}, \citenamefont {Gerstmann}, \citenamefont {Gougoussis}, \citenamefont {Kokalj}, \citenamefont {Lazzeri}, \citenamefont {Martin-Samos}, \citenamefont {Marzari}, \citenamefont {Mauri}, \citenamefont {Mazzarello}, \citenamefont {Paolini}, \citenamefont {Pasquarello}, \citenamefont {Paulatto}, \citenamefont {Sbraccia}, \citenamefont {Scandolo}, \citenamefont {Sclauzero}, \citenamefont {Seitsonen}, \citenamefont {Smogunov}, \citenamefont {Umari},\ and\ \citenamefont {Wentzcovitch}}]{Giannozzi2009}%
  \BibitemOpen
  \bibfield  {author} {\bibinfo {author} {\bibfnamefont {P.}~\bibnamefont {Giannozzi}}, \bibinfo {author} {\bibfnamefont {S.}~\bibnamefont {Baroni}}, \bibinfo {author} {\bibfnamefont {N.}~\bibnamefont {Bonini}}, \bibinfo {author} {\bibfnamefont {M.}~\bibnamefont {Calandra}}, \bibinfo {author} {\bibfnamefont {R.}~\bibnamefont {Car}}, \bibinfo {author} {\bibfnamefont {C.}~\bibnamefont {Cavazzoni}}, \bibinfo {author} {\bibfnamefont {D.}~\bibnamefont {Ceresoli}}, \bibinfo {author} {\bibfnamefont {G.~L.}\ \bibnamefont {Chiarotti}}, \bibinfo {author} {\bibfnamefont {M.}~\bibnamefont {Cococcioni}}, \bibinfo {author} {\bibfnamefont {I.}~\bibnamefont {Dabo}}, \bibinfo {author} {\bibfnamefont {A.~D.}\ \bibnamefont {Corso}}, \bibinfo {author} {\bibfnamefont {S.}~\bibnamefont {de~Gironcoli}}, \bibinfo {author} {\bibfnamefont {S.}~\bibnamefont {Fabris}}, \bibinfo {author} {\bibfnamefont {G.}~\bibnamefont {Fratesi}}, \bibinfo {author} {\bibfnamefont {R.}~\bibnamefont {Gebauer}}, \bibinfo {author} {\bibfnamefont
  {U.}~\bibnamefont {Gerstmann}}, \bibinfo {author} {\bibfnamefont {C.}~\bibnamefont {Gougoussis}}, \bibinfo {author} {\bibfnamefont {A.}~\bibnamefont {Kokalj}}, \bibinfo {author} {\bibfnamefont {M.}~\bibnamefont {Lazzeri}}, \bibinfo {author} {\bibfnamefont {L.}~\bibnamefont {Martin-Samos}}, \bibinfo {author} {\bibfnamefont {N.}~\bibnamefont {Marzari}}, \bibinfo {author} {\bibfnamefont {F.}~\bibnamefont {Mauri}}, \bibinfo {author} {\bibfnamefont {R.}~\bibnamefont {Mazzarello}}, \bibinfo {author} {\bibfnamefont {S.}~\bibnamefont {Paolini}}, \bibinfo {author} {\bibfnamefont {A.}~\bibnamefont {Pasquarello}}, \bibinfo {author} {\bibfnamefont {L.}~\bibnamefont {Paulatto}}, \bibinfo {author} {\bibfnamefont {C.}~\bibnamefont {Sbraccia}}, \bibinfo {author} {\bibfnamefont {S.}~\bibnamefont {Scandolo}}, \bibinfo {author} {\bibfnamefont {G.}~\bibnamefont {Sclauzero}}, \bibinfo {author} {\bibfnamefont {A.~P.}\ \bibnamefont {Seitsonen}}, \bibinfo {author} {\bibfnamefont {A.}~\bibnamefont {Smogunov}}, \bibinfo {author}
  {\bibfnamefont {P.}~\bibnamefont {Umari}},\ and\ \bibinfo {author} {\bibfnamefont {R.~M.}\ \bibnamefont {Wentzcovitch}},\ }\href {https://doi.org/10.1088/0953-8984/21/39/395502} {\bibfield  {journal} {\bibinfo  {journal} {Journal of Physics: Condensed Matter}\ }\textbf {\bibinfo {volume} {21}},\ \bibinfo {pages} {395502} (\bibinfo {year} {2009})}\BibitemShut {NoStop}%
\bibitem [{\citenamefont {Giannozzi}\ \emph {et~al.}(2017)\citenamefont {Giannozzi}, \citenamefont {Andreussi}, \citenamefont {Brumme}, \citenamefont {Bunau}, \citenamefont {Nardelli}, \citenamefont {Calandra}, \citenamefont {Car}, \citenamefont {Cavazzoni}, \citenamefont {Ceresoli}, \citenamefont {Cococcioni}, \citenamefont {Colonna}, \citenamefont {Carnimeo}, \citenamefont {Corso}, \citenamefont {de~Gironcoli}, \citenamefont {Delugas}, \citenamefont {DiStasio}, \citenamefont {Ferretti}, \citenamefont {Floris}, \citenamefont {Fratesi}, \citenamefont {Fugallo}, \citenamefont {Gebauer}, \citenamefont {Gerstmann}, \citenamefont {Giustino}, \citenamefont {Gorni}, \citenamefont {Jia}, \citenamefont {Kawamura}, \citenamefont {Ko}, \citenamefont {Kokalj}, \citenamefont {K\"{u}{\c{c}}\"{u}kbenli}, \citenamefont {Lazzeri}, \citenamefont {Marsili}, \citenamefont {Marzari}, \citenamefont {Mauri}, \citenamefont {Nguyen}, \citenamefont {Nguyen}, \citenamefont {de-la Roza}, \citenamefont {Paulatto}, \citenamefont
  {Ponc{\'{e}}}, \citenamefont {Rocca}, \citenamefont {Sabatini}, \citenamefont {Santra}, \citenamefont {Schlipf}, \citenamefont {Seitsonen}, \citenamefont {Smogunov}, \citenamefont {Timrov}, \citenamefont {Thonhauser}, \citenamefont {Umari}, \citenamefont {Vast}, \citenamefont {Wu},\ and\ \citenamefont {Baroni}}]{Giannozzi2017}%
  \BibitemOpen
  \bibfield  {author} {\bibinfo {author} {\bibfnamefont {P.}~\bibnamefont {Giannozzi}}, \bibinfo {author} {\bibfnamefont {O.}~\bibnamefont {Andreussi}}, \bibinfo {author} {\bibfnamefont {T.}~\bibnamefont {Brumme}}, \bibinfo {author} {\bibfnamefont {O.}~\bibnamefont {Bunau}}, \bibinfo {author} {\bibfnamefont {M.~B.}\ \bibnamefont {Nardelli}}, \bibinfo {author} {\bibfnamefont {M.}~\bibnamefont {Calandra}}, \bibinfo {author} {\bibfnamefont {R.}~\bibnamefont {Car}}, \bibinfo {author} {\bibfnamefont {C.}~\bibnamefont {Cavazzoni}}, \bibinfo {author} {\bibfnamefont {D.}~\bibnamefont {Ceresoli}}, \bibinfo {author} {\bibfnamefont {M.}~\bibnamefont {Cococcioni}}, \bibinfo {author} {\bibfnamefont {N.}~\bibnamefont {Colonna}}, \bibinfo {author} {\bibfnamefont {I.}~\bibnamefont {Carnimeo}}, \bibinfo {author} {\bibfnamefont {A.~D.}\ \bibnamefont {Corso}}, \bibinfo {author} {\bibfnamefont {S.}~\bibnamefont {de~Gironcoli}}, \bibinfo {author} {\bibfnamefont {P.}~\bibnamefont {Delugas}}, \bibinfo {author} {\bibfnamefont
  {R.~A.}\ \bibnamefont {DiStasio}}, \bibinfo {author} {\bibfnamefont {A.}~\bibnamefont {Ferretti}}, \bibinfo {author} {\bibfnamefont {A.}~\bibnamefont {Floris}}, \bibinfo {author} {\bibfnamefont {G.}~\bibnamefont {Fratesi}}, \bibinfo {author} {\bibfnamefont {G.}~\bibnamefont {Fugallo}}, \bibinfo {author} {\bibfnamefont {R.}~\bibnamefont {Gebauer}}, \bibinfo {author} {\bibfnamefont {U.}~\bibnamefont {Gerstmann}}, \bibinfo {author} {\bibfnamefont {F.}~\bibnamefont {Giustino}}, \bibinfo {author} {\bibfnamefont {T.}~\bibnamefont {Gorni}}, \bibinfo {author} {\bibfnamefont {J.}~\bibnamefont {Jia}}, \bibinfo {author} {\bibfnamefont {M.}~\bibnamefont {Kawamura}}, \bibinfo {author} {\bibfnamefont {H.-Y.}\ \bibnamefont {Ko}}, \bibinfo {author} {\bibfnamefont {A.}~\bibnamefont {Kokalj}}, \bibinfo {author} {\bibfnamefont {E.}~\bibnamefont {K\"{u}{\c{c}}\"{u}kbenli}}, \bibinfo {author} {\bibfnamefont {M.}~\bibnamefont {Lazzeri}}, \bibinfo {author} {\bibfnamefont {M.}~\bibnamefont {Marsili}}, \bibinfo {author}
  {\bibfnamefont {N.}~\bibnamefont {Marzari}}, \bibinfo {author} {\bibfnamefont {F.}~\bibnamefont {Mauri}}, \bibinfo {author} {\bibfnamefont {N.~L.}\ \bibnamefont {Nguyen}}, \bibinfo {author} {\bibfnamefont {H.-V.}\ \bibnamefont {Nguyen}}, \bibinfo {author} {\bibfnamefont {A.~O.}\ \bibnamefont {de-la Roza}}, \bibinfo {author} {\bibfnamefont {L.}~\bibnamefont {Paulatto}}, \bibinfo {author} {\bibfnamefont {S.}~\bibnamefont {Ponc{\'{e}}}}, \bibinfo {author} {\bibfnamefont {D.}~\bibnamefont {Rocca}}, \bibinfo {author} {\bibfnamefont {R.}~\bibnamefont {Sabatini}}, \bibinfo {author} {\bibfnamefont {B.}~\bibnamefont {Santra}}, \bibinfo {author} {\bibfnamefont {M.}~\bibnamefont {Schlipf}}, \bibinfo {author} {\bibfnamefont {A.~P.}\ \bibnamefont {Seitsonen}}, \bibinfo {author} {\bibfnamefont {A.}~\bibnamefont {Smogunov}}, \bibinfo {author} {\bibfnamefont {I.}~\bibnamefont {Timrov}}, \bibinfo {author} {\bibfnamefont {T.}~\bibnamefont {Thonhauser}}, \bibinfo {author} {\bibfnamefont {P.}~\bibnamefont {Umari}}, \bibinfo
  {author} {\bibfnamefont {N.}~\bibnamefont {Vast}}, \bibinfo {author} {\bibfnamefont {X.}~\bibnamefont {Wu}},\ and\ \bibinfo {author} {\bibfnamefont {S.}~\bibnamefont {Baroni}},\ }\href {https://doi.org/10.1088/1361-648x/aa8f79} {\bibfield  {journal} {\bibinfo  {journal} {Journal of Physics: Condensed Matter}\ }\textbf {\bibinfo {volume} {29}},\ \bibinfo {pages} {465901} (\bibinfo {year} {2017})}\BibitemShut {NoStop}%
\bibitem [{\citenamefont {Giannozzi}\ \emph {et~al.}(2020)\citenamefont {Giannozzi}, \citenamefont {Baseggio}, \citenamefont {Bonfà}, \citenamefont {Brunato}, \citenamefont {Car}, \citenamefont {Carnimeo}, \citenamefont {Cavazzoni}, \citenamefont {de~Gironcoli}, \citenamefont {Delugas}, \citenamefont {Ferrari~Ruffino}, \citenamefont {Ferretti}, \citenamefont {Marzari}, \citenamefont {Timrov}, \citenamefont {Urru},\ and\ \citenamefont {Baroni}}]{Giannozzi2020}%
  \BibitemOpen
  \bibfield  {author} {\bibinfo {author} {\bibfnamefont {P.}~\bibnamefont {Giannozzi}}, \bibinfo {author} {\bibfnamefont {O.}~\bibnamefont {Baseggio}}, \bibinfo {author} {\bibfnamefont {P.}~\bibnamefont {Bonfà}}, \bibinfo {author} {\bibfnamefont {D.}~\bibnamefont {Brunato}}, \bibinfo {author} {\bibfnamefont {R.}~\bibnamefont {Car}}, \bibinfo {author} {\bibfnamefont {I.}~\bibnamefont {Carnimeo}}, \bibinfo {author} {\bibfnamefont {C.}~\bibnamefont {Cavazzoni}}, \bibinfo {author} {\bibfnamefont {S.}~\bibnamefont {de~Gironcoli}}, \bibinfo {author} {\bibfnamefont {P.}~\bibnamefont {Delugas}}, \bibinfo {author} {\bibfnamefont {F.}~\bibnamefont {Ferrari~Ruffino}}, \bibinfo {author} {\bibfnamefont {A.}~\bibnamefont {Ferretti}}, \bibinfo {author} {\bibfnamefont {N.}~\bibnamefont {Marzari}}, \bibinfo {author} {\bibfnamefont {I.}~\bibnamefont {Timrov}}, \bibinfo {author} {\bibfnamefont {A.}~\bibnamefont {Urru}},\ and\ \bibinfo {author} {\bibfnamefont {S.}~\bibnamefont {Baroni}},\ }\href
  {https://doi.org/10.1063/5.0005082} {\bibfield  {journal} {\bibinfo  {journal} {The Journal of Chemical Physics}\ }\textbf {\bibinfo {volume} {152}},\ \bibinfo {pages} {154105} (\bibinfo {year} {2020})},\ \Eprint {https://arxiv.org/abs/https://doi.org/10.1063/5.0005082} {https://doi.org/10.1063/5.0005082} \BibitemShut {NoStop}%
\bibitem [{PAW()}]{PAWpseudo}%
  \BibitemOpen
  \href@noop {} {}\bibinfo {howpublished} {{\textsc{H.pbek-jpaw\_psl.1.0.0.UPF}} pseudopotential available at \url{http://pseudopotentials.quantum-espresso.org/legacy_tables/ps-library/h}}\BibitemShut {NoStop}%
\bibitem [{\citenamefont {Ricci}\ and\ \citenamefont {Ciccotti}(2003)}]{Ricci2003}%
  \BibitemOpen
  \bibfield  {author} {\bibinfo {author} {\bibfnamefont {A.}~\bibnamefont {Ricci}}\ and\ \bibinfo {author} {\bibfnamefont {G.}~\bibnamefont {Ciccotti}},\ }\href {https://doi.org/10.1080/0026897031000108113} {\bibfield  {journal} {\bibinfo  {journal} {Molecular Physics - MOL PHYS}\ }\textbf {\bibinfo {volume} {101}},\ \bibinfo {pages} {1927} (\bibinfo {year} {2003})}\BibitemShut {NoStop}%
\bibitem [{\citenamefont {Attaccalite}\ and\ \citenamefont {Sorella}(2008)}]{Attaccalite2008}%
  \BibitemOpen
  \bibfield  {author} {\bibinfo {author} {\bibfnamefont {C.}~\bibnamefont {Attaccalite}}\ and\ \bibinfo {author} {\bibfnamefont {S.}~\bibnamefont {Sorella}},\ }\href@noop {} {\bibfield  {journal} {\bibinfo  {journal} {Phys. Rev. Lett.}\ }\textbf {\bibinfo {volume} {100}},\ \bibinfo {pages} {114501} (\bibinfo {year} {2008})}\BibitemShut {NoStop}%
\bibitem [{\citenamefont {Nakano}\ \emph {et~al.}(2020{\natexlab{b}})\citenamefont {Nakano}, \citenamefont {Attaccalite}, \citenamefont {Barborini}, \citenamefont {Capriotti}, \citenamefont {Casula}, \citenamefont {Coccia}, \citenamefont {Dagrada}, \citenamefont {Genovese}, \citenamefont {Luo}, \citenamefont {Mazzola}, \citenamefont {Zen},\ and\ \citenamefont {Sorella}}]{Nakano2020}%
  \BibitemOpen
  \bibfield  {author} {\bibinfo {author} {\bibfnamefont {K.}~\bibnamefont {Nakano}}, \bibinfo {author} {\bibfnamefont {C.}~\bibnamefont {Attaccalite}}, \bibinfo {author} {\bibfnamefont {M.}~\bibnamefont {Barborini}}, \bibinfo {author} {\bibfnamefont {L.}~\bibnamefont {Capriotti}}, \bibinfo {author} {\bibfnamefont {M.}~\bibnamefont {Casula}}, \bibinfo {author} {\bibfnamefont {E.}~\bibnamefont {Coccia}}, \bibinfo {author} {\bibfnamefont {M.}~\bibnamefont {Dagrada}}, \bibinfo {author} {\bibfnamefont {C.}~\bibnamefont {Genovese}}, \bibinfo {author} {\bibfnamefont {Y.}~\bibnamefont {Luo}}, \bibinfo {author} {\bibfnamefont {G.}~\bibnamefont {Mazzola}}, \bibinfo {author} {\bibfnamefont {A.}~\bibnamefont {Zen}},\ and\ \bibinfo {author} {\bibfnamefont {S.}~\bibnamefont {Sorella}},\ }\href@noop {} {\bibfield  {journal} {\bibinfo  {journal} {J. Chem. Phys.}\ }\textbf {\bibinfo {volume} {152}},\ \bibinfo {pages} {204121} (\bibinfo {year} {2020}{\natexlab{b}})}\BibitemShut {NoStop}%
\bibitem [{\citenamefont {Nakano}\ \emph {et~al.}(2023{\natexlab{a}})\citenamefont {Nakano}, \citenamefont {Kohulák}, \citenamefont {Raghav}, \citenamefont {Casula},\ and\ \citenamefont {Sorella}}]{2023turbogenius}%
  \BibitemOpen
  \bibfield  {author} {\bibinfo {author} {\bibfnamefont {K.}~\bibnamefont {Nakano}}, \bibinfo {author} {\bibfnamefont {O.}~\bibnamefont {Kohulák}}, \bibinfo {author} {\bibfnamefont {A.}~\bibnamefont {Raghav}}, \bibinfo {author} {\bibfnamefont {M.}~\bibnamefont {Casula}},\ and\ \bibinfo {author} {\bibfnamefont {S.}~\bibnamefont {Sorella}},\ }\href {https://doi.org/10.1063/5.0179003} {\bibfield  {journal} {\bibinfo  {journal} {J. Chem. Phys.}\ }\textbf {\bibinfo {volume} {159}},\ \bibinfo {pages} {224801} (\bibinfo {year} {2023}{\natexlab{a}})}\BibitemShut {NoStop}%
\bibitem [{\citenamefont {Tiihonen}\ \emph {et~al.}(2021)\citenamefont {Tiihonen}, \citenamefont {Clay~III},\ and\ \citenamefont {Krogel}}]{Tiihonen2021}%
  \BibitemOpen
  \bibfield  {author} {\bibinfo {author} {\bibfnamefont {J.}~\bibnamefont {Tiihonen}}, \bibinfo {author} {\bibfnamefont {R.~C.}\ \bibnamefont {Clay~III}},\ and\ \bibinfo {author} {\bibfnamefont {J.~T.}\ \bibnamefont {Krogel}},\ }\href {https://doi.org/10.1063/5.0052266} {\bibfield  {journal} {\bibinfo  {journal} {J. Chem. Phys.}\ }\textbf {\bibinfo {volume} {154}},\ \bibinfo {pages} {204111} (\bibinfo {year} {2021})}\BibitemShut {NoStop}%
\bibitem [{\citenamefont {Nakano}\ \emph {et~al.}(2022)\citenamefont {Nakano}, \citenamefont {Raghav},\ and\ \citenamefont {Sorella}}]{Nakano2022}%
  \BibitemOpen
  \bibfield  {author} {\bibinfo {author} {\bibfnamefont {K.}~\bibnamefont {Nakano}}, \bibinfo {author} {\bibfnamefont {A.}~\bibnamefont {Raghav}},\ and\ \bibinfo {author} {\bibfnamefont {S.}~\bibnamefont {Sorella}},\ }\href@noop {} {\bibfield  {journal} {\bibinfo  {journal} {The Journal of Chemical Physics}\ }\textbf {\bibinfo {volume} {156}},\ \bibinfo {pages} {034101} (\bibinfo {year} {2022})}\BibitemShut {NoStop}%
\bibitem [{\citenamefont {Nakano}\ \emph {et~al.}(2023{\natexlab{b}})\citenamefont {Nakano}, \citenamefont {Casula},\ and\ \citenamefont {Tenti}}]{Nakano2023}%
  \BibitemOpen
  \bibfield  {author} {\bibinfo {author} {\bibfnamefont {K.}~\bibnamefont {Nakano}}, \bibinfo {author} {\bibfnamefont {M.}~\bibnamefont {Casula}},\ and\ \bibinfo {author} {\bibfnamefont {G.}~\bibnamefont {Tenti}},\ }\href@noop {} {\  (\bibinfo {year} {2023}{\natexlab{b}})},\ \Eprint {https://arxiv.org/abs/arXiv:2312.17608} {arXiv:2312.17608} \BibitemShut {NoStop}%
\bibitem [{SM()}]{SM}%
  \BibitemOpen
  \href@noop {} {}\bibinfo {howpublished} {See Supplemental Material at [\emph{URL will be inserted by publisher}] for additional information about the computational details of QMC calculations, the MLP training and validation, the reference state calculations, finite-size corrections, finite temperature DFT simulations, and comparison with previous results \cite{Casula2005, Nakano2020b, Casula2003, sorella2015geminal, 2017BEC, Tiihonen2021, Nakano2021, Nakano2022, Nakano2023, Umrigar1989,Sorella2010,Claudia2016, Attaccalite2008, vanRhijn2021, Pathak2020, Reynolds1986, Perdew1981, Tirelli2022, Mouhat2017, Giannozzi2009,Giannozzi2017,Giannozzi2020, Lee1988, Kwee2008, Mermin1965, Karasiev2019, Ruggeri2020, Clay2019}}\BibitemShut {NoStop}%
\bibitem [{\citenamefont {Karasiev}\ \emph {et~al.}(2014)\citenamefont {Karasiev}, \citenamefont {Sjostrom}, \citenamefont {Dufty},\ and\ \citenamefont {Trickey}}]{Karasiev2014}%
  \BibitemOpen
  \bibfield  {author} {\bibinfo {author} {\bibfnamefont {V.~V.}\ \bibnamefont {Karasiev}}, \bibinfo {author} {\bibfnamefont {T.}~\bibnamefont {Sjostrom}}, \bibinfo {author} {\bibfnamefont {J.}~\bibnamefont {Dufty}},\ and\ \bibinfo {author} {\bibfnamefont {S.~B.}\ \bibnamefont {Trickey}},\ }\href {https://doi.org/10.1103/PhysRevLett.112.076403} {\bibfield  {journal} {\bibinfo  {journal} {Phys. Rev. Lett.}\ }\textbf {\bibinfo {volume} {112}},\ \bibinfo {pages} {076403} (\bibinfo {year} {2014})}\BibitemShut {NoStop}%
\bibitem [{\citenamefont {Karasiev}\ \emph {et~al.}(2018)\citenamefont {Karasiev}, \citenamefont {Dufty},\ and\ \citenamefont {Trickey}}]{Karasiev2018}%
  \BibitemOpen
  \bibfield  {author} {\bibinfo {author} {\bibfnamefont {V.~V.}\ \bibnamefont {Karasiev}}, \bibinfo {author} {\bibfnamefont {J.~W.}\ \bibnamefont {Dufty}},\ and\ \bibinfo {author} {\bibfnamefont {S.~B.}\ \bibnamefont {Trickey}},\ }\href {https://doi.org/10.1103/PhysRevLett.120.076401} {\bibfield  {journal} {\bibinfo  {journal} {Phys. Rev. Lett.}\ }\textbf {\bibinfo {volume} {120}},\ \bibinfo {pages} {076401} (\bibinfo {year} {2018})}\BibitemShut {NoStop}%
\bibitem [{\citenamefont {Ben~Mahmoud}\ \emph {et~al.}(2022)\citenamefont {Ben~Mahmoud}, \citenamefont {Grasselli},\ and\ \citenamefont {Ceriotti}}]{Mahmoud2022}%
  \BibitemOpen
  \bibfield  {author} {\bibinfo {author} {\bibfnamefont {C.}~\bibnamefont {Ben~Mahmoud}}, \bibinfo {author} {\bibfnamefont {F.}~\bibnamefont {Grasselli}},\ and\ \bibinfo {author} {\bibfnamefont {M.}~\bibnamefont {Ceriotti}},\ }\href {https://doi.org/10.1103/PhysRevB.106.L121116} {\bibfield  {journal} {\bibinfo  {journal} {Phys. Rev. B}\ }\textbf {\bibinfo {volume} {106}},\ \bibinfo {pages} {L121116} (\bibinfo {year} {2022})}\BibitemShut {NoStop}%
\bibitem [{\citenamefont {Dion}\ \emph {et~al.}(2004)\citenamefont {Dion}, \citenamefont {Rydberg}, \citenamefont {Schr\"oder}, \citenamefont {Langreth},\ and\ \citenamefont {Lundqvist}}]{Dion2004}%
  \BibitemOpen
  \bibfield  {author} {\bibinfo {author} {\bibfnamefont {M.}~\bibnamefont {Dion}}, \bibinfo {author} {\bibfnamefont {H.}~\bibnamefont {Rydberg}}, \bibinfo {author} {\bibfnamefont {E.}~\bibnamefont {Schr\"oder}}, \bibinfo {author} {\bibfnamefont {D.~C.}\ \bibnamefont {Langreth}},\ and\ \bibinfo {author} {\bibfnamefont {B.~I.}\ \bibnamefont {Lundqvist}},\ }\href {https://doi.org/10.1103/PhysRevLett.92.246401} {\bibfield  {journal} {\bibinfo  {journal} {Phys. Rev. Lett.}\ }\textbf {\bibinfo {volume} {92}},\ \bibinfo {pages} {246401} (\bibinfo {year} {2004})}\BibitemShut {NoStop}%
\bibitem [{\citenamefont {Berland}\ \emph {et~al.}(2015)\citenamefont {Berland}, \citenamefont {Cooper}, \citenamefont {Lee}, \citenamefont {Schr\"{o}der}, \citenamefont {Thonhauser}, \citenamefont {Hyldgaard},\ and\ \citenamefont {Lundqvist}}]{Berland2015}%
  \BibitemOpen
  \bibfield  {author} {\bibinfo {author} {\bibfnamefont {K.}~\bibnamefont {Berland}}, \bibinfo {author} {\bibfnamefont {V.~R.}\ \bibnamefont {Cooper}}, \bibinfo {author} {\bibfnamefont {K.}~\bibnamefont {Lee}}, \bibinfo {author} {\bibfnamefont {E.}~\bibnamefont {Schr\"{o}der}}, \bibinfo {author} {\bibfnamefont {T.}~\bibnamefont {Thonhauser}}, \bibinfo {author} {\bibfnamefont {P.}~\bibnamefont {Hyldgaard}},\ and\ \bibinfo {author} {\bibfnamefont {B.~I.}\ \bibnamefont {Lundqvist}},\ }\href {https://doi.org/10.1088/0034-4885/78/6/066501} {\bibfield  {journal} {\bibinfo  {journal} {Reports on Progress in Physics}\ }\textbf {\bibinfo {volume} {78}},\ \bibinfo {pages} {066501} (\bibinfo {year} {2015})}\BibitemShut {NoStop}%
\bibitem [{\citenamefont {Kwee}\ \emph {et~al.}(2008)\citenamefont {Kwee}, \citenamefont {Zhang},\ and\ \citenamefont {Krakauer}}]{Kwee2008}%
  \BibitemOpen
  \bibfield  {author} {\bibinfo {author} {\bibfnamefont {H.}~\bibnamefont {Kwee}}, \bibinfo {author} {\bibfnamefont {S.}~\bibnamefont {Zhang}},\ and\ \bibinfo {author} {\bibfnamefont {H.}~\bibnamefont {Krakauer}},\ }\href {https://doi.org/10.1103/PhysRevLett.100.126404} {\bibfield  {journal} {\bibinfo  {journal} {Phys. Rev. Lett.}\ }\textbf {\bibinfo {volume} {100}},\ \bibinfo {pages} {126404} (\bibinfo {year} {2008})}\BibitemShut {NoStop}%
\bibitem [{\citenamefont {Mouhat}\ \emph {et~al.}(2017)\citenamefont {Mouhat}, \citenamefont {Sorella}, \citenamefont {Vuilleumier}, \citenamefont {Saitta},\ and\ \citenamefont {Casula}}]{Mouhat2017}%
  \BibitemOpen
  \bibfield  {author} {\bibinfo {author} {\bibfnamefont {F.}~\bibnamefont {Mouhat}}, \bibinfo {author} {\bibfnamefont {S.}~\bibnamefont {Sorella}}, \bibinfo {author} {\bibfnamefont {R.}~\bibnamefont {Vuilleumier}}, \bibinfo {author} {\bibfnamefont {A.~M.}\ \bibnamefont {Saitta}},\ and\ \bibinfo {author} {\bibfnamefont {M.}~\bibnamefont {Casula}},\ }\href {https://doi.org/10.1021/acs.jctc.7b00017} {\bibfield  {journal} {\bibinfo  {journal} {Journal of Chemical Theory and Computation}\ }\textbf {\bibinfo {volume} {13}},\ \bibinfo {pages} {2400} (\bibinfo {year} {2017})}\BibitemShut {NoStop}%
\bibitem [{\citenamefont {Casula}\ and\ \citenamefont {Sorella}(2003)}]{Casula2003}%
  \BibitemOpen
  \bibfield  {author} {\bibinfo {author} {\bibfnamefont {M.}~\bibnamefont {Casula}}\ and\ \bibinfo {author} {\bibfnamefont {S.}~\bibnamefont {Sorella}},\ }\href@noop {} {\bibfield  {journal} {\bibinfo  {journal} {J. Chem. Phys.}\ }\textbf {\bibinfo {volume} {119}},\ \bibinfo {pages} {6500} (\bibinfo {year} {2003})}\BibitemShut {NoStop}%
\bibitem [{\citenamefont {Sorella}\ \emph {et~al.}(2015)\citenamefont {Sorella}, \citenamefont {Devaux}, \citenamefont {Dagrada}, \citenamefont {Mazzola},\ and\ \citenamefont {Casula}}]{sorella2015geminal}%
  \BibitemOpen
  \bibfield  {author} {\bibinfo {author} {\bibfnamefont {S.}~\bibnamefont {Sorella}}, \bibinfo {author} {\bibfnamefont {N.}~\bibnamefont {Devaux}}, \bibinfo {author} {\bibfnamefont {M.}~\bibnamefont {Dagrada}}, \bibinfo {author} {\bibfnamefont {G.}~\bibnamefont {Mazzola}},\ and\ \bibinfo {author} {\bibfnamefont {M.}~\bibnamefont {Casula}},\ }\href@noop {} {\bibfield  {journal} {\bibinfo  {journal} {J. Chem. Phys.}\ }\textbf {\bibinfo {volume} {143}} (\bibinfo {year} {2015})}\BibitemShut {NoStop}%
\bibitem [{\citenamefont {Becca}\ and\ \citenamefont {Sorella}(2017)}]{2017BEC}%
  \BibitemOpen
  \bibfield  {author} {\bibinfo {author} {\bibfnamefont {F.}~\bibnamefont {Becca}}\ and\ \bibinfo {author} {\bibfnamefont {S.}~\bibnamefont {Sorella}},\ }\href@noop {} {\emph {\bibinfo {title} {{Quantum Monte Carlo approaches for correlated systems}}}}\ (\bibinfo  {publisher} {Cambridge University Press},\ \bibinfo {year} {2017})\BibitemShut {NoStop}%
\bibitem [{\citenamefont {Nakano}\ \emph {et~al.}(2021)\citenamefont {Nakano}, \citenamefont {Morresi}, \citenamefont {Casula}, \citenamefont {Maezono},\ and\ \citenamefont {Sorella}}]{Nakano2021}%
  \BibitemOpen
  \bibfield  {author} {\bibinfo {author} {\bibfnamefont {K.}~\bibnamefont {Nakano}}, \bibinfo {author} {\bibfnamefont {T.}~\bibnamefont {Morresi}}, \bibinfo {author} {\bibfnamefont {M.}~\bibnamefont {Casula}}, \bibinfo {author} {\bibfnamefont {R.}~\bibnamefont {Maezono}},\ and\ \bibinfo {author} {\bibfnamefont {S.}~\bibnamefont {Sorella}},\ }\href {https://doi.org/10.1103/PhysRevB.103.L121110} {\bibfield  {journal} {\bibinfo  {journal} {Phys. Rev. B}\ }\textbf {\bibinfo {volume} {103}},\ \bibinfo {pages} {L121110} (\bibinfo {year} {2021})}\BibitemShut {NoStop}%
\bibitem [{\citenamefont {Umrigar}(1989)}]{Umrigar1989}%
  \BibitemOpen
  \bibfield  {author} {\bibinfo {author} {\bibfnamefont {C.~J.}\ \bibnamefont {Umrigar}},\ }\href@noop {} {\bibfield  {journal} {\bibinfo  {journal} {Int. J. Quantum Chem}\ }\textbf {\bibinfo {volume} {36}},\ \bibinfo {pages} {217} (\bibinfo {year} {1989})}\BibitemShut {NoStop}%
\bibitem [{\citenamefont {Sorella}\ and\ \citenamefont {Capriotti}(2010)}]{Sorella2010}%
  \BibitemOpen
  \bibfield  {author} {\bibinfo {author} {\bibfnamefont {S.}~\bibnamefont {Sorella}}\ and\ \bibinfo {author} {\bibfnamefont {L.}~\bibnamefont {Capriotti}},\ }\href@noop {} {\bibfield  {journal} {\bibinfo  {journal} {J. Chem. Phys.}\ }\textbf {\bibinfo {volume} {133}},\ \bibinfo {pages} {234111} (\bibinfo {year} {2010})}\BibitemShut {NoStop}%
\bibitem [{\citenamefont {Filippi}\ \emph {et~al.}(2016)\citenamefont {Filippi}, \citenamefont {Assaraf},\ and\ \citenamefont {Moroni}}]{Claudia2016}%
  \BibitemOpen
  \bibfield  {author} {\bibinfo {author} {\bibfnamefont {C.}~\bibnamefont {Filippi}}, \bibinfo {author} {\bibfnamefont {R.}~\bibnamefont {Assaraf}},\ and\ \bibinfo {author} {\bibfnamefont {S.}~\bibnamefont {Moroni}},\ }\href {https://doi.org/10.1063/1.4948778} {\bibfield  {journal} {\bibinfo  {journal} {J. Chem. Phys.}\ }\textbf {\bibinfo {volume} {144}},\ \bibinfo {pages} {194105} (\bibinfo {year} {2016})}\BibitemShut {NoStop}%
\bibitem [{\citenamefont {van Rhijn}\ \emph {et~al.}(2021)\citenamefont {van Rhijn}, \citenamefont {Filippi}, \citenamefont {De~Palo},\ and\ \citenamefont {Moroni}}]{vanRhijn2021}%
  \BibitemOpen
  \bibfield  {author} {\bibinfo {author} {\bibfnamefont {J.}~\bibnamefont {van Rhijn}}, \bibinfo {author} {\bibfnamefont {C.}~\bibnamefont {Filippi}}, \bibinfo {author} {\bibfnamefont {S.}~\bibnamefont {De~Palo}},\ and\ \bibinfo {author} {\bibfnamefont {S.}~\bibnamefont {Moroni}},\ }\href@noop {} {\bibfield  {journal} {\bibinfo  {journal} {Journal of chemical theory and computation}\ }\textbf {\bibinfo {volume} {18}},\ \bibinfo {pages} {118} (\bibinfo {year} {2021})}\BibitemShut {NoStop}%
\bibitem [{\citenamefont {Pathak}\ and\ \citenamefont {Wagner}(2020)}]{Pathak2020}%
  \BibitemOpen
  \bibfield  {author} {\bibinfo {author} {\bibfnamefont {S.}~\bibnamefont {Pathak}}\ and\ \bibinfo {author} {\bibfnamefont {L.~K.}\ \bibnamefont {Wagner}},\ }\href@noop {} {\bibfield  {journal} {\bibinfo  {journal} {AIP Adv.}\ }\textbf {\bibinfo {volume} {10}},\ \bibinfo {pages} {085213} (\bibinfo {year} {2020})}\BibitemShut {NoStop}%
\bibitem [{\citenamefont {Reynolds}\ \emph {et~al.}(1986)\citenamefont {Reynolds}, \citenamefont {Barnett}, \citenamefont {Hammond}, \citenamefont {Grimes},\ and\ \citenamefont {Lester~Jr}}]{Reynolds1986}%
  \BibitemOpen
  \bibfield  {author} {\bibinfo {author} {\bibfnamefont {P.}~\bibnamefont {Reynolds}}, \bibinfo {author} {\bibfnamefont {R.}~\bibnamefont {Barnett}}, \bibinfo {author} {\bibfnamefont {B.}~\bibnamefont {Hammond}}, \bibinfo {author} {\bibfnamefont {R.}~\bibnamefont {Grimes}},\ and\ \bibinfo {author} {\bibfnamefont {W.}~\bibnamefont {Lester~Jr}},\ }\href@noop {} {\bibfield  {journal} {\bibinfo  {journal} {Int. J. Quantum Chem.}\ }\textbf {\bibinfo {volume} {29}},\ \bibinfo {pages} {589} (\bibinfo {year} {1986})}\BibitemShut {NoStop}%
\bibitem [{\citenamefont {Lee}\ \emph {et~al.}(1988)\citenamefont {Lee}, \citenamefont {Yang},\ and\ \citenamefont {Parr}}]{Lee1988}%
  \BibitemOpen
  \bibfield  {author} {\bibinfo {author} {\bibfnamefont {C.}~\bibnamefont {Lee}}, \bibinfo {author} {\bibfnamefont {W.}~\bibnamefont {Yang}},\ and\ \bibinfo {author} {\bibfnamefont {R.~G.}\ \bibnamefont {Parr}},\ }\href {https://doi.org/10.1103/PhysRevB.37.785} {\bibfield  {journal} {\bibinfo  {journal} {Phys. Rev. B}\ }\textbf {\bibinfo {volume} {37}},\ \bibinfo {pages} {785} (\bibinfo {year} {1988})}\BibitemShut {NoStop}%
\bibitem [{\citenamefont {Mermin}(1965)}]{Mermin1965}%
  \BibitemOpen
  \bibfield  {author} {\bibinfo {author} {\bibfnamefont {N.~D.}\ \bibnamefont {Mermin}},\ }\href {https://doi.org/10.1103/PhysRev.137.A1441} {\bibfield  {journal} {\bibinfo  {journal} {Phys. Rev.}\ }\textbf {\bibinfo {volume} {137}},\ \bibinfo {pages} {A1441} (\bibinfo {year} {1965})}\BibitemShut {NoStop}%
\end{thebibliography}%


\begin{thebibliography}{28}%
\makeatletter
\providecommand \@ifxundefined [1]{%
 \@ifx{#1\undefined}
}%
\providecommand \@ifnum [1]{%
 \ifnum #1\expandafter \@firstoftwo
 \else \expandafter \@secondoftwo
 \fi
}%
\providecommand \@ifx [1]{%
 \ifx #1\expandafter \@firstoftwo
 \else \expandafter \@secondoftwo
 \fi
}%
\providecommand \natexlab [1]{#1}%
\providecommand \enquote  [1]{``#1''}%
\providecommand \bibnamefont  [1]{#1}%
\providecommand \bibfnamefont [1]{#1}%
\providecommand \citenamefont [1]{#1}%
\providecommand \href@noop [0]{\@secondoftwo}%
\providecommand \href [0]{\begingroup \@sanitize@url \@href}%
\providecommand \@href[1]{\@@startlink{#1}\@@href}%
\providecommand \@@href[1]{\endgroup#1\@@endlink}%
\providecommand \@sanitize@url [0]{\catcode `\\12\catcode `\$12\catcode `\&12\catcode `\#12\catcode `\^12\catcode `\_12\catcode `\%12\relax}%
\providecommand \@@startlink[1]{}%
\providecommand \@@endlink[0]{}%
\providecommand \url  [0]{\begingroup\@sanitize@url \@url }%
\providecommand \@url [1]{\endgroup\@href {#1}{\urlprefix }}%
\providecommand \urlprefix  [0]{URL }%
\providecommand \Eprint [0]{\href }%
\providecommand \doibase [0]{https://doi.org/}%
\providecommand \selectlanguage [0]{\@gobble}%
\providecommand \bibinfo  [0]{\@secondoftwo}%
\providecommand \bibfield  [0]{\@secondoftwo}%
\providecommand \translation [1]{[#1]}%
\providecommand \BibitemOpen [0]{}%
\providecommand \bibitemStop [0]{}%
\providecommand \bibitemNoStop [0]{.\EOS\space}%
\providecommand \EOS [0]{\spacefactor3000\relax}%
\providecommand \BibitemShut  [1]{\csname bibitem#1\endcsname}%
\let\auto@bib@innerbib\@empty
\bibitem [{\citenamefont {Casula}\ \emph {et~al.}(2005)\citenamefont {Casula}, \citenamefont {Filippi},\ and\ \citenamefont {Sorella}}]{Casula2005}%
  \BibitemOpen
  \bibfield  {author} {\bibinfo {author} {\bibfnamefont {M.}~\bibnamefont {Casula}}, \bibinfo {author} {\bibfnamefont {C.}~\bibnamefont {Filippi}},\ and\ \bibinfo {author} {\bibfnamefont {S.}~\bibnamefont {Sorella}},\ }\bibfield  {title} {\bibinfo {title} {Diffusion monte carlo method with lattice regularization},\ }\href {https://doi.org/10.1103/PhysRevLett.95.100201} {\bibfield  {journal} {\bibinfo  {journal} {Phys. Rev. Lett.}\ }\textbf {\bibinfo {volume} {95}},\ \bibinfo {pages} {100201} (\bibinfo {year} {2005})}\BibitemShut {NoStop}%
\bibitem [{\citenamefont {Nakano}\ \emph {et~al.}(2020)\citenamefont {Nakano}, \citenamefont {Attaccalite}, \citenamefont {Barborini}, \citenamefont {Capriotti}, \citenamefont {Casula}, \citenamefont {Coccia}, \citenamefont {Dagrada}, \citenamefont {Genovese}, \citenamefont {Luo}, \citenamefont {Mazzola}, \citenamefont {Zen},\ and\ \citenamefont {Sorella}}]{Nakano2020}%
  \BibitemOpen
  \bibfield  {author} {\bibinfo {author} {\bibfnamefont {K.}~\bibnamefont {Nakano}}, \bibinfo {author} {\bibfnamefont {C.}~\bibnamefont {Attaccalite}}, \bibinfo {author} {\bibfnamefont {M.}~\bibnamefont {Barborini}}, \bibinfo {author} {\bibfnamefont {L.}~\bibnamefont {Capriotti}}, \bibinfo {author} {\bibfnamefont {M.}~\bibnamefont {Casula}}, \bibinfo {author} {\bibfnamefont {E.}~\bibnamefont {Coccia}}, \bibinfo {author} {\bibfnamefont {M.}~\bibnamefont {Dagrada}}, \bibinfo {author} {\bibfnamefont {C.}~\bibnamefont {Genovese}}, \bibinfo {author} {\bibfnamefont {Y.}~\bibnamefont {Luo}}, \bibinfo {author} {\bibfnamefont {G.}~\bibnamefont {Mazzola}}, \bibinfo {author} {\bibfnamefont {A.}~\bibnamefont {Zen}},\ and\ \bibinfo {author} {\bibfnamefont {S.}~\bibnamefont {Sorella}},\ }\bibfield  {title} {\bibinfo {title} {Turborvb: A many-body toolkit for ab initio electronic simulations by quantum monte carlo},\ }\href@noop {} {\bibfield  {journal} {\bibinfo  {journal} {J. Chem. Phys.}\ }\textbf {\bibinfo {volume}
  {152}},\ \bibinfo {pages} {204121} (\bibinfo {year} {2020})}\BibitemShut {NoStop}%
\bibitem [{\citenamefont {Casula}\ and\ \citenamefont {Sorella}(2003)}]{Casula2003}%
  \BibitemOpen
  \bibfield  {author} {\bibinfo {author} {\bibfnamefont {M.}~\bibnamefont {Casula}}\ and\ \bibinfo {author} {\bibfnamefont {S.}~\bibnamefont {Sorella}},\ }\bibfield  {title} {\bibinfo {title} {Geminal wave functions with jastrow correlation: A first application to atoms},\ }\href@noop {} {\bibfield  {journal} {\bibinfo  {journal} {J. Chem. Phys.}\ }\textbf {\bibinfo {volume} {119}},\ \bibinfo {pages} {6500} (\bibinfo {year} {2003})}\BibitemShut {NoStop}%
\bibitem [{\citenamefont {Sorella}\ \emph {et~al.}(2015)\citenamefont {Sorella}, \citenamefont {Devaux}, \citenamefont {Dagrada}, \citenamefont {Mazzola},\ and\ \citenamefont {Casula}}]{sorella2015geminal}%
  \BibitemOpen
  \bibfield  {author} {\bibinfo {author} {\bibfnamefont {S.}~\bibnamefont {Sorella}}, \bibinfo {author} {\bibfnamefont {N.}~\bibnamefont {Devaux}}, \bibinfo {author} {\bibfnamefont {M.}~\bibnamefont {Dagrada}}, \bibinfo {author} {\bibfnamefont {G.}~\bibnamefont {Mazzola}},\ and\ \bibinfo {author} {\bibfnamefont {M.}~\bibnamefont {Casula}},\ }\bibfield  {title} {\bibinfo {title} {Geminal embedding scheme for optimal atomic basis set construction in correlated calculations},\ }\href@noop {} {\bibfield  {journal} {\bibinfo  {journal} {J. Chem. Phys.}\ }\textbf {\bibinfo {volume} {143}} (\bibinfo {year} {2015})}\BibitemShut {NoStop}%
\bibitem [{\citenamefont {Becca}\ and\ \citenamefont {Sorella}(2017)}]{2017BEC}%
  \BibitemOpen
  \bibfield  {author} {\bibinfo {author} {\bibfnamefont {F.}~\bibnamefont {Becca}}\ and\ \bibinfo {author} {\bibfnamefont {S.}~\bibnamefont {Sorella}},\ }\href@noop {} {\emph {\bibinfo {title} {{Quantum Monte Carlo approaches for correlated systems}}}}\ (\bibinfo  {publisher} {Cambridge University Press},\ \bibinfo {year} {2017})\BibitemShut {NoStop}%
\bibitem [{\citenamefont {Tiihonen}\ \emph {et~al.}(2021)\citenamefont {Tiihonen}, \citenamefont {Clay~III},\ and\ \citenamefont {Krogel}}]{Tiihonen2021}%
  \BibitemOpen
  \bibfield  {author} {\bibinfo {author} {\bibfnamefont {J.}~\bibnamefont {Tiihonen}}, \bibinfo {author} {\bibfnamefont {R.~C.}\ \bibnamefont {Clay~III}},\ and\ \bibinfo {author} {\bibfnamefont {J.~T.}\ \bibnamefont {Krogel}},\ }\bibfield  {title} {\bibinfo {title} {Toward quantum monte carlo forces on heavier ions: Scaling properties},\ }\href {https://doi.org/10.1063/5.0052266} {\bibfield  {journal} {\bibinfo  {journal} {J. Chem. Phys.}\ }\textbf {\bibinfo {volume} {154}},\ \bibinfo {pages} {204111} (\bibinfo {year} {2021})}\BibitemShut {NoStop}%
\bibitem [{\citenamefont {Nakano}\ \emph {et~al.}(2022)\citenamefont {Nakano}, \citenamefont {Raghav},\ and\ \citenamefont {Sorella}}]{Nakano2022}%
  \BibitemOpen
  \bibfield  {author} {\bibinfo {author} {\bibfnamefont {K.}~\bibnamefont {Nakano}}, \bibinfo {author} {\bibfnamefont {A.}~\bibnamefont {Raghav}},\ and\ \bibinfo {author} {\bibfnamefont {S.}~\bibnamefont {Sorella}},\ }\bibfield  {title} {\bibinfo {title} {Space-warp coordinate transformation for efficient ionic force calculations in quantum monte carlo},\ }\href@noop {} {\bibfield  {journal} {\bibinfo  {journal} {The Journal of Chemical Physics}\ }\textbf {\bibinfo {volume} {156}},\ \bibinfo {pages} {034101} (\bibinfo {year} {2022})}\BibitemShut {NoStop}%
\bibitem [{\citenamefont {Nakano}\ \emph {et~al.}(2023)\citenamefont {Nakano}, \citenamefont {Casula},\ and\ \citenamefont {Tenti}}]{Nakano2023}%
  \BibitemOpen
  \bibfield  {author} {\bibinfo {author} {\bibfnamefont {K.}~\bibnamefont {Nakano}}, \bibinfo {author} {\bibfnamefont {M.}~\bibnamefont {Casula}},\ and\ \bibinfo {author} {\bibfnamefont {G.}~\bibnamefont {Tenti}},\ }\bibfield  {title} {\bibinfo {title} {Unbiased and affordable atomic forces in ab initio variational monte carlo},\ }\href@noop {} {\  (\bibinfo {year} {2023})},\ \Eprint {https://arxiv.org/abs/arXiv:2312.17608} {arXiv:2312.17608} \BibitemShut {NoStop}%
\bibitem [{\citenamefont {Nakano}\ \emph {et~al.}(2021)\citenamefont {Nakano}, \citenamefont {Morresi}, \citenamefont {Casula}, \citenamefont {Maezono},\ and\ \citenamefont {Sorella}}]{Nakano2021}%
  \BibitemOpen
  \bibfield  {author} {\bibinfo {author} {\bibfnamefont {K.}~\bibnamefont {Nakano}}, \bibinfo {author} {\bibfnamefont {T.}~\bibnamefont {Morresi}}, \bibinfo {author} {\bibfnamefont {M.}~\bibnamefont {Casula}}, \bibinfo {author} {\bibfnamefont {R.}~\bibnamefont {Maezono}},\ and\ \bibinfo {author} {\bibfnamefont {S.}~\bibnamefont {Sorella}},\ }\bibfield  {title} {\bibinfo {title} {Atomic forces by quantum monte carlo: Application to phonon dispersion calculations},\ }\href {https://doi.org/10.1103/PhysRevB.103.L121110} {\bibfield  {journal} {\bibinfo  {journal} {Phys. Rev. B}\ }\textbf {\bibinfo {volume} {103}},\ \bibinfo {pages} {L121110} (\bibinfo {year} {2021})}\BibitemShut {NoStop}%
\bibitem [{\citenamefont {Umrigar}(1989)}]{Umrigar1989}%
  \BibitemOpen
  \bibfield  {author} {\bibinfo {author} {\bibfnamefont {C.~J.}\ \bibnamefont {Umrigar}},\ }\bibfield  {title} {\bibinfo {title} {Two aspects of quantum monte carlo: determination of accurate wavefunctions and determination of potential energy surfaces of molecules},\ }\href@noop {} {\bibfield  {journal} {\bibinfo  {journal} {Int. J. Quantum Chem}\ }\textbf {\bibinfo {volume} {36}},\ \bibinfo {pages} {217} (\bibinfo {year} {1989})}\BibitemShut {NoStop}%
\bibitem [{\citenamefont {Sorella}\ and\ \citenamefont {Capriotti}(2010)}]{Sorella2010}%
  \BibitemOpen
  \bibfield  {author} {\bibinfo {author} {\bibfnamefont {S.}~\bibnamefont {Sorella}}\ and\ \bibinfo {author} {\bibfnamefont {L.}~\bibnamefont {Capriotti}},\ }\bibfield  {title} {\bibinfo {title} {Algorithmic differentiation and the calculation of forces by quantum monte carlo},\ }\href@noop {} {\bibfield  {journal} {\bibinfo  {journal} {J. Chem. Phys.}\ }\textbf {\bibinfo {volume} {133}},\ \bibinfo {pages} {234111} (\bibinfo {year} {2010})}\BibitemShut {NoStop}%
\bibitem [{\citenamefont {Filippi}\ \emph {et~al.}(2016)\citenamefont {Filippi}, \citenamefont {Assaraf},\ and\ \citenamefont {Moroni}}]{Claudia2016}%
  \BibitemOpen
  \bibfield  {author} {\bibinfo {author} {\bibfnamefont {C.}~\bibnamefont {Filippi}}, \bibinfo {author} {\bibfnamefont {R.}~\bibnamefont {Assaraf}},\ and\ \bibinfo {author} {\bibfnamefont {S.}~\bibnamefont {Moroni}},\ }\bibfield  {title} {\bibinfo {title} {Simple formalism for efficient derivatives and multi-determinant expansions in quantum monte carlo},\ }\href {https://doi.org/10.1063/1.4948778} {\bibfield  {journal} {\bibinfo  {journal} {J. Chem. Phys.}\ }\textbf {\bibinfo {volume} {144}},\ \bibinfo {pages} {194105} (\bibinfo {year} {2016})}\BibitemShut {NoStop}%
\bibitem [{\citenamefont {Attaccalite}\ and\ \citenamefont {Sorella}(2008)}]{Attaccalite2008}%
  \BibitemOpen
  \bibfield  {author} {\bibinfo {author} {\bibfnamefont {C.}~\bibnamefont {Attaccalite}}\ and\ \bibinfo {author} {\bibfnamefont {S.}~\bibnamefont {Sorella}},\ }\bibfield  {title} {\bibinfo {title} {Stable liquid hydrogen at high pressure by a novel ab initio molecular-dynamics calculation},\ }\href@noop {} {\bibfield  {journal} {\bibinfo  {journal} {Phys. Rev. Lett.}\ }\textbf {\bibinfo {volume} {100}},\ \bibinfo {pages} {114501} (\bibinfo {year} {2008})}\BibitemShut {NoStop}%
\bibitem [{\citenamefont {van Rhijn}\ \emph {et~al.}(2021)\citenamefont {van Rhijn}, \citenamefont {Filippi}, \citenamefont {De~Palo},\ and\ \citenamefont {Moroni}}]{vanRhijn2021}%
  \BibitemOpen
  \bibfield  {author} {\bibinfo {author} {\bibfnamefont {J.}~\bibnamefont {van Rhijn}}, \bibinfo {author} {\bibfnamefont {C.}~\bibnamefont {Filippi}}, \bibinfo {author} {\bibfnamefont {S.}~\bibnamefont {De~Palo}},\ and\ \bibinfo {author} {\bibfnamefont {S.}~\bibnamefont {Moroni}},\ }\bibfield  {title} {\bibinfo {title} {Energy derivatives in real-space diffusion monte carlo},\ }\href@noop {} {\bibfield  {journal} {\bibinfo  {journal} {Journal of chemical theory and computation}\ }\textbf {\bibinfo {volume} {18}},\ \bibinfo {pages} {118} (\bibinfo {year} {2021})}\BibitemShut {NoStop}%
\bibitem [{\citenamefont {Pathak}\ and\ \citenamefont {Wagner}(2020)}]{Pathak2020}%
  \BibitemOpen
  \bibfield  {author} {\bibinfo {author} {\bibfnamefont {S.}~\bibnamefont {Pathak}}\ and\ \bibinfo {author} {\bibfnamefont {L.~K.}\ \bibnamefont {Wagner}},\ }\bibfield  {title} {\bibinfo {title} {A light weight regularization for wave function parameter gradients in quantum monte carlo},\ }\href@noop {} {\bibfield  {journal} {\bibinfo  {journal} {AIP Adv.}\ }\textbf {\bibinfo {volume} {10}},\ \bibinfo {pages} {085213} (\bibinfo {year} {2020})}\BibitemShut {NoStop}%
\bibitem [{\citenamefont {Reynolds}\ \emph {et~al.}(1986)\citenamefont {Reynolds}, \citenamefont {Barnett}, \citenamefont {Hammond}, \citenamefont {Grimes},\ and\ \citenamefont {Lester~Jr}}]{Reynolds1986}%
  \BibitemOpen
  \bibfield  {author} {\bibinfo {author} {\bibfnamefont {P.}~\bibnamefont {Reynolds}}, \bibinfo {author} {\bibfnamefont {R.}~\bibnamefont {Barnett}}, \bibinfo {author} {\bibfnamefont {B.}~\bibnamefont {Hammond}}, \bibinfo {author} {\bibfnamefont {R.}~\bibnamefont {Grimes}},\ and\ \bibinfo {author} {\bibfnamefont {W.}~\bibnamefont {Lester~Jr}},\ }\bibfield  {title} {\bibinfo {title} {Quantum chemistry by quantum monte carlo: Beyond ground-state energy calculations},\ }\href@noop {} {\bibfield  {journal} {\bibinfo  {journal} {Int. J. Quantum Chem.}\ }\textbf {\bibinfo {volume} {29}},\ \bibinfo {pages} {589} (\bibinfo {year} {1986})}\BibitemShut {NoStop}%
\bibitem [{\citenamefont {Perdew}\ and\ \citenamefont {Zunger}(1981)}]{Perdew1981}%
  \BibitemOpen
  \bibfield  {author} {\bibinfo {author} {\bibfnamefont {J.~P.}\ \bibnamefont {Perdew}}\ and\ \bibinfo {author} {\bibfnamefont {A.}~\bibnamefont {Zunger}},\ }\bibfield  {title} {\bibinfo {title} {Self-interaction correction to density-functional approximations for many-electron systems},\ }\href {https://doi.org/10.1103/PhysRevB.23.5048} {\bibfield  {journal} {\bibinfo  {journal} {Phys. Rev. B}\ }\textbf {\bibinfo {volume} {23}},\ \bibinfo {pages} {5048} (\bibinfo {year} {1981})}\BibitemShut {NoStop}%
\bibitem [{\citenamefont {Tirelli}\ \emph {et~al.}(2022)\citenamefont {Tirelli}, \citenamefont {Tenti}, \citenamefont {Nakano},\ and\ \citenamefont {Sorella}}]{Tirelli2022}%
  \BibitemOpen
  \bibfield  {author} {\bibinfo {author} {\bibfnamefont {A.}~\bibnamefont {Tirelli}}, \bibinfo {author} {\bibfnamefont {G.}~\bibnamefont {Tenti}}, \bibinfo {author} {\bibfnamefont {K.}~\bibnamefont {Nakano}},\ and\ \bibinfo {author} {\bibfnamefont {S.}~\bibnamefont {Sorella}},\ }\bibfield  {title} {\bibinfo {title} {High-pressure hydrogen by machine learning and quantum monte carlo},\ }\href {https://doi.org/10.1103/PhysRevB.106.L041105} {\bibfield  {journal} {\bibinfo  {journal} {Phys. Rev. B}\ }\textbf {\bibinfo {volume} {106}},\ \bibinfo {pages} {L041105} (\bibinfo {year} {2022})}\BibitemShut {NoStop}%
\bibitem [{\citenamefont {Mouhat}\ \emph {et~al.}(2017)\citenamefont {Mouhat}, \citenamefont {Sorella}, \citenamefont {Vuilleumier}, \citenamefont {Saitta},\ and\ \citenamefont {Casula}}]{Mouhat2017}%
  \BibitemOpen
  \bibfield  {author} {\bibinfo {author} {\bibfnamefont {F.}~\bibnamefont {Mouhat}}, \bibinfo {author} {\bibfnamefont {S.}~\bibnamefont {Sorella}}, \bibinfo {author} {\bibfnamefont {R.}~\bibnamefont {Vuilleumier}}, \bibinfo {author} {\bibfnamefont {A.~M.}\ \bibnamefont {Saitta}},\ and\ \bibinfo {author} {\bibfnamefont {M.}~\bibnamefont {Casula}},\ }\bibfield  {title} {\bibinfo {title} {Fully quantum description of the zundel ion: Combining variational quantum monte carlo with path integral langevin dynamics},\ }\href {https://doi.org/10.1021/acs.jctc.7b00017} {\bibfield  {journal} {\bibinfo  {journal} {Journal of Chemical Theory and Computation}\ }\textbf {\bibinfo {volume} {13}},\ \bibinfo {pages} {2400} (\bibinfo {year} {2017})}\BibitemShut {NoStop}%
\bibitem [{\citenamefont {Giannozzi}\ \emph {et~al.}(2009)\citenamefont {Giannozzi}, \citenamefont {Baroni}, \citenamefont {Bonini}, \citenamefont {Calandra}, \citenamefont {Car}, \citenamefont {Cavazzoni}, \citenamefont {Ceresoli}, \citenamefont {Chiarotti}, \citenamefont {Cococcioni}, \citenamefont {Dabo}, \citenamefont {Corso}, \citenamefont {de~Gironcoli}, \citenamefont {Fabris}, \citenamefont {Fratesi}, \citenamefont {Gebauer}, \citenamefont {Gerstmann}, \citenamefont {Gougoussis}, \citenamefont {Kokalj}, \citenamefont {Lazzeri}, \citenamefont {Martin-Samos}, \citenamefont {Marzari}, \citenamefont {Mauri}, \citenamefont {Mazzarello}, \citenamefont {Paolini}, \citenamefont {Pasquarello}, \citenamefont {Paulatto}, \citenamefont {Sbraccia}, \citenamefont {Scandolo}, \citenamefont {Sclauzero}, \citenamefont {Seitsonen}, \citenamefont {Smogunov}, \citenamefont {Umari},\ and\ \citenamefont {Wentzcovitch}}]{Giannozzi2009}%
  \BibitemOpen
  \bibfield  {author} {\bibinfo {author} {\bibfnamefont {P.}~\bibnamefont {Giannozzi}}, \bibinfo {author} {\bibfnamefont {S.}~\bibnamefont {Baroni}}, \bibinfo {author} {\bibfnamefont {N.}~\bibnamefont {Bonini}}, \bibinfo {author} {\bibfnamefont {M.}~\bibnamefont {Calandra}}, \bibinfo {author} {\bibfnamefont {R.}~\bibnamefont {Car}}, \bibinfo {author} {\bibfnamefont {C.}~\bibnamefont {Cavazzoni}}, \bibinfo {author} {\bibfnamefont {D.}~\bibnamefont {Ceresoli}}, \bibinfo {author} {\bibfnamefont {G.~L.}\ \bibnamefont {Chiarotti}}, \bibinfo {author} {\bibfnamefont {M.}~\bibnamefont {Cococcioni}}, \bibinfo {author} {\bibfnamefont {I.}~\bibnamefont {Dabo}}, \bibinfo {author} {\bibfnamefont {A.~D.}\ \bibnamefont {Corso}}, \bibinfo {author} {\bibfnamefont {S.}~\bibnamefont {de~Gironcoli}}, \bibinfo {author} {\bibfnamefont {S.}~\bibnamefont {Fabris}}, \bibinfo {author} {\bibfnamefont {G.}~\bibnamefont {Fratesi}}, \bibinfo {author} {\bibfnamefont {R.}~\bibnamefont {Gebauer}}, \bibinfo {author} {\bibfnamefont
  {U.}~\bibnamefont {Gerstmann}}, \bibinfo {author} {\bibfnamefont {C.}~\bibnamefont {Gougoussis}}, \bibinfo {author} {\bibfnamefont {A.}~\bibnamefont {Kokalj}}, \bibinfo {author} {\bibfnamefont {M.}~\bibnamefont {Lazzeri}}, \bibinfo {author} {\bibfnamefont {L.}~\bibnamefont {Martin-Samos}}, \bibinfo {author} {\bibfnamefont {N.}~\bibnamefont {Marzari}}, \bibinfo {author} {\bibfnamefont {F.}~\bibnamefont {Mauri}}, \bibinfo {author} {\bibfnamefont {R.}~\bibnamefont {Mazzarello}}, \bibinfo {author} {\bibfnamefont {S.}~\bibnamefont {Paolini}}, \bibinfo {author} {\bibfnamefont {A.}~\bibnamefont {Pasquarello}}, \bibinfo {author} {\bibfnamefont {L.}~\bibnamefont {Paulatto}}, \bibinfo {author} {\bibfnamefont {C.}~\bibnamefont {Sbraccia}}, \bibinfo {author} {\bibfnamefont {S.}~\bibnamefont {Scandolo}}, \bibinfo {author} {\bibfnamefont {G.}~\bibnamefont {Sclauzero}}, \bibinfo {author} {\bibfnamefont {A.~P.}\ \bibnamefont {Seitsonen}}, \bibinfo {author} {\bibfnamefont {A.}~\bibnamefont {Smogunov}}, \bibinfo {author}
  {\bibfnamefont {P.}~\bibnamefont {Umari}},\ and\ \bibinfo {author} {\bibfnamefont {R.~M.}\ \bibnamefont {Wentzcovitch}},\ }\bibfield  {title} {\bibinfo {title} {{QUANTUM} {ESPRESSO}: a modular and open-source software project for quantum simulations of materials},\ }\href {https://doi.org/10.1088/0953-8984/21/39/395502} {\bibfield  {journal} {\bibinfo  {journal} {Journal of Physics: Condensed Matter}\ }\textbf {\bibinfo {volume} {21}},\ \bibinfo {pages} {395502} (\bibinfo {year} {2009})}\BibitemShut {NoStop}%
\bibitem [{\citenamefont {Giannozzi}\ \emph {et~al.}(2017)\citenamefont {Giannozzi}, \citenamefont {Andreussi}, \citenamefont {Brumme}, \citenamefont {Bunau}, \citenamefont {Nardelli}, \citenamefont {Calandra}, \citenamefont {Car}, \citenamefont {Cavazzoni}, \citenamefont {Ceresoli}, \citenamefont {Cococcioni}, \citenamefont {Colonna}, \citenamefont {Carnimeo}, \citenamefont {Corso}, \citenamefont {de~Gironcoli}, \citenamefont {Delugas}, \citenamefont {DiStasio}, \citenamefont {Ferretti}, \citenamefont {Floris}, \citenamefont {Fratesi}, \citenamefont {Fugallo}, \citenamefont {Gebauer}, \citenamefont {Gerstmann}, \citenamefont {Giustino}, \citenamefont {Gorni}, \citenamefont {Jia}, \citenamefont {Kawamura}, \citenamefont {Ko}, \citenamefont {Kokalj}, \citenamefont {K\"{u}{\c{c}}\"{u}kbenli}, \citenamefont {Lazzeri}, \citenamefont {Marsili}, \citenamefont {Marzari}, \citenamefont {Mauri}, \citenamefont {Nguyen}, \citenamefont {Nguyen}, \citenamefont {de-la Roza}, \citenamefont {Paulatto}, \citenamefont
  {Ponc{\'{e}}}, \citenamefont {Rocca}, \citenamefont {Sabatini}, \citenamefont {Santra}, \citenamefont {Schlipf}, \citenamefont {Seitsonen}, \citenamefont {Smogunov}, \citenamefont {Timrov}, \citenamefont {Thonhauser}, \citenamefont {Umari}, \citenamefont {Vast}, \citenamefont {Wu},\ and\ \citenamefont {Baroni}}]{Giannozzi2017}%
  \BibitemOpen
  \bibfield  {author} {\bibinfo {author} {\bibfnamefont {P.}~\bibnamefont {Giannozzi}}, \bibinfo {author} {\bibfnamefont {O.}~\bibnamefont {Andreussi}}, \bibinfo {author} {\bibfnamefont {T.}~\bibnamefont {Brumme}}, \bibinfo {author} {\bibfnamefont {O.}~\bibnamefont {Bunau}}, \bibinfo {author} {\bibfnamefont {M.~B.}\ \bibnamefont {Nardelli}}, \bibinfo {author} {\bibfnamefont {M.}~\bibnamefont {Calandra}}, \bibinfo {author} {\bibfnamefont {R.}~\bibnamefont {Car}}, \bibinfo {author} {\bibfnamefont {C.}~\bibnamefont {Cavazzoni}}, \bibinfo {author} {\bibfnamefont {D.}~\bibnamefont {Ceresoli}}, \bibinfo {author} {\bibfnamefont {M.}~\bibnamefont {Cococcioni}}, \bibinfo {author} {\bibfnamefont {N.}~\bibnamefont {Colonna}}, \bibinfo {author} {\bibfnamefont {I.}~\bibnamefont {Carnimeo}}, \bibinfo {author} {\bibfnamefont {A.~D.}\ \bibnamefont {Corso}}, \bibinfo {author} {\bibfnamefont {S.}~\bibnamefont {de~Gironcoli}}, \bibinfo {author} {\bibfnamefont {P.}~\bibnamefont {Delugas}}, \bibinfo {author} {\bibfnamefont
  {R.~A.}\ \bibnamefont {DiStasio}}, \bibinfo {author} {\bibfnamefont {A.}~\bibnamefont {Ferretti}}, \bibinfo {author} {\bibfnamefont {A.}~\bibnamefont {Floris}}, \bibinfo {author} {\bibfnamefont {G.}~\bibnamefont {Fratesi}}, \bibinfo {author} {\bibfnamefont {G.}~\bibnamefont {Fugallo}}, \bibinfo {author} {\bibfnamefont {R.}~\bibnamefont {Gebauer}}, \bibinfo {author} {\bibfnamefont {U.}~\bibnamefont {Gerstmann}}, \bibinfo {author} {\bibfnamefont {F.}~\bibnamefont {Giustino}}, \bibinfo {author} {\bibfnamefont {T.}~\bibnamefont {Gorni}}, \bibinfo {author} {\bibfnamefont {J.}~\bibnamefont {Jia}}, \bibinfo {author} {\bibfnamefont {M.}~\bibnamefont {Kawamura}}, \bibinfo {author} {\bibfnamefont {H.-Y.}\ \bibnamefont {Ko}}, \bibinfo {author} {\bibfnamefont {A.}~\bibnamefont {Kokalj}}, \bibinfo {author} {\bibfnamefont {E.}~\bibnamefont {K\"{u}{\c{c}}\"{u}kbenli}}, \bibinfo {author} {\bibfnamefont {M.}~\bibnamefont {Lazzeri}}, \bibinfo {author} {\bibfnamefont {M.}~\bibnamefont {Marsili}}, \bibinfo {author}
  {\bibfnamefont {N.}~\bibnamefont {Marzari}}, \bibinfo {author} {\bibfnamefont {F.}~\bibnamefont {Mauri}}, \bibinfo {author} {\bibfnamefont {N.~L.}\ \bibnamefont {Nguyen}}, \bibinfo {author} {\bibfnamefont {H.-V.}\ \bibnamefont {Nguyen}}, \bibinfo {author} {\bibfnamefont {A.~O.}\ \bibnamefont {de-la Roza}}, \bibinfo {author} {\bibfnamefont {L.}~\bibnamefont {Paulatto}}, \bibinfo {author} {\bibfnamefont {S.}~\bibnamefont {Ponc{\'{e}}}}, \bibinfo {author} {\bibfnamefont {D.}~\bibnamefont {Rocca}}, \bibinfo {author} {\bibfnamefont {R.}~\bibnamefont {Sabatini}}, \bibinfo {author} {\bibfnamefont {B.}~\bibnamefont {Santra}}, \bibinfo {author} {\bibfnamefont {M.}~\bibnamefont {Schlipf}}, \bibinfo {author} {\bibfnamefont {A.~P.}\ \bibnamefont {Seitsonen}}, \bibinfo {author} {\bibfnamefont {A.}~\bibnamefont {Smogunov}}, \bibinfo {author} {\bibfnamefont {I.}~\bibnamefont {Timrov}}, \bibinfo {author} {\bibfnamefont {T.}~\bibnamefont {Thonhauser}}, \bibinfo {author} {\bibfnamefont {P.}~\bibnamefont {Umari}}, \bibinfo
  {author} {\bibfnamefont {N.}~\bibnamefont {Vast}}, \bibinfo {author} {\bibfnamefont {X.}~\bibnamefont {Wu}},\ and\ \bibinfo {author} {\bibfnamefont {S.}~\bibnamefont {Baroni}},\ }\bibfield  {title} {\bibinfo {title} {Advanced capabilities for materials modelling with quantum {ESPRESSO}},\ }\href {https://doi.org/10.1088/1361-648x/aa8f79} {\bibfield  {journal} {\bibinfo  {journal} {Journal of Physics: Condensed Matter}\ }\textbf {\bibinfo {volume} {29}},\ \bibinfo {pages} {465901} (\bibinfo {year} {2017})}\BibitemShut {NoStop}%
\bibitem [{\citenamefont {Giannozzi}\ \emph {et~al.}(2020)\citenamefont {Giannozzi}, \citenamefont {Baseggio}, \citenamefont {Bonfà}, \citenamefont {Brunato}, \citenamefont {Car}, \citenamefont {Carnimeo}, \citenamefont {Cavazzoni}, \citenamefont {de~Gironcoli}, \citenamefont {Delugas}, \citenamefont {Ferrari~Ruffino}, \citenamefont {Ferretti}, \citenamefont {Marzari}, \citenamefont {Timrov}, \citenamefont {Urru},\ and\ \citenamefont {Baroni}}]{Giannozzi2020}%
  \BibitemOpen
  \bibfield  {author} {\bibinfo {author} {\bibfnamefont {P.}~\bibnamefont {Giannozzi}}, \bibinfo {author} {\bibfnamefont {O.}~\bibnamefont {Baseggio}}, \bibinfo {author} {\bibfnamefont {P.}~\bibnamefont {Bonfà}}, \bibinfo {author} {\bibfnamefont {D.}~\bibnamefont {Brunato}}, \bibinfo {author} {\bibfnamefont {R.}~\bibnamefont {Car}}, \bibinfo {author} {\bibfnamefont {I.}~\bibnamefont {Carnimeo}}, \bibinfo {author} {\bibfnamefont {C.}~\bibnamefont {Cavazzoni}}, \bibinfo {author} {\bibfnamefont {S.}~\bibnamefont {de~Gironcoli}}, \bibinfo {author} {\bibfnamefont {P.}~\bibnamefont {Delugas}}, \bibinfo {author} {\bibfnamefont {F.}~\bibnamefont {Ferrari~Ruffino}}, \bibinfo {author} {\bibfnamefont {A.}~\bibnamefont {Ferretti}}, \bibinfo {author} {\bibfnamefont {N.}~\bibnamefont {Marzari}}, \bibinfo {author} {\bibfnamefont {I.}~\bibnamefont {Timrov}}, \bibinfo {author} {\bibfnamefont {A.}~\bibnamefont {Urru}},\ and\ \bibinfo {author} {\bibfnamefont {S.}~\bibnamefont {Baroni}},\ }\bibfield  {title} {\bibinfo {title}
  {Quantum espresso toward the exascale},\ }\href {https://doi.org/10.1063/5.0005082} {\bibfield  {journal} {\bibinfo  {journal} {The Journal of Chemical Physics}\ }\textbf {\bibinfo {volume} {152}},\ \bibinfo {pages} {154105} (\bibinfo {year} {2020})},\ \Eprint {https://arxiv.org/abs/https://doi.org/10.1063/5.0005082} {https://doi.org/10.1063/5.0005082} \BibitemShut {NoStop}%
\bibitem [{\citenamefont {Lee}\ \emph {et~al.}(1988)\citenamefont {Lee}, \citenamefont {Yang},\ and\ \citenamefont {Parr}}]{Lee1988}%
  \BibitemOpen
  \bibfield  {author} {\bibinfo {author} {\bibfnamefont {C.}~\bibnamefont {Lee}}, \bibinfo {author} {\bibfnamefont {W.}~\bibnamefont {Yang}},\ and\ \bibinfo {author} {\bibfnamefont {R.~G.}\ \bibnamefont {Parr}},\ }\bibfield  {title} {\bibinfo {title} {Development of the colle-salvetti correlation-energy formula into a functional of the electron density},\ }\href {https://doi.org/10.1103/PhysRevB.37.785} {\bibfield  {journal} {\bibinfo  {journal} {Phys. Rev. B}\ }\textbf {\bibinfo {volume} {37}},\ \bibinfo {pages} {785} (\bibinfo {year} {1988})}\BibitemShut {NoStop}%
\bibitem [{\citenamefont {Kwee}\ \emph {et~al.}(2008)\citenamefont {Kwee}, \citenamefont {Zhang},\ and\ \citenamefont {Krakauer}}]{Kwee2008}%
  \BibitemOpen
  \bibfield  {author} {\bibinfo {author} {\bibfnamefont {H.}~\bibnamefont {Kwee}}, \bibinfo {author} {\bibfnamefont {S.}~\bibnamefont {Zhang}},\ and\ \bibinfo {author} {\bibfnamefont {H.}~\bibnamefont {Krakauer}},\ }\bibfield  {title} {\bibinfo {title} {Finite-size correction in many-body electronic structure calculations},\ }\href {https://doi.org/10.1103/PhysRevLett.100.126404} {\bibfield  {journal} {\bibinfo  {journal} {Phys. Rev. Lett.}\ }\textbf {\bibinfo {volume} {100}},\ \bibinfo {pages} {126404} (\bibinfo {year} {2008})}\BibitemShut {NoStop}%
\bibitem [{\citenamefont {Mermin}(1965)}]{Mermin1965}%
  \BibitemOpen
  \bibfield  {author} {\bibinfo {author} {\bibfnamefont {N.~D.}\ \bibnamefont {Mermin}},\ }\bibfield  {title} {\bibinfo {title} {Thermal properties of the inhomogeneous electron gas},\ }\href {https://doi.org/10.1103/PhysRev.137.A1441} {\bibfield  {journal} {\bibinfo  {journal} {Phys. Rev.}\ }\textbf {\bibinfo {volume} {137}},\ \bibinfo {pages} {A1441} (\bibinfo {year} {1965})}\BibitemShut {NoStop}%
\bibitem [{\citenamefont {Karasiev}\ \emph {et~al.}(2019)\citenamefont {Karasiev}, \citenamefont {Hu}, \citenamefont {Zaghoo},\ and\ \citenamefont {Boehly}}]{Karasiev2019}%
  \BibitemOpen
  \bibfield  {author} {\bibinfo {author} {\bibfnamefont {V.~V.}\ \bibnamefont {Karasiev}}, \bibinfo {author} {\bibfnamefont {S.~X.}\ \bibnamefont {Hu}}, \bibinfo {author} {\bibfnamefont {M.}~\bibnamefont {Zaghoo}},\ and\ \bibinfo {author} {\bibfnamefont {T.~R.}\ \bibnamefont {Boehly}},\ }\bibfield  {title} {\bibinfo {title} {{Exchange-correlation thermal effects in shocked deuterium: Softening the principal Hugoniot and thermophysical properties}},\ }\href {https://doi.org/10.1103/PhysRevB.99.214110} {\bibfield  {journal} {\bibinfo  {journal} {Physical Review B}\ }\textbf {\bibinfo {volume} {99}},\ \bibinfo {pages} {1} (\bibinfo {year} {2019})}\BibitemShut {NoStop}%
\bibitem [{\citenamefont {Ruggeri}\ \emph {et~al.}(2020)\citenamefont {Ruggeri}, \citenamefont {Holzmann}, \citenamefont {Ceperley},\ and\ \citenamefont {Pierleoni}}]{Ruggeri2020}%
  \BibitemOpen
  \bibfield  {author} {\bibinfo {author} {\bibfnamefont {M.}~\bibnamefont {Ruggeri}}, \bibinfo {author} {\bibfnamefont {M.}~\bibnamefont {Holzmann}}, \bibinfo {author} {\bibfnamefont {D.~M.}\ \bibnamefont {Ceperley}},\ and\ \bibinfo {author} {\bibfnamefont {C.}~\bibnamefont {Pierleoni}},\ }\bibfield  {title} {\bibinfo {title} {{Quantum Monte Carlo determination of the principal Hugoniot of deuterium}},\ }\href {https://doi.org/10.1103/PhysRevB.102.144108} {\bibfield  {journal} {\bibinfo  {journal} {Physical Review B}\ }\textbf {\bibinfo {volume} {102}},\ \bibinfo {pages} {144108} (\bibinfo {year} {2020})},\ \Eprint {https://arxiv.org/abs/2008.00269} {2008.00269} \BibitemShut {NoStop}%
\bibitem [{\citenamefont {Clay}\ \emph {et~al.}(2019)\citenamefont {Clay}, \citenamefont {Desjarlais},\ and\ \citenamefont {Shulenburger}}]{Clay2019}%
  \BibitemOpen
  \bibfield  {author} {\bibinfo {author} {\bibfnamefont {R.~C.}\ \bibnamefont {Clay}}, \bibinfo {author} {\bibfnamefont {M.~P.}\ \bibnamefont {Desjarlais}},\ and\ \bibinfo {author} {\bibfnamefont {L.}~\bibnamefont {Shulenburger}},\ }\bibfield  {title} {\bibinfo {title} {{Deuterium Hugoniot: Pitfalls of thermodynamic sampling beyond density functional theory}},\ }\href {https://doi.org/10.1103/PhysRevB.100.075103} {\bibfield  {journal} {\bibinfo  {journal} {Physical Review B}\ }\textbf {\bibinfo {volume} {100}},\ \bibinfo {pages} {75103} (\bibinfo {year} {2019})}\BibitemShut {NoStop}%
\end{thebibliography}%
\end{document}


\title{Supplemental material: {Principal deuterium Hugoniot via Quantum Monte Carlo and $\Delta$-Learning}}
\author{Giacomo Tenti} 
\email{gtenti@sissa.it}
\affiliation{International School for Advanced Studies (SISSA),
Via Bonomea 265, 34136 Trieste, Italy}

\author{Kousuke Nakano} 
\email{kousuke\_1123@icloud.com}
\affiliation{Center for Basic Research on Materials, National Institute for Materials Science (NIMS), Tsukuba, Ibaraki 305-0047, Japan}

\author{Andrea Tirelli} 
\affiliation{International School for Advanced Studies (SISSA)\\
Via Bonomea 265, 34136 Trieste, Italy}

\author{Sandro Sorella} 
\affiliation{International School for Advanced Studies (SISSA),
Via Bonomea 265, 34136 Trieste, Italy}

\author{Michele Casula}
\affiliation{Institut de Minéralogie, de Physique des Matériaux et de Cosmochimie (IMPMC), Sorbonne Université, CNRS UMR 7590, MNHN, 4 Place Jussieu, 75252 Paris, France}

\date{\today}
\maketitle

\section{Computational details of QMC calculations}
The Variational Monte Carlo (VMC) and lattice regularized diffusion Monte Carlo (LRDMC)~{\cite{Casula2005}} calculations in this study were performed by the {\textsc{TurboRVB}} package~{\cite{Nakano2020}}.
The package employs a many-body WF ansatz $\Psi$ which can be written as the product of two terms, i.e.,
$
\Psi  =  \Phi _\text{AS} \times \exp J \,,
$
where the term $\exp J$ and $\Phi _\text{AS}$ are conventionally called Jastrow and antisymmetric parts, respectively.
The antisymmetric part is expressed as an Antisymmetrized Geminal Power (AGP) that reads:
$
{\Phi_{{\text{AGP}}}}\left( {{{\mathbf{r}}_1}, \ldots ,{{\mathbf{r}}_N}} \right) = {\hat A} \left[ {\Phi \left( {{\mathbf{r}}_1^ \uparrow ,{\mathbf{r}}_1^ \downarrow } \right)\Phi \left( {{\mathbf{r}}_2^ \uparrow ,{\mathbf{r}}_2^ \downarrow } \right) \cdots \Phi \left( {{\mathbf{r}}_{N/2}^ \uparrow ,{\mathbf{r}}_{N/2}^ \downarrow } \right)} \right],
$
where ${\hat A}$ is the antisymmetrization operator, and $\Phi \left( {{\mathbf{r}}_{}^ \uparrow ,{\mathbf{r}}_{}^ \downarrow } \right)$ is called the paring (geminal) function~\cite{Casula2003}. The spatial part of the geminal function is expanded over the Gaussian-type atomic orbitals (GTOs):
$
{\Phi}\left( {{{\mathbf{r}}_i},{{\mathbf{r}}_j}} \right) = \sum\limits_{a,b, l,m} {{f_{\left\{ {a,l} \right\},\left\{ {b,m} \right\}}}{\psi _{a,l}}\left( {{{\mathbf{r}}_i}} \right){\psi _{b,m}}( {{{\mathbf{r}}_j}} )}
\label{agp_expansion}
$
where ${\psi _{a,l}}$ and ${\psi _{b,m}}$ are primitive Gaussian atomic orbitals, their indices $l$ and $m$ indicate different orbitals centered on atoms $a$ and $b$, $i$ and $j$ are coordinates of spin up and down electrons, respectively, and ${{f_{\left\{ {a,l} \right\},\left\{ {b,m} \right\}}}}$ are the variational parameters.
The pairing function can be also written as
$
{\Phi }\left( {{{\mathbf{r}}_i},{{\mathbf{r}}_j}} \right) = 
\sum_{k=1}^M \lambda_k \phi_k(\mathbf{r}_i) \phi_k(\mathbf{r}_j)
$
with $\lambda_{k} > 0$, where $\phi_k(\mathbf{r})$ is a molecular orbital, i.e., $\phi_k(\mathbf{r}) = \sum_{i=1}^{L}{c_{i,k}{\psi_{i}}(\mathbf{r})}$ and $L$ is the total number of atomic orbitals.
When the paring function is expanded over $M$ molecular orbitals where $M$ is equal to half of the total number of electrons ($N/2$), the AGP coincides with the Slater-Determinant ansatz. In this work, we took $M=N/2$.
\vspace{2mm}

The Jastrow term is composed of one-body, two-body and three/four-body factors ($J = {J_1}+{J_2}+{J_{3/4}}$). The one-body and two-body factors are essentially used to fulfill the electron-ion and electron-electron cusp conditions, respectively, and the three/four-body factor is employed to consider further electron-electron correlations (e.g., electron-nucleus-electron). The one-body Jastrow is decomposed into the so-called homogeneous and inhomogeneous parts, i.e., ${J_1 = {J_1^{{\rm{hom}}}} + {J_1^{{\rm{inh}}}}}$.
The homogeneous one-body Jastrow factor is
$
{J_1}^{{\rm{hom}}}
\left( {{{\mathbf{r}}_1}, \ldots ,{{\mathbf{r}}_N}} \right) 
= 
\sum\limits_{i,I} \left({ - {{\left( {2{Z_I}} \right)}^{3/4}}u\left( {\left(2{Z_I}\right)^{1/4}\left| {{\mathbf{r}_{i}} - {{\mathbf{R}}_I}} \right|} \right)} \right) 
$
where ${{{\mathbf{r}}_i}}$ are the electron positions, ${{{\mathbf{R}}_I}}$ are the atomic positions with corresponding atomic number $Z_I$, and ${u\left( r \right)}$ is a short-range function containing a variational parameter $b$:
$
u\left( r \right) = \frac{b}{2}\left( {1 - {e^{ - r/b}}} \right).
\label{onebody_u}
$
The inhomogeneous one-body Jastrow factor ${J_1^{inh}}$ is represented as:

$
{J_1^{{\rm{inh}}}}\left( {{{\mathbf{r}}_1}, \ldots, {{\mathbf{r}}_N}} \right) =
\sum_{i=1}^N \sum_{a=1}^{N_\text{atom}} \left( {\sum\limits_{l} {M_{a,l} \chi_{a,l}\left( {{{\mathbf{r}}_i}} \right)} } \right),
$
where ${{{\mathbf{r}}_i}}$ are the electron positions, ${{{\mathbf{R}}_a}}$ are the atomic positions with corresponding atomic number $Z_a$, $l$ runs over atomic orbitals $\chi _{a,l}$ ({\it e.g.}, GTO) centered on the atom $a$, ${N_\text{atom}}$ is the total number of atoms in a system, and $\{ M_{a,l} \}$ are variational parameters.
The two-body Jastrow factor is defined as:
$
{J_2}\left( {{{\mathbf{r}}_1}, \ldots {{\mathbf{r}}_N}} \right) = \exp \left( {\sum\limits_{i < j} {v\left( {\left| {{{\mathbf{r}}_i} - {{\mathbf{r}}_j}} \right|} \right)} } \right),
$
where
$
v\left( r \right) = \frac{1}{2}r \cdot {\left( {1 - F \cdot r} \right)^{ - 1}}
$
and $F$ is a variational parameter. The three-body Jastrow factor is:
$
{J_{3/4}}\left( {{{\mathbf{r}}_1}, \ldots {{\mathbf{r}}_N}} \right) = \exp \left( {\sum\limits_{i < j} {{\Phi _{{\text{Jas}}}}\left( {{{\mathbf{r}}_i},{{\mathbf{r}}_j}} \right)} } \right),
$
and
$
{\Phi _{{\text{Jas}}}}\left( {{{\mathbf{r}}_i},{{\mathbf{r}}_j}} \right) = \sum\limits_{l,m,a,b} {g_{a,l,b,m}^{}\chi _{a,l}\left( {{{\mathbf{r}}_i}} \right)\chi _{b,m} \left( {{{\mathbf{r}}_j}} \right)},
$
where, as before, the indices $l$ and $m$ indicate different orbitals centered on
corresponding atoms $a$ and $b$. Here, the coefficients of the three/four-body Jastrow factor were set to zero for $a \neq b$; this significantly decreases the number of variational parameters and rarely affects variational energies.
In this study, we used a basis set of [4s2p1d] and [2s2p1d] GTOs for the antisymmetric part and Jastrow part, respectively. The exponents in the gaussian orbitals were optimized beforehand for our particular system and thermodynamic conditions. We verified that the optimal exponents do not vary significantly in the density and temperature range explored and thus we fixed their value for all the configurations in the dataset. 
For the antisymmetric part, we further reduced the number of variational parameters thanks to contracted orbitals. In particular, the WF was first initialized with the Density Functional theory (DFT) built-in code of the {\textsc{TurboRVB}} package at the $\Gamma$ point and then projected into a basis of $6$ hybrid orbitals (geminal embedded orbitals or GEOs)~{\cite{sorella2015geminal}}. The resulting geminal WF is then expressed as $\Phi \left(\mathbf{r}_i , \mathbf{r}_j \right) = \sum\limits_{a, b,\mu,\nu} {{f_{\left\{ {a,\mu} \right\},\left\{ {b,\nu} \right\}}}{\tilde{\psi} _{a,\mu}}\left( {{{\mathbf{r}}_i}} \right){\tilde{\psi} _{b,\nu}}( {{{\mathbf{r}}_j}} )}$ were the sum now runs over the GEOs $\tilde{\psi}_{a,\mu}$ (in general, for $a \neq b$, $\psi _{a,\mu} \neq \psi _{b,\mu}$ ).
The WF in the new basis was initialized by running a DFT-LDA calculation at the Baldereschi point $k = (0.25,0.25,0.25)$ (in crystal units).
Both the Jastrow factor and the antisymmetric part of the WF were optimized using the Hessian matrix method. 
Finally, only the variational parameters $f_{\left\{ {a,\mu} \right\},\left\{ {b,\nu} \right\}}$ corresponding to atoms closer than a cutoff $r_c = 4.0$~Bohr were optimized. The details of the constrained optimization technique are written in Ref.~{\onlinecite{2017BEC}}. For stable optimizations, we projected the AGP into $N/2$ molecular orbitals at each step~{\cite{2017BEC}}, such that the resulting optimized WF is equivalent to an optimal JSD WF. We found that the explicit optimization of the determinantal part of the WF is not only important for getting more accurate energies and forces, but also mitigates the self-consistent error \cite{Tiihonen2021, Nakano2022, Nakano2023} present in the JSD ansatz.
Total energies and forces are calculated at the VMC and the LRDMC levels with the optimized wavefunctions. The LRDMC calculations were performed by the original single-grid scheme~{\cite{Casula2005}} with the discretization grid size $a = 0.20$ Bohr. 
The resulting WF is a clear improvement with respect to JSD ansatz commonly employed in QMC, where only the Jastrow factor parameters are optimized, and the SD is defined with frozen DFT orbitals. This can be appreciated by comparing the reference state average potential energy based on our optimized workflow and the one calculated using a standard JSD WF (see Tab.\ref{tab:reference_state_energy comparison }). 
To obtain a statistically meaningful value of VMC and LRDMC forces with finite variance~{\cite{Nakano2021}}, the so-called
reweighting techniques are needed because the Hellmann--Feynman (HF) and Pulay terms may diverge when the minimum electron–nucleus distance vanishes and when an electronic configuration is close to the nodal surface, respectively~{\cite{Nakano2022}}. The infinite variance of the first term is cured by applying the so-called space-warp coordinate transformation (SWCT) algorithm~{\cite{Umrigar1989,Sorella2010,Claudia2016,Nakano2022}}, whereas that of the second term can be alleviated by modifying the VMC sampling distribution using a modified trial wave function that differs from the original trial wave function only in the vicinity of the nodal surface~{\cite{Attaccalite2008}}, which we dub the Attaccalite and Sorella (AS) regularization.
The AS regularization is not an optimal regularization for this purpose because it enforces a finite density of walkers on the nodal surface~{\cite{vanRhijn2021}}. Therefore, in this study, we employed the regularization technique recently proposed by Pathak and Wagner~{\cite{Pathak2020}} combined with mixed-averaged forces proposed by Reynolds~{\cite{Reynolds1986}}.

The final statistical noise on energies, forces and pressures was of the order of $7 \times 10^{-6}$~Ha/atom, $1$~mHa/Bohr and $0.05$~GPa, for VMC,  and $2 \times 10^{-5}$~Ha/atom, $2$~mHa/Bohr and $0.1$~GPa, for LRDMC.

\begin{table}[b]
    \centering
    \begin{tabular}{l|cc}
    \toprule
    & JSD    & optimized JSD  \\
    \midrule
    $e_{VMC}$ (Ha / atom )  & -0.58465(3) &  -0.58622(2)  \\
    $e_{LRDMC}$ (Ha / atom) & -0.58653(2) & -0.58660(2)  \\
     \bottomrule
    \end{tabular}
    \caption{Average VMC and LRDMC energy for the reference state, calculated on a Jastrow-Slater ansatz with Jastrow optimization and $4\times4 \times4$ k points integration and the optimized JSD WF obtained with our workflow at the Baldereschi point.}
    \label{tab:reference_state_energy comparison }
\end{table}

\section{MLP training and validation}
\subsection{Dataset construction}
\begin{figure}
     \centering
     \includegraphics[width = .7\textwidth]{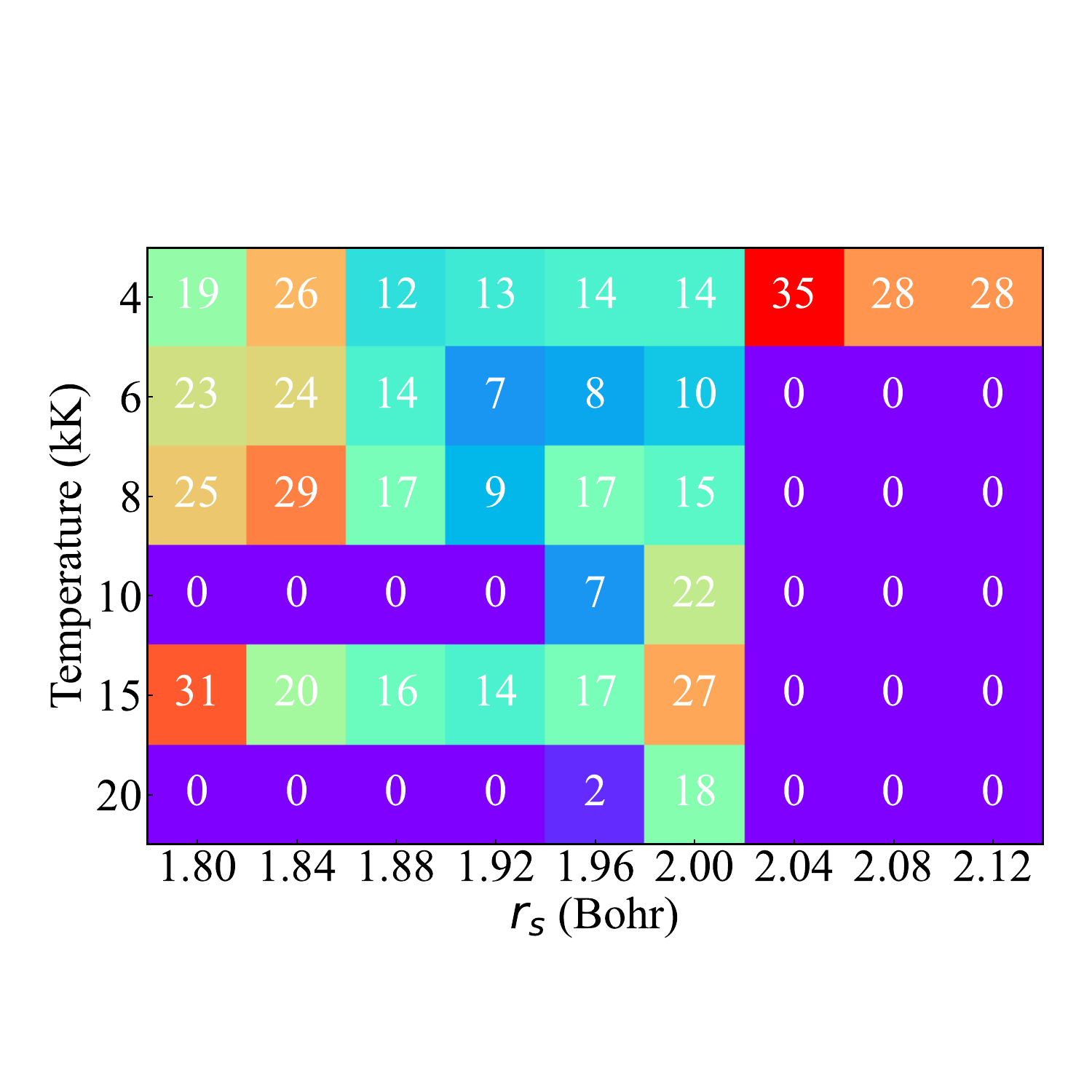}
     \caption{Distribution in temperature and Wigner-Seitz radius of the configurations belonging to the final dataset used. The exact number of configurations is also shown.}
     \label{fig:distribution dataset}
 \end{figure}

To construct our dataset, we performed a first set of ab initio DFT MD simulations with the PBE functional \cite{Perdew1981} on a system of $N = 128$ atoms for temperatures in the range $[4 $kK$, 20$kK$]$ and densities in the range $[1.80$ Bohr, $2.12 $ Bohr$]$, from which we extracted an initial set of decorrelated snapshots. We then added other configurations according to an active learning scheme: with a model trained using this first dataset we ran MD simulations and iteratively selected new points where the MLP performances were expected to be poor. In particular we did this by monitoring, for each unseen configuration, the quantity
\begin{align}
    \chi = \frac{1}{N} \sum_{i = 1}^N \min_{\mu \in \textit{training set}} K ( R_i , R_{\mu}) \label{eq: training set similarity}
\end{align}

where $K(R_i, R_{\mu})$ is the normalized SOAP kernel between the $i$-th local environment of the configuration $R_i$ and the $\mu$-th local environment  in the training set $R_{\mu}$. 
The number $\chi$ defined in \eqref{eq: training set similarity} gives a quantitative measure of "how far" the unknown configuration is from what is already included in the training set. In particular, we stopped adding configurations when $\chi$ did not drop under a fixed threshold of $0.80$ during the dynamics.
The final dataset comprised $561$ configurations of $128$ atoms in total. The final range of temperatures and Wigner-Seitz radii $r_s$ spanned by these configurations was $[4$ kK : $20$ kK $]$ and $[1.80$ Bohr : $2.12$ Bohr $]$, respectively. The distribution of these configurations in the $r_s - T$ space is shown in Fig. \ref{fig:distribution dataset}.

\subsection{Training details}
 For the training procedure we followed the strategy outlined in \cite[\S I.B]{Tirelli2022}. 
The energy is decomposed into local contributions of the form:
\[
e(R,  c_{\mu} ) = e_0 + e_{2body}(R,\{ \alpha_i \}) + \sum_{j = 1}^{N_{envs}} \beta_j K \left( R , \bar{R}_j \right) 
\] 
where  $e_{2body}$ is a pairwise term written as a linear combination of spline functions, $K(R,\bar{R}_j)$ is the SOAP kernel defined in \cite{Tirelli2022} between the local environment to be predicted and one of the $N_{envs}$ environments belonging to training set, and $(e_0,\{ \alpha_i \}, \{ \beta_j\} ) \equiv  c_{\mu} $ are the variational parameter to optimize.
The cost function $C$ employed is the regularized weighted sum of the MSE on the observables, \textit{i.e.} 
\[
C = C(c_\mu) = \alpha \mathrm{MSE}(E, \hat{E}(c_\mu)) +
\beta\mathrm{MSE}(F, \hat{F}(c_\mu)) + 
\gamma\mathrm{MSE}(P, \hat{P}(c_\mu))
+ \lambda ||c_\mu||^2,
\]
where $E$, $F$, $P$ and $\hat{E}$, $\hat{F}$, $\hat{P}$ are the vectors representing energy, forces and pressures obtained through QMC simulations and GKR, respectively. For the choice of the model hyperparameters, a cross-validation test led the following: 
\begin{itemize}
\item the cutoff radius used to compute local environments has been set to $r_c=4.0$ Bohr.
\item the term $e_{2body}$ was expressed as a linear combination of cubic splines defined on a grid of $10$ equally spaced points, extending from $r_{min} = 0.3$~Bohr and $r_{max} = 5.5$~Bohr.
\item the parameters $\alpha, \beta, \gamma$ and $\lambda$ determining the cost function $C(c_{\mu})$ have been set to $20, 10, 10^7$, $10^{-5}$ and   and $20, 10, 2 \times 10^7, 10^{-5}$ (in the appropriate inverse atomic units) for the VMC and LRDMC models respectively.
\item The number of environments selected among the full $128$ atoms configurations was $N_{envs} = 6000$.
\end{itemize}

\subsection{Models performance}
The performances of the models employed are measured through the root mean squared error (RMSE) on the observables on which the models were trained, avaluated on the test set. Such RMSEs are reported in Table~\ref{tab: model performances }. 
\begin{table}[h!]
    \centering
    \begin{tabular}{l|ccc}
    \toprule
          & $\textrm{RMSE}_E $ (mHa / atom) & $\textrm{RMSE}_f$ (mHa / Bohr) & $\textrm{RMSE}_p$ (GPa)\\
    \midrule
    VMC - PBE   & $0.26$  & $4.9$ & $0.76$ \\
    VMC - PBE (with FSC)  & $0.29$ & $5.1$ & $1.1$\\
    VMC - LDA &  $0.35$ & $5.5$ &  $0.80$\\
    LRDMC - PBE (with FSC) & $0.36$ & $5.5$ & $0.72$\\
    \bottomrule
    \end{tabular}
    \caption{Value of the RMSE on different observables as calculated on the test set for the final models used in the simulations.}
    \label{tab: model performances }
\end{table}

\subsection{Effect of different baselines}\label{sec: baselines}

\begin{figure*}[!h]
  \centering
  \includegraphics[width=0.8\linewidth]{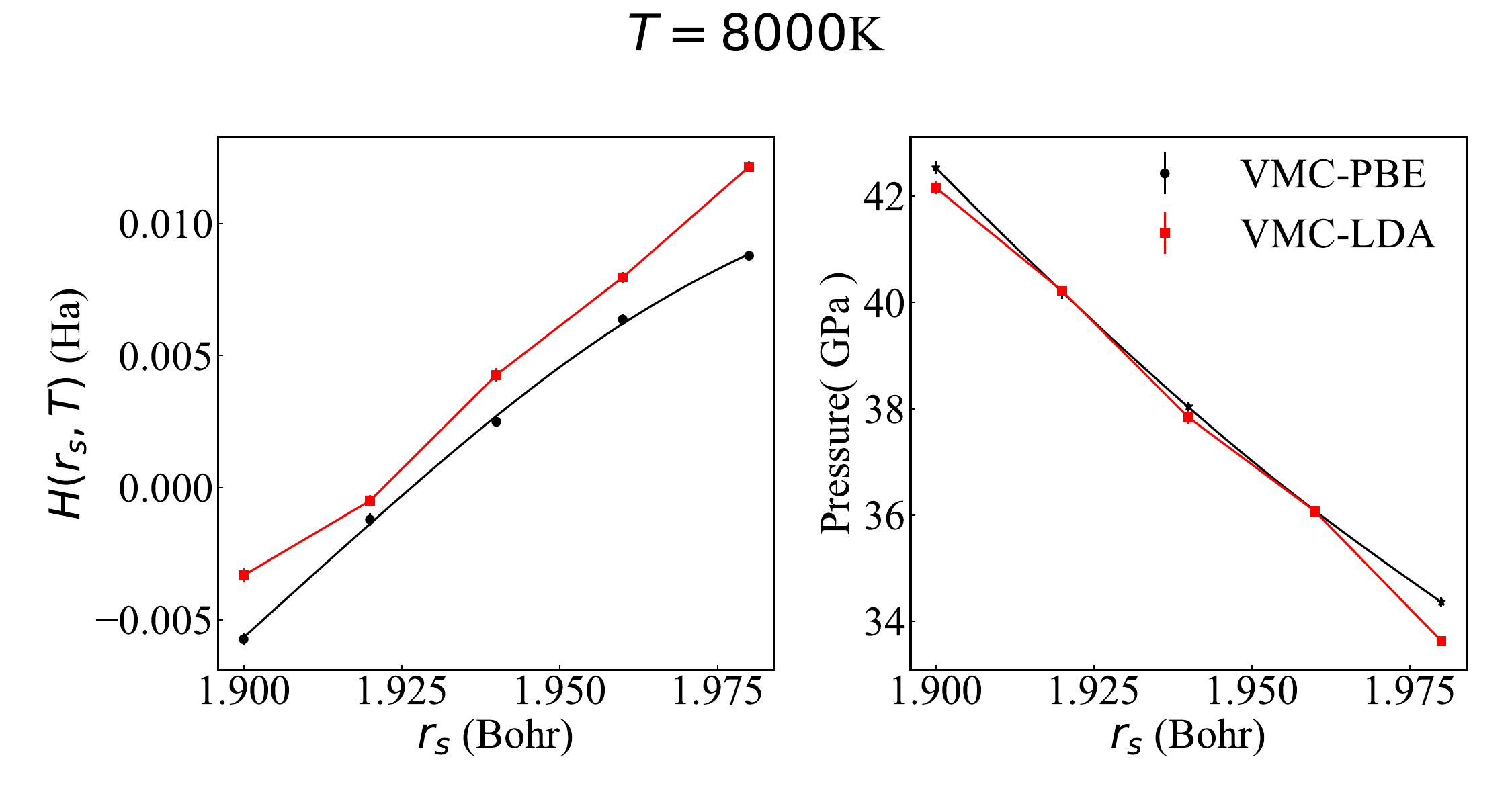}
\caption{Results for the Hugoniot function $H(r_s,T)$ and pressure for $T = 8$~kK with different MLPs trained on VMC data, using different baselines potentials.}
\label{fig: baseline}
\end{figure*}

In order to further validate the accuracy of an MLP, a common strategy is to compare the results of the dynamics obtained using the trained model with those obtained with the target ab initio method directly, at least for some small system sizes. 
In our case this is not an easy task, given the large computational time that would be needed for computing energies and forces at each step with QMC. An alternative way to establish the performances of the models is to look at the variance of the results obtained with MLPs trained using different baselines.
The Hugoniot function $H(\rho,T)$ and pressure at $T=8$~kK, calculated using models trained on the difference between VMC and two DFT functionals (i.e., PBE and LDA),  are shown in Fig.~\ref{fig: baseline}. We observe that the variability of the Hugoniot relative density and pressure is compatible with the value estimated from the RMSE (i.e., $\Delta_{\rho / \rho_0} = 0.06$ and $\Delta_p = 1$~GPa).

\section{Reference state calculations}

\begin{figure}
     \centering
     \includegraphics[width = .5\textwidth]{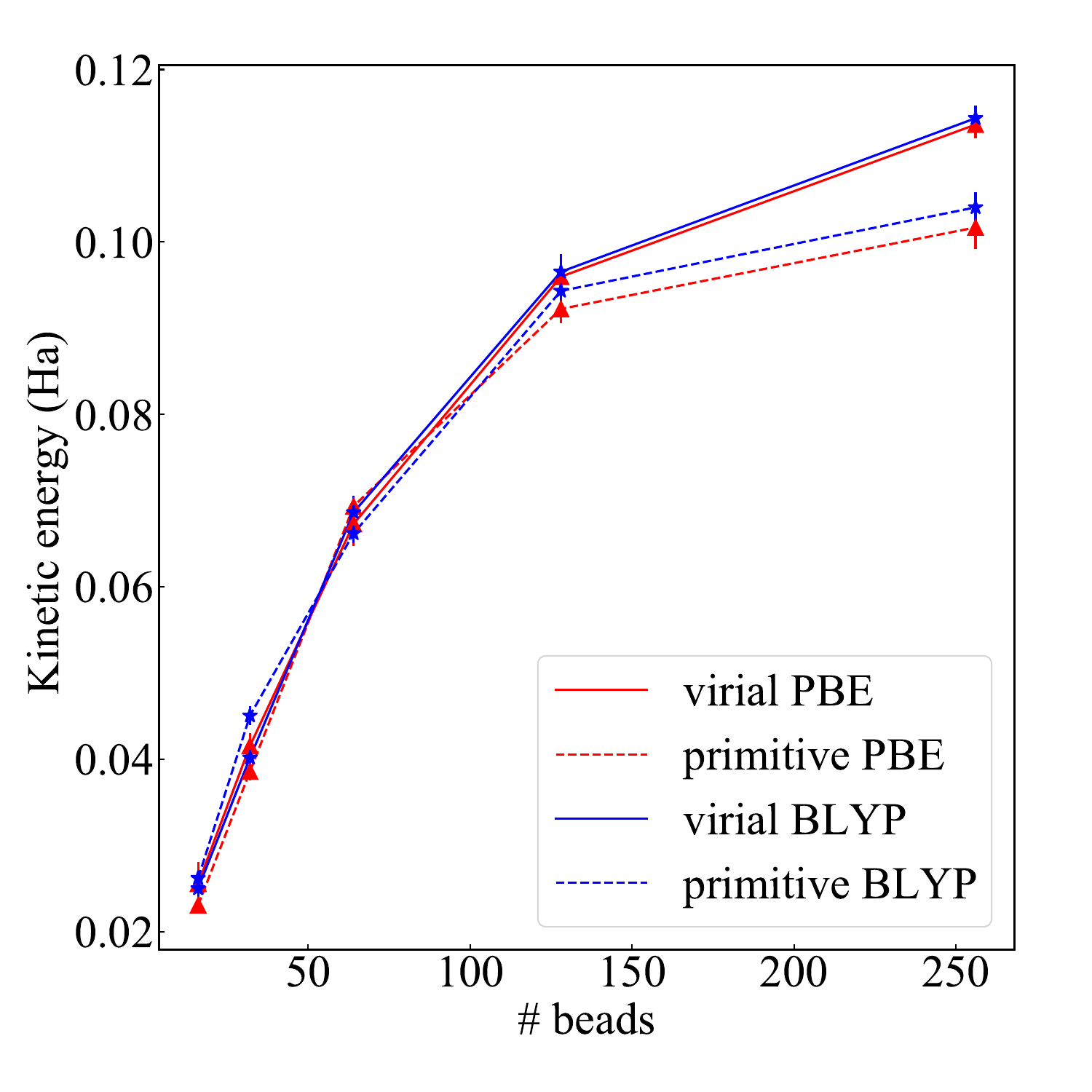}
     \caption{Convergence of virial and primitive estimators for the quantum kinetic energy, as computed with Eqs. \eqref{eq: quantum kin virial} \eqref{eq: quantum kin primitive}, with the number of replicas used in the PIMD simulation.}
     \label{fig:convergence_beads_kinE}
 \end{figure}

As explained in the main text, a crucial part in the numerical determination of the Hugoniot is to estimate the reference state energy per atom $e_0$ and pressure $p_0$. In particular, having a precise value of $e_0$ within the target method is important to take advantage of possible error cancellation effects and remove biases related to finite basis sets. We considered a system of $N = 64$ deuterium atoms at $T_0 = 22$~K and $\rho_0 = 0.167$ g/cm$^{-3}$ and ran a path integral Ornstein-Uhlenbeck molecular dynamics \cite{Mouhat2017} (PIOUD) simulation to account for quantum effects, which are required because of the light deuterium mass and low temperature. 
Forces and energy were calculated with DFT through the Quantum Espresso package\cite{Giannozzi2009,Giannozzi2017,Giannozzi2020}. We checked the dependence of thermodynamic quantities on the number of replicas (or beads) $M$ and on the choice of the DFT functional by studying the quantum kinetic energy $T$ for several values of $M$ using the BLYP \cite{Lee1988} and PBE functionals. In particular we considered two estimators for $T$, namely the virial and primitive (or Barker) estimator, given respectively by 
\begin{align}
    T_{M,vir} & = \frac{N}{2\beta} + \frac{1}{2M}\sum_{i=1}^{3N}\sum_{j=1}^M \left( x_i^{(j)} - \Bar{x}_i \right) \partial_{x_i^{(j)}}V \label{eq: quantum kin virial}\\
    T_{M,pri} & = \frac{3NM}{2\beta} - \frac{mM}{2\beta^2 \hbar^2} \sum_{j=1}^M \left( x_i^{(j)} - x_i^{(j-1)}\right)^2 \label{eq: quantum kin primitive}
\end{align}

where $M$ is the number of replicas used in the PIOUD simulation, $\mathbf{x}^{(j)} =\left( x_1^{(j)} , \dots , x_{3N}^{(j)}\right) $ are the coordinates of the system belonging to the $j$-th bead,  $\Bar{\mathbf{x}}_i = \frac{1}{M} \sum_{j=1}^M \mathbf{x}_i^{(j)}$ is the centroid position and $\beta = k_B T_0$. The results are shown in Fig.~\ref{fig:convergence_beads_kinE}. 

We noticed that a very large number of replicas is necessary for having a sufficiently converged result, while the value obtained with the two functionals is extremely similar for all values of $M$. At the end we chose to use the PBE functional and $M = 128$ replicas to have a reasonable trade-off between convergence and computational cost. For the DFT calculation we used a $60$ Ry plane waves cutoff and a $2\times 2 \times 2$ Monkhorst-Pack $k$ point mesh; for the dynamics we used a time step of $0.3$ fm and let the system thermalize for $0.3$ ps. We then extracted one configuration from a randomly chosen bead every $10$ MD steps, for a total of $N_{sample} = 170$ snapshots. Finally, the potential energy of these configurations was calculated using the appropriate method (PBE, VMC or LRDMC). We then estimated $e_0$ for each method as 

\begin{align}
    e_0 = \frac{1}{N} \left( \frac{1}{N_{sample}} \sum_{sample} E_{pot} \left(\mathbf{x}_i\right) + T_{256,pri}^{PBE} \right)
\end{align}

using the value of the primitive estimator at $M = 256$ beads as the best guess for the converged value of the kinetic energy. 
The approximation for the potential energy was checked by running PBE simulations on this set and confirming that the "true" mean value (as calculated by averaging over the beads and the trajectory) was consistent with our estimate obtained by averaging over the sample.

\section{Finite size corrections}

\begin{figure*}[!h]
\centering
\includegraphics[width = 1.0\textwidth]{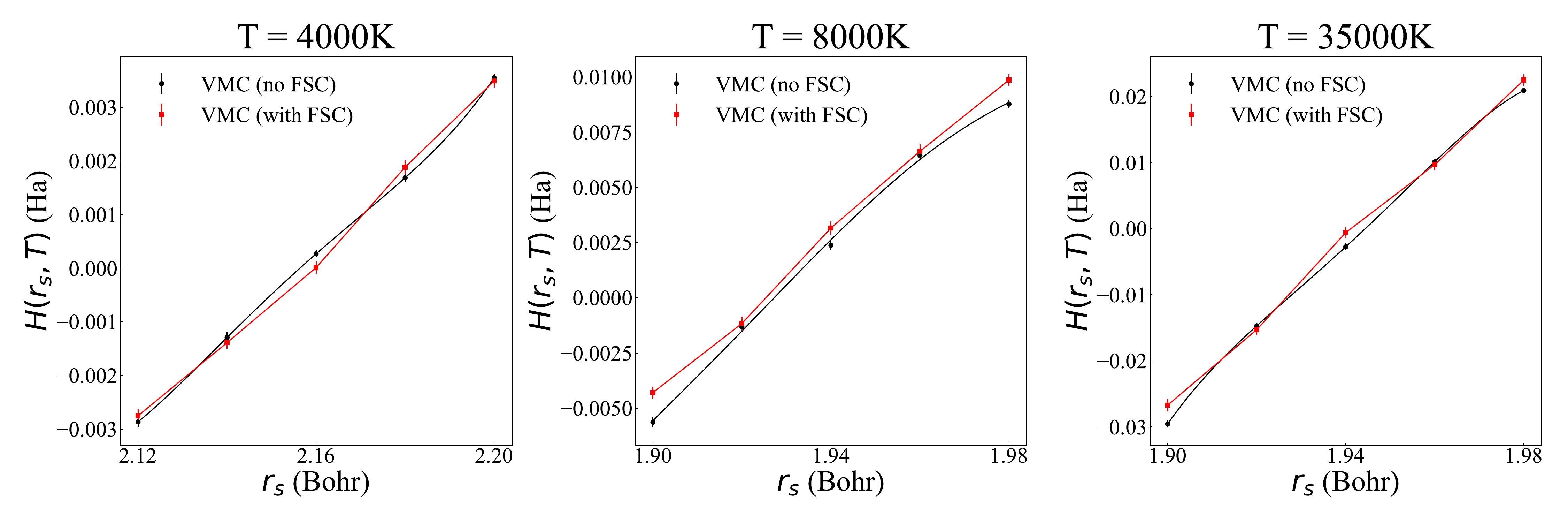}
\caption{Hugoniot functions obtained using two MLPs trained on the difference between PBE and VMC with and without finite size corrections respectively, for $T = 4$~kK, $T = 8$~kK and $T = 35$~kK.}
\label{fig: convergence finite size corrections}
\end{figure*}

In this section we investigate the effect of finite size corrections (as estimated using the KZK functional \cite{Kwee2008}) on our results. In Fig.~\ref{fig: convergence finite size corrections} we show the Hugoniot function ( at $T = 4$~kK, $T = 8$~kK and $T = 35$~kK) given by two models trained on VMC and VMC with finite size corrections respectively, both with a PBE baseline potential. The difference between the two turns out to be similar to the prediction error discussed in Sec. \ref{sec: baselines}, for the system size used in the simulations (i.e., $N = 128$). 
At the end we chose to apply finite size correction only for the model trained with LRDMC data. 

\section{Finite temperature DFT simulations}

\begin{figure*}[!ht]
    \centering
    \begin{subfigure}{1\textwidth}
  \centering
  \includegraphics[width=0.8\linewidth]{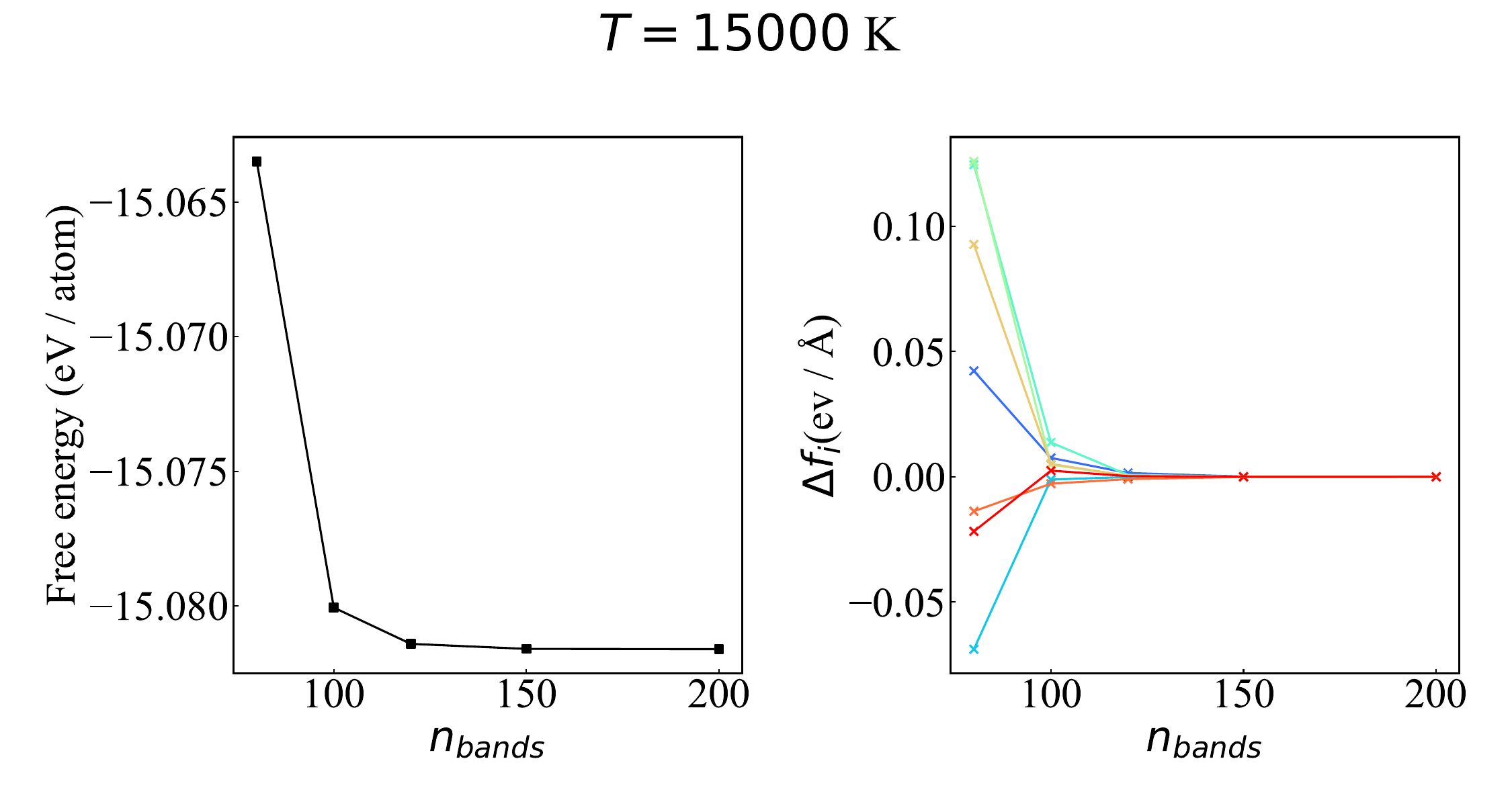}
  \caption{}
  \label{subfig: convergence with bands T = 15kK}
\end{subfigure}

    \begin{subfigure}{1\textwidth}
  \centering
  \includegraphics[width=0.8\linewidth]{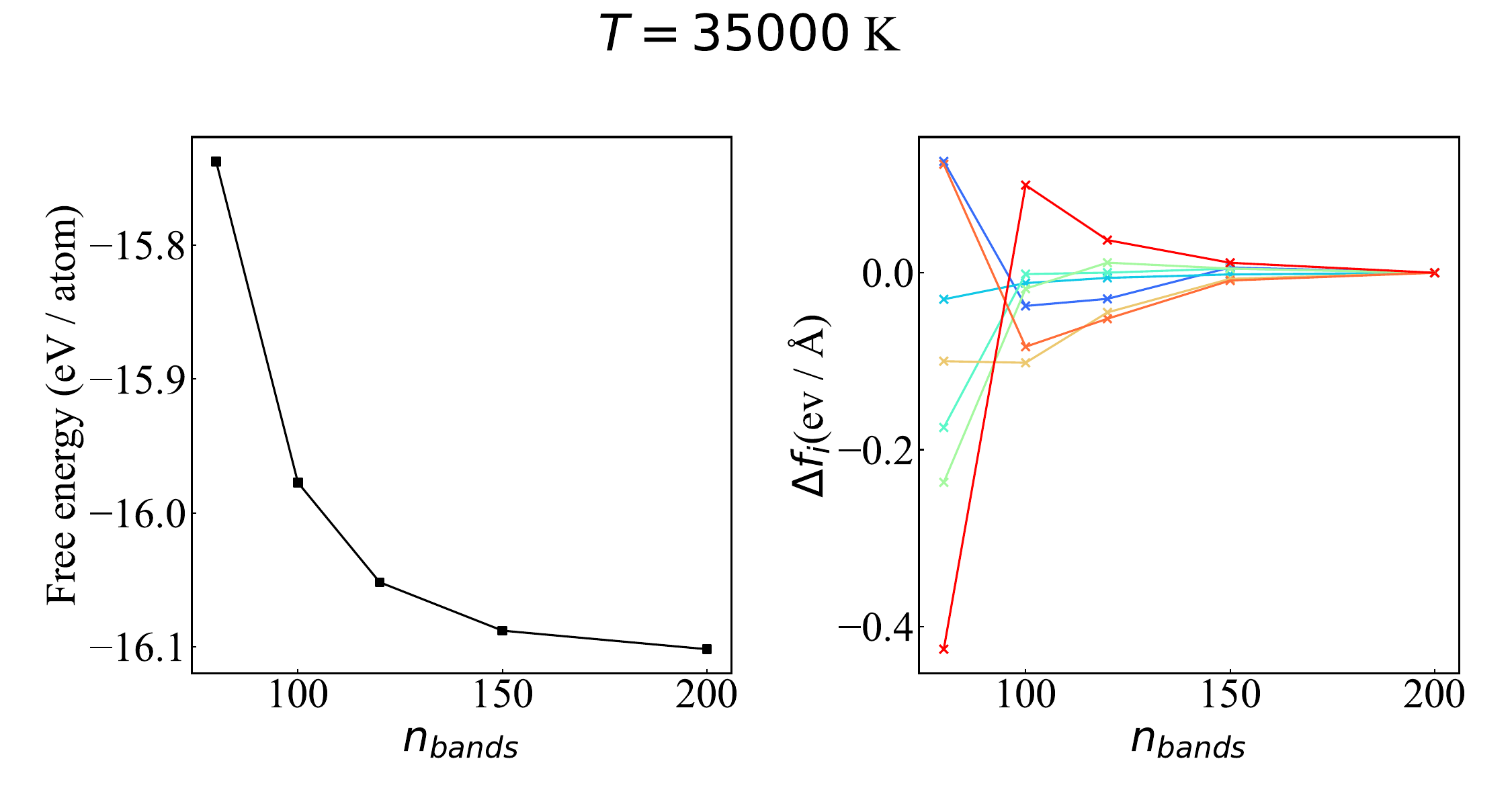}
  \caption{}
  \label{subfig: convergence with bands T = 35kK}
\end{subfigure}

\caption{Convergence of the free energy $A = E - TS$  and some force components vs the number of bands calculated with DFT-PBE for $T = 15$~kK (\ref{subfig: convergence with bands T = 15kK}),  and $T= 35$~kK (\ref{subfig: convergence with bands T = 35kK}). For the force components the difference $\Delta f_i$ between the force obtained with $N = n_{bands}$ bands and $N = 200$ is shown. Bands are occupied using the Fermi-Dirac distribution at the appropriate temperature. }
\label{fig: convergence finite temperature DFT }
\end{figure*}

Using Mermin's extension of the Hohenberg and Kohn theorems to non-zero temperature \cite{Mermin1965} we can treat finite temperature electrons in DFT by appropriately occupying the bands of the system according to the Fermi-Dirac distribution and minimizing the Helmholtz free energy functional $A = E - TS$. In this work we performed finite temperature DFT (FT-DFT) simulations to obtain the PBE Hugoniot and estimating the effect on the QMC Hugoniot. In both cases we used the zero temperature PBE functional for the simulations. Even if this is not rigorous, recent FT-DFT results on the Hugoniot using a temperature dependent GGA functional \cite{Karasiev2019} have shown that for $T \lesssim 40$~kK this approximation provides consistent results. 
In Fig. \ref{fig: convergence finite temperature DFT } we show the convergence of the free energy and some force components with the number of bands calculated for the FT-PBE functional at two values of temperature. In the simulations we decided to use $120 $ bands for $T = 10$~kK,  $T = 15$~kK  and $150$ bands for $T= 20$~kK  and $T = 35$~kK. 

\section{Comparison between equations of state}
\begin{figure}[!h]
    \centering
    \includegraphics[width = 0.6\textwidth]{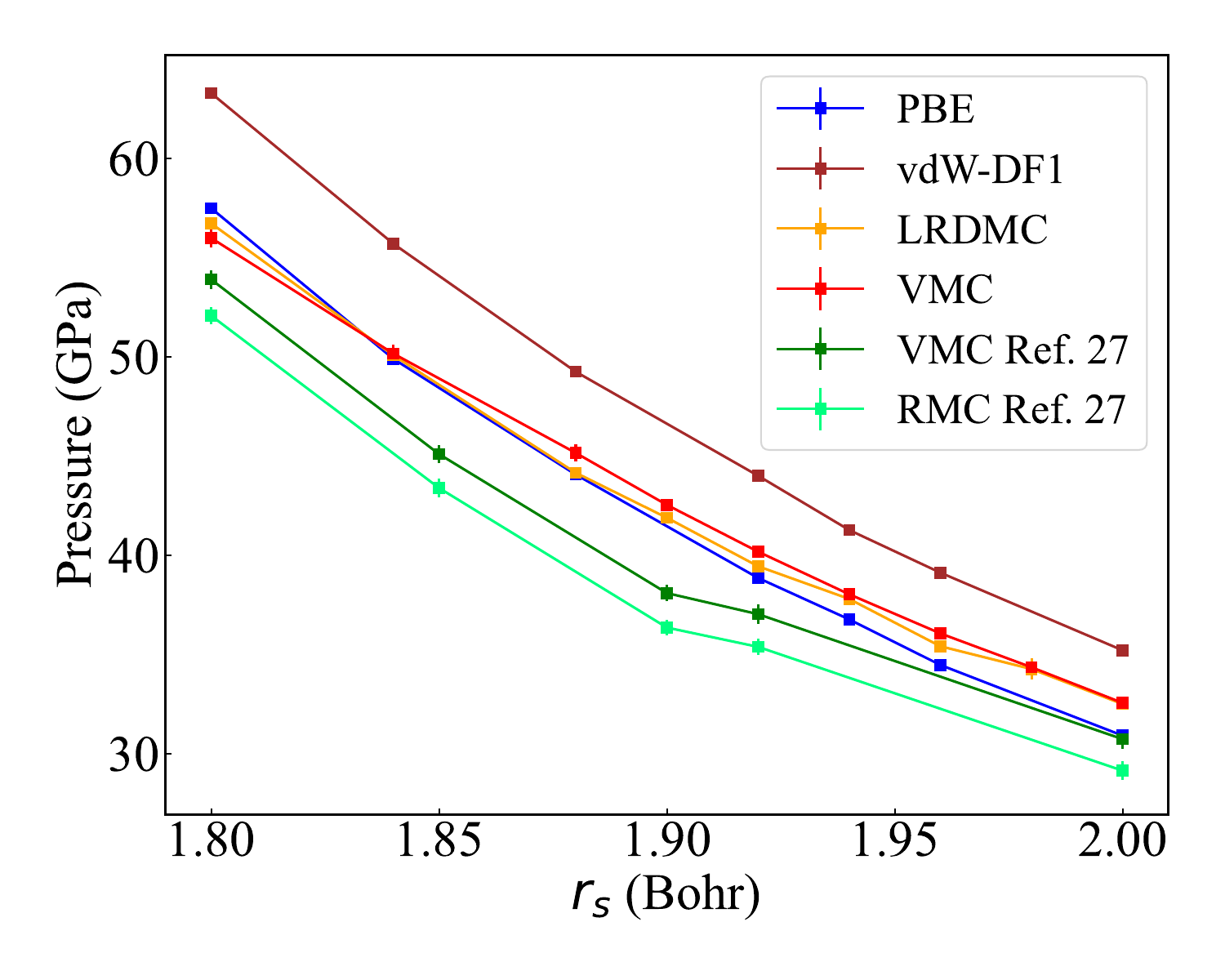}
    \caption{Average pressure vs $r_s$ for $T = 8$~kK. Results obtained with our MLPs are shown together with the ones reported in \cite{Ruggeri2020} and those calculated with the PBE and vdW-DF1 functionals.}
    \label{fig: Comparison pressures Ruggeri} 
\end{figure}

\begin{figure}[!h]
    \centering
    \includegraphics[width = 0.6\textwidth]{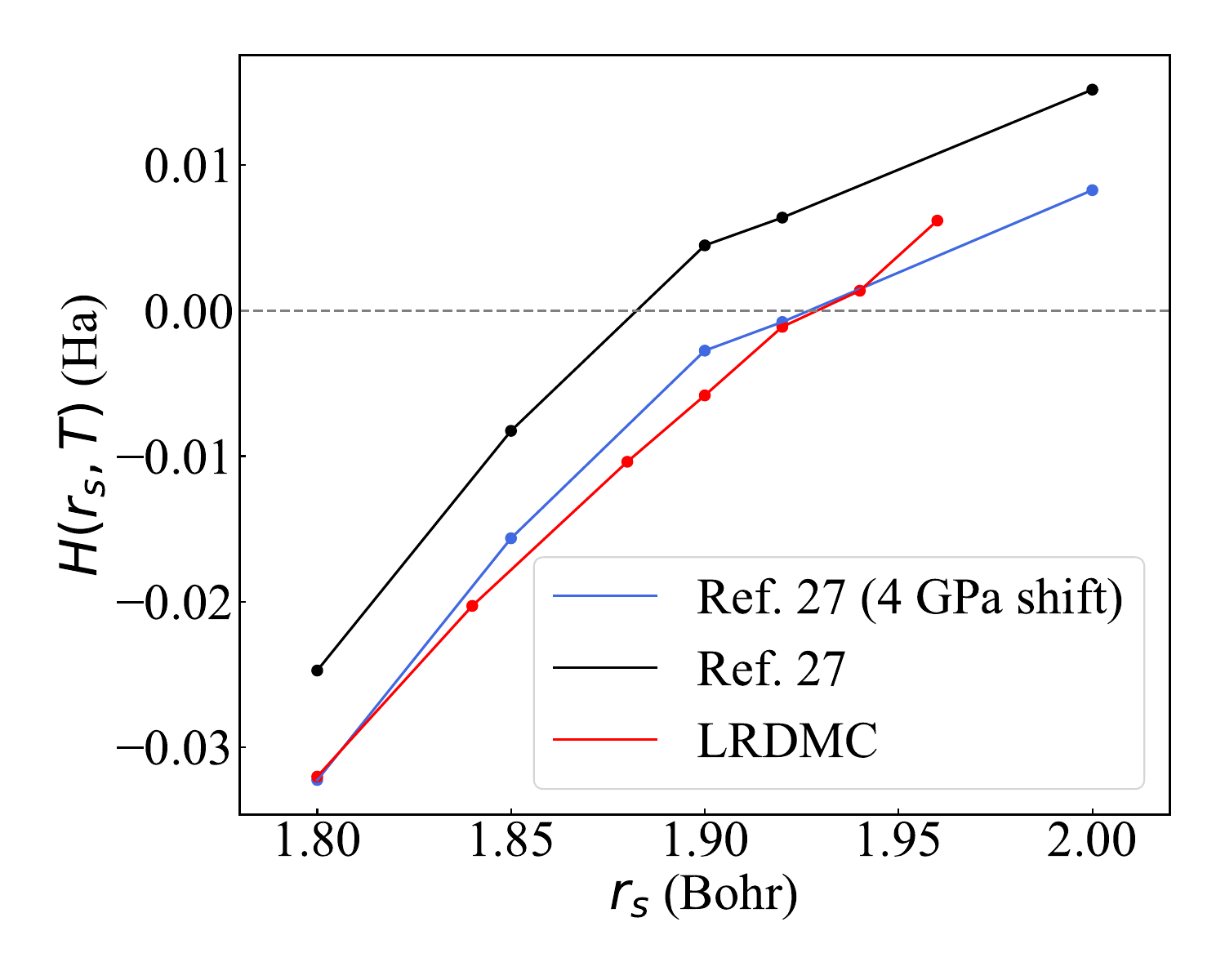}
    \caption{LRDMC (this work) and RMC (Ref.\cite{Ruggeri2020}) Hugoniot curves for $T = 8$~kK. For the latter, we also show the result of applying a costant $4$~GPa positive pressure shift.}
    \label{fig: Comparison hugoniot Ruggeri} 
\end{figure}

\begin{figure}[!h]
    \centering
    \includegraphics[width = 1.0\textwidth]{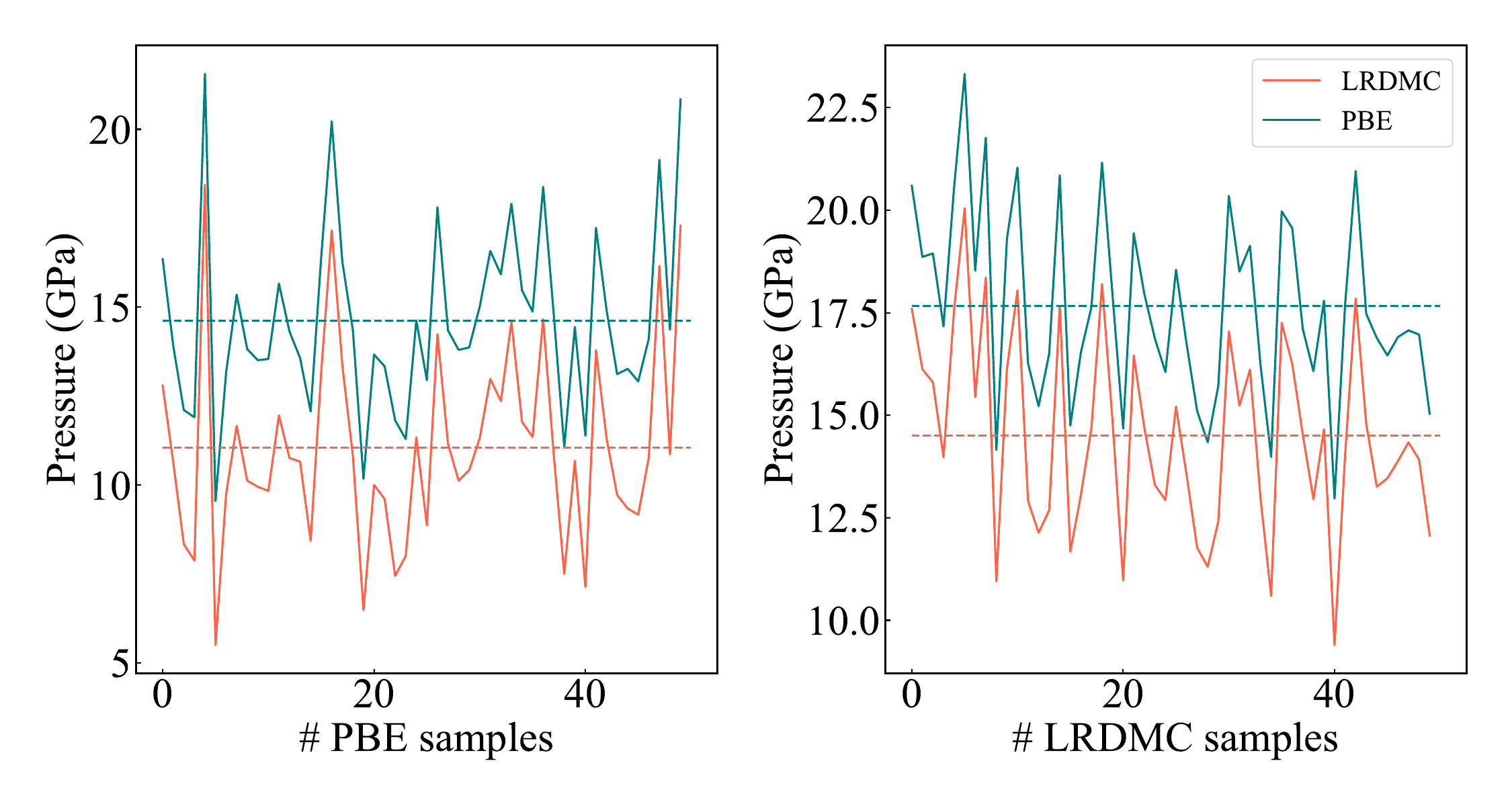}
    \caption{Virial pressure calculated by DFT-PBE and by the LRDMC MLP, on sampled trajectories obtained using the two methods, at thermodynamic conditions $r_s = 1.92$~Bohr and $T = 8$~kK.}
    \label{fig: trajectory sample} 
\end{figure}

The equations of state at $T = 8$~kK reported in Ref.~\cite{Ruggeri2020} for both variational and reptation Monte Carlo are shown in Fig.~\ref{fig: Comparison pressures Ruggeri}, together with the VMC-MLP, LRDMC-MLP (with PBE baseline) and the ab initio PBE and vdW-DF1 ones. From this figure we can observe a $\sim4$ GPa discrepancy between the pressure estimated with our MLPs and the one in Ref.~\cite{Ruggeri2020}, which is one of the causes of the slightly different Hugoniot position. In fact, by shifting the data reported in \cite{Ruggeri2020} by this quantity, we can see an almost perfect match of the Hugoniot positions (Fig.~\ref{fig: Comparison hugoniot Ruggeri}).
In our case, the VMC and LRDMC pressures, which we used for training, were calculated using the adjoint algorithmic differentiation method to obtain directly the derivative of the total energy with respect to the cell parameters. As mentioned in the main text, the self consistency error can, in principle, spoil the consistency between the pressure and the potential energy surface of the system. We mitigated this by directly optimizing the WF nodal surface, as previously described. Nevertheless, we identified the pressure as a particularly sensitive quantity, difficult to reliably obtain, and we took particular care when including it in the training. 
In Ref.~\cite{Ruggeri2020} a virial estimator was used, which can in principle produce discrepancies of the order of magnitude observed here, as shown in Ref. \cite{Clay2019}. 
We also point out that both our models and the vdW-DF1 functional predict pressures slightly larger than those given by the PBE in the proximity of the Hugoniot position, in contrast with the EOS reported in \cite{Ruggeri2020}, that predicts lower pressures. 
If it is true that QMC pressures are consistently lower than PBE ones on a given configuration, here we found that the different dynamics of the system samples different points in the phase space, such that the resulting average pressure is, in some cases, even larger for our MLPs. To demonstrate this, we extracted two samples of configurations from an ab initio PBE and a LRDMC-MLP trajectory, respectively, and then computed both PBE pressures and $\Delta$ corrections. The results show how the increased PBE pressures for the LRDMC sampled trajectory can compensate the negative  contribution given by the MLP correction (Tab.~\ref{tab:trajectory sample PBE vs LRDMC} and Fig.~\ref{fig: trajectory sample}).

\begin{table}[h!]
    \centering
    \begin{tabular}{l|cc}
    \toprule
   Virial pressure (GPa)  & PBE sample \qquad& LRDMC sample \\
    \midrule
PBE & $14.6(4)$ & $17.7(4)$ \\
LRDMC & $11.1(4)$ & $14.5(5)$\\
$\Delta$ correction & $-3.57(4)$ & $-3.17(4)$\\
     \bottomrule
    \end{tabular} 
    \caption{Average value of the virial pressure calculated by DFT-PBE and by the LRDMC MLP, on sampled trajectories obtained using the two methods, at thermodynamic conditions $r_s = 1.92$~Bohr and $T = 8$~kK. The average $\Delta$ correction is also reported.}
    \label{tab:trajectory sample PBE vs LRDMC}
\end{table}

\newpage
\bibliography{Hugoniot_paper.bib}